\documentclass[useAMS,usenatbib]{mn2e}

\usepackage{url}
\usepackage{graphicx}
\usepackage{amsmath}
\usepackage{amssymb}
\usepackage{wasysym}
\usepackage{epstopdf}
\usepackage{natbib}
\usepackage{multirow}
\usepackage{booktabs}

\title[Bow-shock evolution along the orbit]{Effect of an isotropic outflow from the Galactic centre on
the bow-shock evolution along the orbit}
\author[M. Zaja\v{c}ek et al.]{M. Zaja\v{c}ek$^{1,2,3}$\thanks{E-mail:
zajacek@ph1.uni-koeln.de (MZ)}, A. Eckart$^{2,1}$, V. Karas$^{3}$, D. Kunneriath$^{3}$,\and B. Shahzamanian$^{2,1}$, N. Sabha$^{2}$, K. Mu\v{z}i\'c$^{4,5}$, M. Valencia-S.$^{2}$\\
$^{1}$Max-Planck-Institut f\"ur Radioastronomie (MPIfR), Auf dem H\"ugel 69, D-53121 Bonn, Germany,\\
$^{2}$I. Physikalisches Institut der Universit\"at zu K\"oln, Z\"ulpicher Strasse 77, D-50937 K\"oln, Germany,\\
$^{3}$Astronomical Institute, Academy of Sciences, Bo\v{c}n\'{\i}~II 1401, CZ-14131~Prague, Czech Republic,\\ 
$^{4}$European Southern Observatory, Alonso de C\'ordova 3107, Casilla 19001, Santiago 19, Chile,\\
$^{5}$Nucleo de Astronom\'ia, Facultad de Ingenier\'ia, Universidad Diego Portales, Av. Ejercito 441, Santiago, Chile.
}
\begin{document}

\date{Accepted 2015 October 7. Received 2015 October 7; in original form 2015 May 26}

\pagerange{\pageref{firstpage}--\pageref{lastpage}} \pubyear{2015}

\maketitle

\label{firstpage}

\begin{abstract}

 Motivated by the observations of several infrared-excess bow-shock sources and proplyd-like objects near the Galactic centre, we analyse the effect of a potential outflow from the centre on bow shock properties. We show that due to the non-negligible isotropic central outflow the bow-shock evolution along the orbit becomes asymmetric between the pre-peribothron and post-peribothron phases. This is demonstrated by the calculation of the bow-shock size evolution, the velocity along the shocked layer, the surface density of the bow-shock, and by emission-measure maps close to the peribothron passage. Within the ambient velocity range of $\lesssim 2000\,{\rm km\, s^{-1}}$ the asymmetry is profound and the changes are considerable for different outflow velocities. As a case study we perform model calculations for the Dusty S-cluster Object (DSO/G2) as a potential young stellar object that is currently being monitored and has passed the pericentre at $\sim 2000$ Schwarzschild radii from  the supermassive black hole (Sgr~A*) in 2014. We show that the velocity field of the shocked layer can contribute to the observed increasing line width of the DSO source up to the peribothron. Subsequently, supposing that the line emission originates in the bow shock, a decrease of the line width is expected. Furthermore, the decline of the bow-shock emission measure in the post-peribothron phase could help to reveal the emission of the putative star. The dominant contribution of circumstellar matter (either inflow or outflow) is consistent with the observed stable luminosity and compactness of the DSO/G2 source during its pericentre passage.

\end{abstract}


\section{Introduction}

The Galactic centre of the Milky Way contains a large number of young stars ($\sim 10$ -- $100\,{\rm Myr}$) that orbit the compact radio source Sgr~A* associated with the supermassive black hole of $\sim 4\times 10^6\,M_{\odot}$ \citep[hereafter denoted as SMBH,][]{2010RvMP...82.3121G,2005bhcm.book.....E,2003ApJ...586L.127G}. These stars of spectral type O/B are concentrated partially in the coherent clockwise disk structure with a sharp inner edge at $\sim 0.03\,{\rm pc}$ and extending out to $0.5\,{\rm pc}$ \citep{2003ApJ...590L..33L}. High-velocity stars of spectral type B, which have orbital periods of $\lesssim 300\,{\rm yr}$, are located in the S-cluster in the innermost $0.05\,{\rm pc}$  with isotropic orientation of orbits \citep{1996Natur.383..415E, 1997MNRAS.284..576E, 2009ApJ...692.1075G,2012A&A...545A..70S} and high eccentricities approximately following the thermal distribution, $f(e)\mathrm{d}e \approx 2e\mathrm{d}e$.  Some young stars of spectral type O/WR do not form any clear kinematic structure \citep{2010RvMP...82.3121G} and they have rather random orbital orientations \citep{2014A&A...567A..21S}.

\begin{figure}
  \centering
  \includegraphics[width=0.5\textwidth]{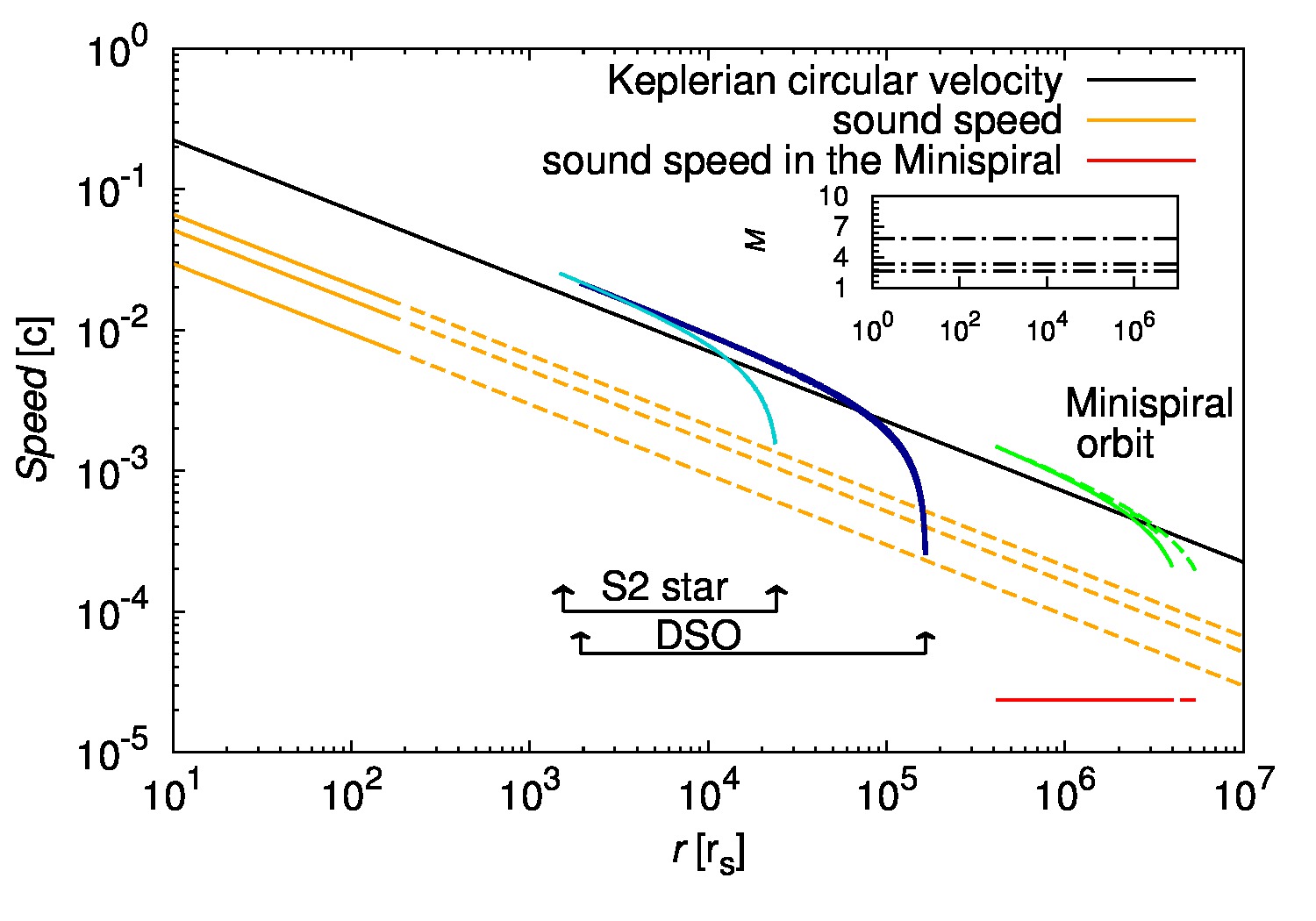}
  \caption{Comparison of Keplerian orbital speed and sound speed towards the Galactic centre, both expressed in terms of the speed of light. Three lines for the sound speed (orange) stand for possible different temperatures of electron and ion components. The dashed lines mark the extrapolation of the original fit (solid lines). The distance-orbital speed plots for the DSO and S2 are also labelled and the arrows denote the distance range for both objects. The estimate of the sound speed for the gas phase of the Minispiral region is marked by the red line. The set of green lines corresponds to the estimate of the orbital velocity for three prominent streams (arms) of the Minispiral. The plot inset shows the estimates of the Mach number $M$, $M=v_{{\rm orb}}/c_{{\rm s}}$.}
  \label{fig_velocity_comparison}
\end{figure}

The radius of the sphere of the gravitational influence of the supermassive black hole depends on the central black hole mass $M_{\bullet}$ and the stellar velocity dispersion $\sigma$ \citep{2013degn.book.....M} in the following way,

\begin{equation}
r_{\rm{SI}}\approx 1.7 \left(\frac{M_{\bullet}}{4.0\times 10^6\,\rm{M_{\odot}}}\right)\left(\frac{\sigma}{100\,\rm{km\,s^{-1}}}\right)^{-2}\,\rm{pc},
\label{eq_sphereofinfluence}
\end{equation}
This coincides approximately with the radius of the central cavity, which is filled with ionised hot and sparse gas mostly supplied by stellar winds of massive OB stars. The density and temperature profiles of this ambient corona are fitted by semi-analytical radial profiles based on the model of radiatively inefficient accretion flows \citep{1999MNRAS.303L...1B,2006ApJ...636L.109B,2011ApJ...738...38B},

\begin{align}
n_{{\rm a}} &\approx n_{{\rm a}}^0\left(\frac{r}{r_{{\rm s}}}\right)^{-1}\,, \label{eq_density}\\
T_{{\rm a}} &\approx T_{{\rm a}}^0\left(\frac{r}{r_{{\rm s}}}\right)^{-1}\,, \label{eq_temperature}  
\end{align}
where $r_{{\rm s}}$ is the Schwarzschild radius ($r_{{\rm s}} \equiv 2GM_{\bullet}/c^2 \doteq 2.95\times 10^5\,M_{\bullet}/M_{\odot}\,{\rm cm}$). The normalisation parameters are set to $n_{{\rm a}}^{0}=1.3\times 10^7\,{\rm cm}^{-3}$ and $T_{{\rm a}}^0=9.5 \times 10^{10}\,{\rm{K}}$. At a small distance from the SMBH, electrons are expected to be decoupled from ions. The ion temperature $T_{{\rm i}}$ can be $\sim 1$--$5$ times higher than the electron temperature $T_{{\rm e}}$ as was found by comparing MHD simulations with VLBI millimetre data \citep{2010ApJ...717.1092D}. This is also expressed in Fig. \ref{fig_velocity_comparison} by three lines for the sound speed profile corresponding to the temperature of $T_{{\rm e}}$, $3\,T_{{\rm e}}$, and $5\,T_{{\rm e}}$.

The profile for the sound speed in Fig. \ref{fig_velocity_comparison} is a simple extrapolation of the semi-analytic fits, eqs. \eqref{eq_density} and \eqref{eq_temperature}, to VLBI millimetre measurements of the emission in the inner $\sim 100\,r_{\rm{s}}$, while the orbits of the observed S-stars lie at least an order of magnitude further from the SMBH. However, the extrapolation of the one-dimensional fit of \textit{Chandra} X-ray measurements \citep{2004ApJ...613..322Q} towards smaller distances leads to similar values of density and temperature in the region of our interest (a factor of $3$ difference, see \citet{2012ApJ...759..130P} for discussion). In any case the density increase towards the centre naturally leads to higher ambient ram pressure, $P_{\rm{a}}=\rho_{\rm{a}}v_{\star}^2$, acting on the outflows of propagating stars.

When the stellar motion with respect to the surrounding environment is supersonic, bow-shock structures are formed. The distance range, where the Keplerian orbital velocity $v_{{\rm orb}}/c=\sqrt{2}/2(r/r_{\rm s})^{-1/2}$ is greater than the sound speed $c_{{\rm s}}\approx \sqrt{k_{{\rm B}}T_{{\rm a}}/(\mu m_{{\rm H}})}$, calculated from profiles given by eqs. \eqref{eq_density} and \eqref{eq_temperature}, may be inferred from the comparison in Fig. \ref{fig_velocity_comparison}, where we plot both velocities (in units of speed of light $c$) as functions of the distance $r$ from the SMBH. The motion seems to be supersonic for a large span of distances, from $\sim 10\,r_{{\rm s}}$ up to $\sim 10^6\,r_{{\rm s}}$, where the orbits of the S-stars lie \citep[see also][]{2012ApJ...759..130P}. In fact, the radial profile of Mach number, $M\equiv v_{\rm{orb}}/c_{\rm{s}}$, is approximately flat, since both the orbital velocity and the sound speed fall off as $(r/r_{{\rm s}})^{-1/2}$ for the assumed ambient temperature profile adopted here, which leads to $M \approx (GM_{\bullet}\mu m_{{\rm H}}/k_{{\rm B}} T_{{\rm a}}^0 r_{{\rm s}})$ and approximate values of $2.6$, $3.4$, and $5.8$ (see also the plot inset in Fig. \ref{fig_velocity_comparison}).    

Nowadays we have a great deal of observational information for the interaction of wind-blowing stars with their surroundings near the Galactic centre \citep[see e.g.,][]{2014A&A...567A..21S}. The sources IRS~1W, IRS~3, IRS~5, IRS~8, IRS~10W, and IRS~21, which interact mainly with the warm ionised gas in the Minispiral region \citep{2002ApJ...575..860T,2005ApJ...624..742T,2005A&A...433..117V,2013A&A...557A..82B,2014A&A...567A..21S}, are clear evidence of bow-shock formation in the near- and mid-infrared parts of the spectrum. The Minispiral region is in general colder and denser than the radial profiles (eqs. \eqref{eq_density} and \eqref{eq_temperature}) of the ambient medium would imply at its distance of $\sim 10^6\,r_{{\rm s}}$ \citep{2012A&A...538A.127K}. This can be understood in the framework of a multiphase environment where cool and hot phases co-exist \citep{2014MNRAS.445.4385R}. The ionised component of the high-density regions of the northern and eastern arm has the mean electron temperature of $\overline{T}_{\rm e}\approx 6 \times 10^3\,{\rm K}$, corresponding to a sound speed of $\sim 7\,{\rm km\,s}^{-1}$, whereas the Keplerian circular velocity at this distance is of the order of a few $100\,{\rm km\, s}^{-1}$. Hence, unless the sources are co-moving with the Minispiral streams, the relative velocities are typically supersonic.

 Bow-shock sources appear to be extended and brighter at longer wavelengths, which is in agreement with the dust emission in the bow shocks. \citet{2006ApJ...642..861V} showed that the peak of the continuum emission of typical bow-shock sources lies in the wavelength range $\sim 4$ to $10\,\mu {\rm m}$, corresponding to the dust temperature of a few $100\,{\rm K}$. Moreover, \citet{2013A&A...557A..82B} pointed out the strong polarisation degree of IRS 1W and 21 in the NIR wavelengths that increases towards longer wavelengths, which highlights the importance of dust for explaining the emission properties of bow-shock sources in this region. Besides the continuum emission, bow-shock sources associated with young stellar objects in HII regions are observed in hydrogen recombination lines \citep{2000AJ....119.2919B}. \citet{2013ApJ...768..108S} show that the dense, shocked layer typically associated with a young star of T Tauri type may significantly contribute to observed Br$\gamma$ emission of the dusty S-cluster object (DSO/G2), which is intensively monitored.      

The detection and analysis of the comet-shaped sources named X3 and X7 indicates that they interact with the outflowing medium from the direction of the Galactic centre \citep{2007A&A...469..993M,2010A&A...521A..13M}. This outflow originates within the central $\sim 0.2\,{\rm pc}$ from the direction of Sgr~A* and its estimated terminal velocities are of the order of $\sim 1000\,{\rm km\, s}^{-1}$. The discovery of a new MIR bow shock suggests the presence of this outflow at even larger distances, $\sim 0.68\,{\rm pc}$ \citep{2014IAUS..303..150S}. 

The presence of an outflow from the Galactic centre is consistent with the low-luminosity state of Sgr~A* \citep{2013Sci...341..981W}. The observed unabsorbed X-ray luminosity in the $2$--$10\,{\rm keV}$ range is $\sim 10^{33}\,{\rm erg\,s^{-1}}$, which is about eight orders of magnitude less than the X-ray luminosity expected from the accretion of stellar winds at the Bondi capture radius, $r_{{\rm B}} \approx 4''(T_{{\rm a}}/10^7\,{\rm K})^{-1}$. This has been often explained by the presence of radiatively inefficient accretion flows \citep[RIAFs,][]{1999MNRAS.303L...1B}, which involve a powerful wind that is responsible for the loss of mass, angular momentum, and energy from accreting material. 
    
 Besides the \textit{RIAF} mechanism of accretion the outflow may be supported by stellar winds from $\sim 100$ hot, massive OB/WR stars that are known to be present at $\sim 1$--$10''$, or it may be launched by the activity of compact remnants indicated by diffuse hard X-ray emission \citep{2015...Perez}. It is also not clear whether the outflow is isotropic or rather anisotropic.  However, the recent analysis of radio interferometric observations showed that the closure phase remained zero within uncertainties, which indicates that the outflow is rather symmetric outside $\sim 100\,r_{\rm s}$ from Sgr~A* \citep{park2015}.

 In this paper we investigate the effect of an isotropic outflow from the Galactic centre with different terminal velocities ($\mathbf{v_{{\rm a}}}$) on the stellar bow-shock properties. The fundamental feature studied here is the temporal asymmetry of a bow-shock evolution due to the outflow from the centre. Using an analytic solution of \citet{1996ApJ...459L..31W} we construct a semi-analytical toy model that can be used to investigate the bow-shock properties (equilibrium stand-off distance, shell velocity, surface density, emission measure maps) for different outflow velocities and a large span of distances from Sgr~A*.


 The basic qualitative properties of the asymmetry are general and are applicable to any wind-blowing source moving in the potential of the SMBH. To show specific results, we apply the model to the Dusty S-cluster object (hereafter denoted as DSO, \citeauthor{2013A&A...551A..18E}, \citeyear{2013A&A...551A..18E}), also named G2 \citep{2012Natur.481...51G,2012ApJ...750...58B}, which is being monitored in detail and shows signs of being an infrared-excess dust-enshrouded star with the basic characteristics of a young T Tauri star \citep[see][for details]{2012NatCo...3E1049M,2013A&A...551A..18E,2013ApJ...768..108S,2013ApJ...776...13B,2014A&A...565A..17Z,2014ApJ...789L..33D,2015ApJ...800..125V}. Our basic set-up is similar to the 2D hydrodynamic simulations of \citet{2013ApJ...776...13B} and 3D hydrodynamic simulations of \citet{2014ApJ...789L..33D}. We complement their work by analysing the effect of the outflow from the Galactic centre on bow-shock properties. Although our semi-analytic approach does not allow us to trace microphysical processes along the shocked layer in detail, it can be easily used to track the evolutionary trends of the bow-shock source along the whole orbit and a large span of distances from the Galactic centre, which would be computationally highly demanding using 3D hydrodynamical calculations. The qualitative results can be applied to any bow-shock source that is followed along the orbit and its bow-shock emission is at least partially resolved. Even if the orbit of the source is not determined, the internal properties of the bow shock depend not only on its distance from Sgr~A*, but are also influenced by the sense of motion with respect to the source of the outflow in the Galactic centre.

The structure of the paper is as follows. In Section \ref{sec_model_setup} we describe the model set-up and the calculation of characteristics of stellar bow shocks. We continue with the description of the main results in Section \ref{section_results}, where we focus on the formation of asymmetry in the bow-shock evolution along the orbit (Subsection \ref{subsection_asymmetry}) and calculate the shell velocity (Subsection \ref{subsection_shell_velocity}) and density profiles (Subsection \ref{subsection_surface_density}) as functions of both the distance from the SMBH and the angle measured along the bow shock. We calculate the emission measure maps in Subsection \ref{subsection_emission_measure}. In Section \ref{subsection_comparison_emission} we compare the bow-shock emission measure to the other possible sources of emission -- the stellar wind and pre-main-sequence accretion. The analysis of the Doppler contribution of the shell velocity field to the emission of bow shock sources is presented in Subsection \ref{subsection_Doppler}. In Subsection \ref{subsection_nonthermal} we revisit the previous estimates of non-thermal emission for a stellar bow shock at the peribothron and compare it with the passage of a core-less gas cloud. We discuss the results in Section \ref{sec_discussion} and conclude in Section \ref{sec_conclusions}.

\section{Description of the bow-shock model set-up}
\label{sec_model_setup}

In this section we describe the mathematical details of an adopted bow-shock model. We consider a momentum-supported bow-shock model originally proposed by \citet{1971SPhD...15..791B}. The analytical solution and the formalism of the momentum-supported, geometrically thin bow shock was developed by \citet{1996ApJ...459L..31W,2000ApJ...532..400W}, where the reader can find the full derivation. Here we extend the analytical scenario by a semi-analytical treatment of the bow-shock evolution along the orbit.

\begin{figure*}
\centering
\begin{tabular}{cc}
\includegraphics[width=0.45\textwidth]{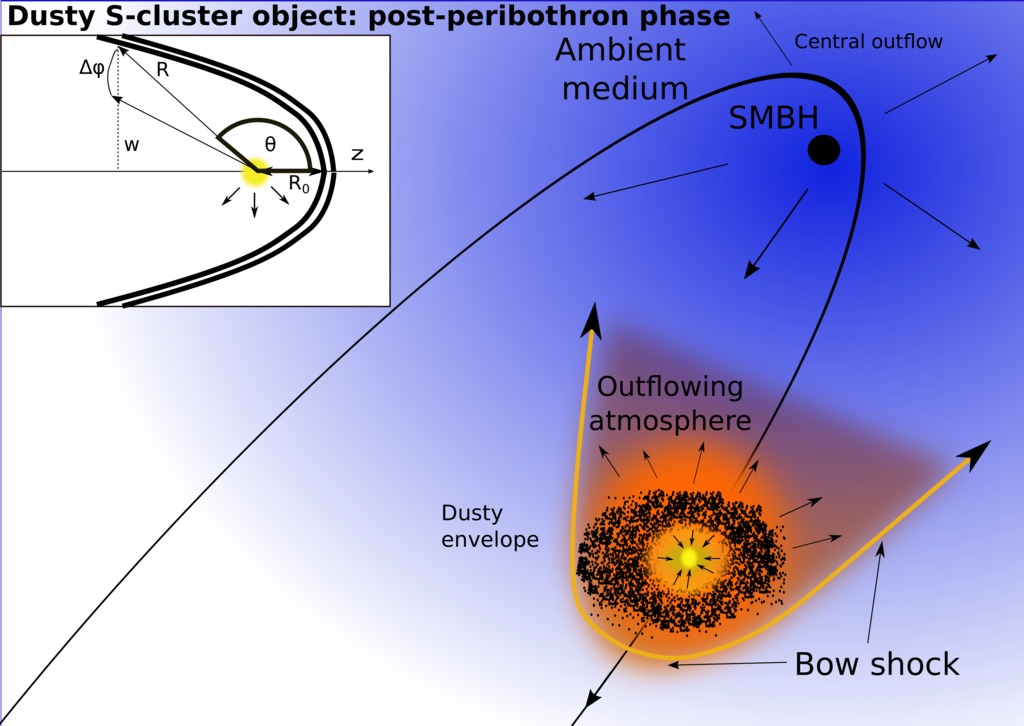} & \includegraphics[width=0.55\textwidth]{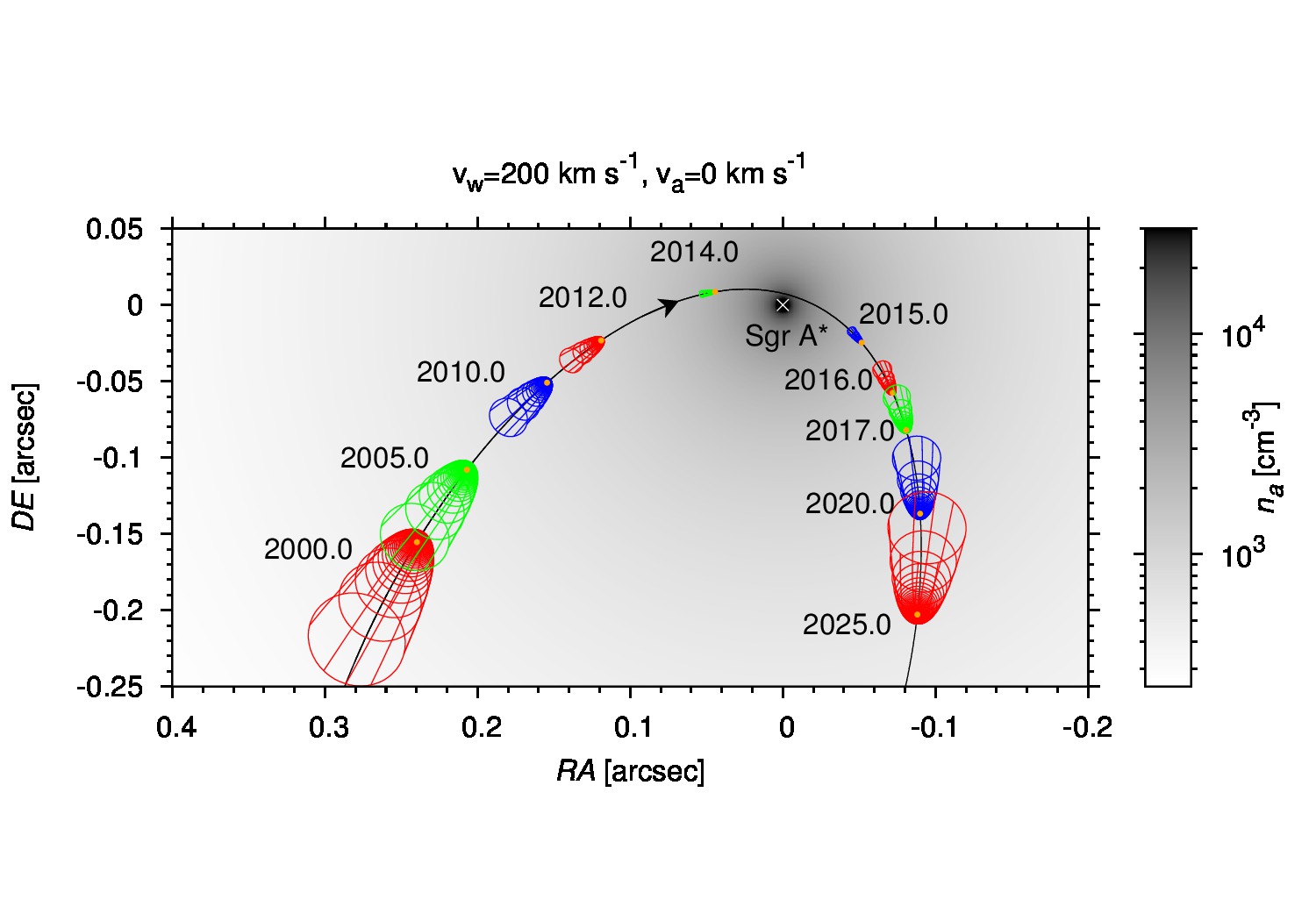}
\end{tabular}
\caption{\textit{Left}: Illustration of the model set-up: wind-blowing Dusty S-cluster Object (DSO) on an eccentric trajectory around the supermassive black hole in the post-peribothron phase. The figure inset depicts the geometry of the axisymmetric bow shock. The stand-off distance $R_0$ is labelled as well as the distance $R=R(\theta)$ of the bow shock at angle $\theta$ from the axis of symmetry. \textit{Right}: The exemplary evolution of the bow-shock geometry  along an eccentric trajectory. In this case the ambient flow is considered negligible with respect to the stellar motion. The color axis shows the particle density per ${\rm cm}^{-3}$ according to the profile given by eq. \eqref{eq_density}.}
\label{fig_star_dso_illustration}
\end{figure*}

We consider a star on a nearly radial, highly eccentric orbit $(e\gtrsim 0.9)$ around the SMBH. As a first approximation, the ambient medium is described by the spherically symmetric temperature and density profiles, eqs. \eqref{eq_density} and \eqref{eq_temperature}. The outflow is assumed to be isotropic with the characteristic terminal outflow velocity $\mathbf{v_{{\rm a}}}$. The basic set-up is illustrated in Fig. \ref{fig_star_dso_illustration}.

Young, low-mass stars in the Galactic centre region, which the DSO source may belong to, seem to have a proplyd-like, bow-shock appearance \citep{2015ApJ...801L..26Y}. Several sources exhibit strong near-infrared excess \citep{2013A&A...551A..18E}, and the spectral decomposition implies that they are probably young stars surrounded by dust shells  \citep{2004ApJ...602..760E}. They are embedded within the photoionised HII region of the central cavity. The central region ionised by massive OB stars is similar to the expanding M42 region in the Orion nebula. Young, low-mass stars of LL Ori type \citep{2000AJ....119.2919B,2001ApJ...546..299B} have radiative winds that drive a shock into the ambient medium. In order to apply the exact solution of \citet{1996ApJ...459L..31W,2000ApJ...532..400W} to study the evolution and properties of stellar wind bow shocks near the Galactic centre, we assume the following properties:

\begin{itemize}
\item[-] the bow shock is momentum-supported, i.e. the internal momentum is conserved within the shell,
\item[-] the postshock cooling is efficient. Consequently, the shell of shocked gas becomes \textit{geometrically} thin\footnote{This does not imply that the material is optically thin. The term thin bow shock is used to denote a bow shock with negligible geometrical thickness in comparison with the stagnation radius.}. In other words, the cooling parameter 
$\chi=t_{\rm{cool}}/t_{\rm{dyn}}$, where $t_{\rm{cool}}$ is the cooling timescale of the shocked gas and $t_{\rm{dyn}}$ is the dynamical timescale of the system, is assumed to be smaller than one. Especially slower winds $(v_{\rm{w}}\lesssim 300\,{\rm km\,s^{-1}})$ seem to be susceptible to radiative cooling in the high-pressure environment of Sgr~A* \citep{2005MNRAS.360L..55C},
\item[-] the stellar wind with the terminal velocity $v_{{\rm w}}$ is isotropic and stationary. The density profile of the circumstellar medium is then given simply by the relation $\rho_{{\rm w}}=\dot{m}_{{\rm w}}/(4 \pi r_{\star}^2 v_{{\rm w}})$, where $\dot{m}_{{\rm w}}$ is the stellar mass-loss rate and $r_{\star}$ is the distance from the star,
\item[-] at each point of the stellar orbit the equilibrium bow-shock structure is approximately reached, see Appendix \ref{appendix_eq} for estimates. The density of the ambient medium $\rho_{{\rm a}}$ is given by eq. \eqref{eq_density}. We neglect the density gradient at a given position of the star from the centre as well as the velocity divergence of the ambient flow, which are small for a fixed distance and do not effect the main features of orbital asymmetry studied here. See Appendix \ref{appendix_divgrad} for details.    
\end{itemize}

The star passing through the interstellar medium around the SMBH is characterised by the mass-loss rate $\dot{m}_{{\rm w}}$ and terminal stellar wind velocity $v_{{\rm w}}$. An important variable is the relative velocity of the star with respect to the surrounding medium $\mathbf{v_{{\rm rel}}}=\mathbf{v_{\star}}-\mathbf{v_{{\rm a}}}$. In general, the bow shock geometry may be described by spherical coordinates $(R,\theta,\phi)$, where $R$ is the distance of the shock from the star, $\theta$ is the angle from the $z$-axis (the axis of symmetry for axisymmetric bow shocks, see Fig. \ref{fig_star_dso_illustration}), and $\phi$ is the polar angle in $xy$-plane. We also define the cylindrical radius $w=R\sin{\theta}$ and denote the velocity of the flow in the shell as $v_{{\rm t}}$. The relative velocity of the star has the direction of the $z$-axis, $\mathbf{v_{{\rm rel}}}=v_{{\rm rel}}\mathbf{e_{z}}$. 

 The material within the shell of a stellar axisymmetric bow shock flows along a slice with constant $\phi$ \citep{1996ApJ...459L..31W}. The flux of the mass along any slice of angle $\Delta \phi$ may be expressed as,

\begin{equation}
  \Delta\dot{m}=wv_{{\rm t}}\sigma \Delta \phi\,,
\label{eq_mass_flux1}  
\end{equation} 
where $\sigma$ is the surface density within the shock.
In an axisymmetric bow-shock model, the flux of the mass from the stellar wind and the ambient medium (along any slice) may be also expressed as,
\begin{equation}
  \Delta\dot{m} \equiv \frac{\dot{m}_{{\rm w}}}{4\pi}f_{{\rm m}}(\theta) \Delta\phi\,,
\label{eq_mass_flux2}  
\end{equation} 
where the function $f_{{\rm m}}(\theta)$ depends on an adopted axisymmetric bow shock model.

The momentum flux in the shell resulting from the momentum imparted by the stellar wind and ambient medium may be expressed using model-dependent functions $f_{w}(\theta)$, $f_{z}(\theta)$ (separately in $w$ and $z$ direction):

\begin{align}
  \Delta \dot{\Pi}_z &\equiv \frac{\dot{m}_{{\rm w}}v_{{\rm w}}}{4\pi}f_z(\theta) \Delta \phi\,,\\
   \Delta \dot{\Pi}_w &\equiv \frac{\dot{m}_{{\rm w}}v_{{\rm w}}}{4\pi}f_w(\theta) \Delta \phi.
\end{align}

Consequently, the velocity in the shell $v_{{\rm t}}$ may be expressed in terms of its components $v_w$ and $v_z$,
\begin{align}
 v_w &= v_{{\rm w}}\frac{f_w(\theta)}{f_m(\theta)}\,,\\
 v_z &= v_{{\rm w}}\frac{f_z(\theta)}{f_m(\theta)}\,,
\end{align}
as $v_{{\rm t}}=\sqrt{v_w^2+v_z^2}$, leading to
\begin{equation}
v_{{\rm t}}=v_{{\rm w}}\frac{\sqrt{f_w(\theta)^2+f_z(\theta)^2}}{f_{{\rm m}}(\theta)}\,.
\label{eq_shell_velocity}
\end{equation}

Let us denote the coordinate across the shell of the bow shock by $l$ and the total thickness of the shell by $h$. The flow in the bow shock creates centrifugal pressure and the final shape of the shock is determined by the balance of the ram pressure of both the stellar wind and ambient medium flows on one hand and the centrifugal pressure on the other hand. Then, assuming that the flow is characterised by isothermal sound speed $c_{\rm{s}}$, the surface density $\sigma$ of the shell of the bow shock can be calculated as \citep{2005RMxAA..41..101C},

\begin{equation}
   \sigma=\int_0^h \rho_{l}\mathrm{d}l=\frac{H}{c_{\rm{s}}^2}(P_{{\rm a}}-P_{{\rm w}})\,,
\label{eq_surface_density}   
\end{equation}
where $H\equiv c_{{\rm s}}^2/g$ is the pressure scale height; $g$ is the centrifugal acceleration of the flow. The pressure of the ambient flow, $P_{\rm{a}}=\rho_{\rm{a}}(r)v_{\rm{rel,n}}^2$, and the pressure of the stellar wind $P_{\rm{w}}=\rho_{{\rm w}}(R)v_{\rm{w,n}}^2$, depend on normal components of corresponding velocities, which can be expressed in an axisymmetric form as \citep{2005RMxAA..41..101C},

\begin{align}
  v_{\rm{rel,n}} &= v_{\rm{rel}}(r)\frac{\sin{\theta}[R\cos{\theta}+(\partial R/\partial \theta)\sin{\theta}]}{\sqrt{[R^2+(\partial R/\partial \theta)^2]\sin^2{\theta}}}\,,\\
  v_{\rm{w,n}} &= v_{\rm{w}}\frac{R \sin{\theta}}{\sqrt{[R^2+(\partial R/\partial \theta)^2]\sin^2{\theta}}}\,.
\end{align}

The surface density $\sigma$ can be also calculated by combining the expressions for the mass flow and the velocity along the shell, eqs. \eqref{eq_mass_flux1} and \eqref{eq_shell_velocity}, respectively. 

In order to produce the emission maps of hydrogen line emission or free-free continuum emission we follow \citet{2005RMxAA..41..101C} who calculate the emission measure by integrating the square of the density profile across the shell  in the following way,

\begin{equation}
   EM=\int_0^h (\rho/\overline{m})^2 \mathrm{d}l=\frac{\sigma(P_{\rm{a}}+P_{\rm{w}})}{2\overline{m}^2c_{\rm{s}}^2}\,,
   \label{eq_emission_measure}
\end{equation}
where $\overline{m}\approx 1.4 m_{\rm{h}}$ is the average mass per ionized hydrogen particle. 

\subsection{Application to a single axisymmetric bow shock around the SMBH}
\label{subsec_application}

An exact solution for a thin shell bow shock was found by \citet{1996ApJ...459L..31W} for the case of an isotropic stellar wind and homogeneous ambient medium. We apply this model for every point of the stellar orbit around the SMBH and calculate the properties of the equilibrium stellar bow shock along the trajectory \citep[a similar approach was already used in][]{2014A&A...565A..17Z}. In this way we create a semi-analytic toy model that can be used to easily investigate the changes of basic bow-shock characteristics along the orbit. 

The bow-shock shape $R=R(\theta)$ in the axisymmetric thin shell model can be expressed analytically as follows \citep{1996ApJ...459L..31W},

\begin{equation}
  R=R_0 \csc{\theta} \sqrt{3(1-\theta\cot{\theta})}\,,
  \label{eq_R_theta}
\end{equation}
where $R_0$ is the stagnation radius given by the equilibrium of the ambient and the stellar wind ram pressure,

\begin{equation}
  R_0=\sqrt{\frac{\dot{m}_{\rm{w}}v_{\rm{w}}}{4\pi \rho_{\rm{a}} v_{\rm{rel}}^2}}\,.
 \label{eq_stagnation_radius} 
\end{equation}

The 3D model is generated by rotation around the $z$-axis:
\begin{equation}
R_{\rm{3D}}(\theta,\phi)=\rm{rot}\{R(\theta),\phi\},\,0^{\circ} \leq \phi \leq 180^{\circ}\,.
\end{equation}

In the following expressions we will use the dimensionless cylindrical radius $\overline{w}=w/R_0$. Let us define the ratio of relative velocity and stellar wind velocity, $\alpha\equiv v_{\rm{rel}}/v_{\rm{w}}$. Then, in the framework of the thin-shell axisymmetric model \citep{1996ApJ...459L..31W}, the model-dependent functions $f_{\rm{m}}(\theta)$, $f_w(\theta)$, and $f_{z}(\theta)$ may be expressed in terms of the angle $\theta$ along the shocked layer, the cylindrical radius $\overline{w}$, and the ratio of the relative and stellar wind velocities $\alpha$:

\begin{align}
 f_{\rm{m}}(\theta) &= (1-\cos{\theta})+\frac{\overline{w}^2}{2\alpha}\,, \label{eq_fm}\\
 f_w(\theta) &= \frac{1}{2}(\theta-\sin{\theta}\cos{\theta})\,,\label{eq_fw}\\
 f_z(\theta) &= \frac{1}{2}(\sin^2{\theta}-\overline{w}^2)\,. \label{eq_fz}
\end{align}  

Using equations \eqref{eq_mass_flux1}, \eqref{eq_shell_velocity}, and \eqref{eq_emission_measure} and the functions for the axisymmetric thin bow shock model, eqs. \eqref{eq_fm}, \eqref{eq_fw}, and \eqref{eq_fz}, we may calculate basic characteristics of stellar bow shocks that can be tested observationally.

 Finally, the axisymmetric bow-shock model applied throughout the paper gives the following relations for the thin-shell velocity $v_{\rm{t}}$ and surface density $\sigma$ \citep{1996ApJ...459L..31W}, respectively,

\begin{align}
    v_{\rm{t}} &=& v_{\rm{rel}}\frac{\sqrt{(\theta-\sin{\theta}\cos{\theta})^2-(\sin^2{\theta}-\overline{w}^2)^2}}{2\alpha(1-\cos{\theta})+\overline{w}^2}\,,\\
    \sigma     &=& R_0 \rho_{\rm{a}} \frac{[2\alpha (1-\cos{\theta})+\overline{w}^2]^2}{2\overline{w}\sqrt{(\theta-\sin{\theta}\cos{\theta})^2-(\sin^2{\theta}-\overline{w}^2)^2}}\,.
\end{align}
For a given angle $\theta$ along the shocked layer, the emission measure in the axisymmetric case is computed using eq. \eqref{eq_emission_measure}. The exact formula is as follows,
\begin{align}
   EM(\theta) &= \frac{\sigma(\theta)}{2\overline{m}^2c_{\rm{s}}^2} \times \notag\\ &\times \frac{\rho_{\rm{a}}v_{\rm{rel}}^2 [R\cos{\theta}+(\mathrm{d}R/\mathrm{d}\theta)\sin{\theta}]^2+\rho_{\rm{w}}v_{\rm{w}}^2R^2}{R^2+(\mathrm{d}R/\mathrm{d}\theta)^2}\,,
   \label{eq_EM_axisym}
\end{align}   
where the radial distance $R$ of the shock from the source is given by eq. \eqref{eq_R_theta} and the derivative with respect to the angle $\theta$ may be expressed in the following way,
\begin{equation}
    \frac{\mathrm{d}R}{\mathrm{d}\theta}=R_0 \csc{\theta} \left[\frac{3(\theta-\cos{\theta}\sin{\theta})}{2\overline{w}\sin^2{\theta}}-\overline{w}\cot{\theta}\right]\,.
    \label{eq_R_derivative}
\end{equation}

The model is applied for the case of a single star on an eccentric trajectory around the SMBH. We set the mass-loss rate of the star, $\dot{m}_{\rm{w}}$, to $10^{-8}\,M_{\odot}\,{\rm yr}^{-1}$ and the terminal stellar wind speed, $v_{\rm{w}}$, to $200\,\rm{km\, s^{-1}}$. These are based on a comparison of the observed emission of the DSO with the model of a pre-main-sequence accreting star \citep{2015ApJ...800..125V} (see also \citet{2013ApJ...768..108S} for a discussion of the T Tauri star model of DSO. 

 An exemplary evolution of the bow-shock geometry along the trajectory of a stellar source is depicted in Fig. \ref{fig_star_dso_illustration} (right panel), where the ambient flow is negligible with respect to the motion of the star and the density distribution of the ambient medium is colour-coded. In our model calculations we assume the mass of the SMBH of $4\times 10^6\,M_{\odot}$ and its distance from the Sun is $8\,{\rm kpc}$, which is in the range of values inferred by independent measurements \citep{2010RvMP...82.3121G}.

\section{Qualitative properties of temporal asymmetry for an equilibrium bow shock: Dusty S-cluster object as case study}
\label{section_results}

The near-infrared excess source DSO has been highly monitored during its passage around Sgr~A* \citep[][and references therein]{2015ApJ...798..111P, 2015ApJ...800..125V, 2014ApJ...796L...8W} since its discovery \citep{2012Natur.481...51G}. The DSO shows several characteristics of a dust-enshrouded young star, namely emission lines with a large FWHM of the order of $1\,\rm{nm}$ \citep[see][for detailed discussion]{2015ApJ...800..125V}. Since the source has been reported to remain spatially compact during the pericentre passage\footnote{We use \textit{pericentre} and \textit{peribothron} interchangeably.} and immediate post-pericentre phase \citep{2015ApJ...800..125V, 2014ApJ...796L...8W}, it can serve as a probe of the properties of the ambient medium during the post-peribothron phase of its orbital evolution.

 The estimate of accretion rate $\dot{m}_{{\rm acc}}$ is determined using the tight correlation between Br$\gamma$ emission-line luminosity and accretion luminosity for young stellar objects \citep[see][for details]{2015ApJ...800..125V}. The mass-loss rate is then estimated using the ratio $\dot{m}_{{\rm w}}/\dot{m}_{{\rm acc}} \approx 0.01-0.1$ that is verified both theoretically and observationally for several T Tauri stars \citep{2006ApJ...646..319E,2006A&A...453..785F}. The adopted terminal wind velocity of $200\,{\rm km\,s}^{-1}$ lies in the range $\sim 50$--$400\,\rm{km\,s}^{-1}$ expected for supersonic outflows of T Tauri stars \citep[extended disc winds, X-winds or stellar winds, see][for a derived diagnostic diagram]{2006A&A...453..785F}. A typical Mach number for the stellar wind is then $M_{{\rm w}}\approx 8.52\, v_{2}\, T_{4}^{-1/2}$, where $v_{2}=v_{{\rm w}}/100\,{\rm km\,s}^{-1}$ is the terminal wind velocity and $T_4=T/10^4\,{\rm K}$ is the isothermal temperature of the wind. Given the uncertainties the computed bow-shock characteristics presented here can differ by a factor of a few in absolute terms, but their qualitative evolution is not affected. 

 We study the evolution of basic bow-shock characteristics, mainly the size, orientation, surface density, and emission measure, for different velocities of the isotropic ambient outflow, specifically for a negligible outflow velocity and a terminal outflow velocity of $500\,\rm{km\,s^{-1}}$, $1000\,\rm{km\,s^{-1}}$, and $2000\,\rm{km\,s^{-1}}$. The last value is close to the one found from the analysis of X3 and X7 comet-shaped sources \citep{2010A&A...521A..13M}. We assume an isotropic outflow, whose origin can be the ensemble of stars closer to the SMBH or an outflow from Sgr~A* itself. The ambient density distribution is kept the same in all cases (see eq. \eqref{eq_density}). So for increasing outflow velocities the ambient outflow rate $\dot{M}_{{\rm a}}$ needs to be changed accordingly to match the ambient density profile, for the hydrostatic equilibrium case $\dot{M}_{\rm{a}}\propto \rho_{{\rm a}} v_{{\rm a}}$.   


 Whereas the spherically symmetric stationary inflow \citep{1952MNRAS.112..195B} represents one of the first detailed scenarios of SMBH accretion with zero angular momentum near the horizon \citep{1999MNRAS.303L...1B,1992ApJ...387L..25M,1999ApJ...511..750M,2000ApJ...545L.117M}, it has to be matched with an outflow at larger radii. Analytical estimates from the Bondi flow can be employed to relate the accretion rate to temperature and density measured near the Bondi capture radius, $r_{\rm B} \simeq 4^{\prime\prime} T_7^{-1}$, where $T_7$ is temperature in units of $10^7$ Kelvin. Subsequently, these can be tested in the X-ray regime  \citep[e.g.,][]{2003ApJ...591..891B,2013Sci...341..981W}. An outflow prevails above the characteristic Bondi radius, where the material is supplied by hot-star winds present in the region \citep{2004ApJ...613..322Q,2007A&A...468..233M}. The outflow exceeds the amount of the inflowing material by orders of magnitude \citep{2006ApJ...640..308M}. Substantial uncertainties still persist regarding the interpretation of the quiescent state of Sgr A*, nevertheless, many of them can be addressed within the radiatively inefficient scenario of RIAF models \citep{1994ApJ...428L..13N,2000ApJ...545..842Q,2001A&A...376..697D}.

\subsection{Formation of temporal asymmetry of bow-shock properties due to central outflow}
\label{subsection_asymmetry}

\begin{figure*}
  \centering
  \begin{tabular}{cc}
    \includegraphics[width=0.5\textwidth]{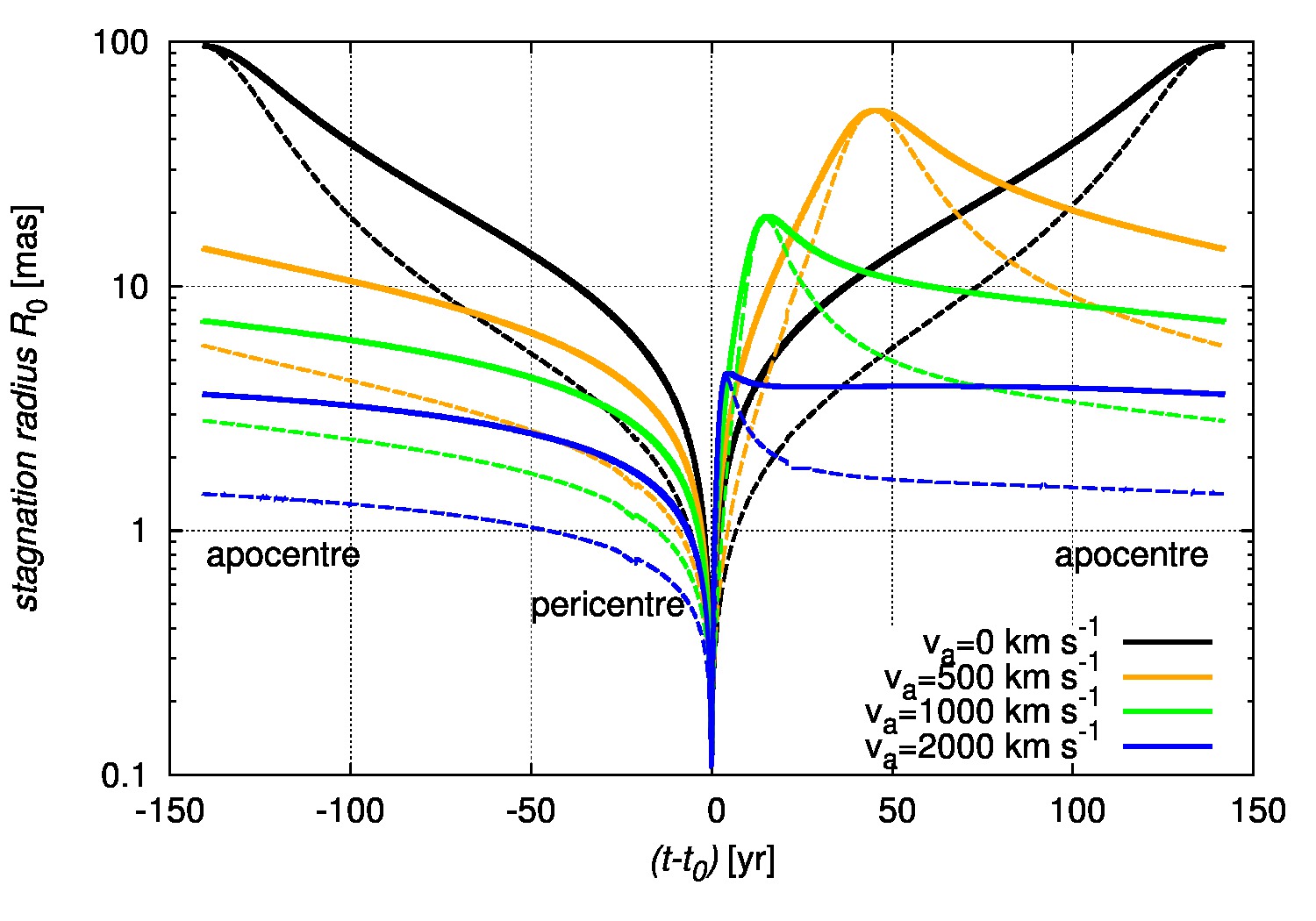} & \includegraphics[width=0.5\textwidth]{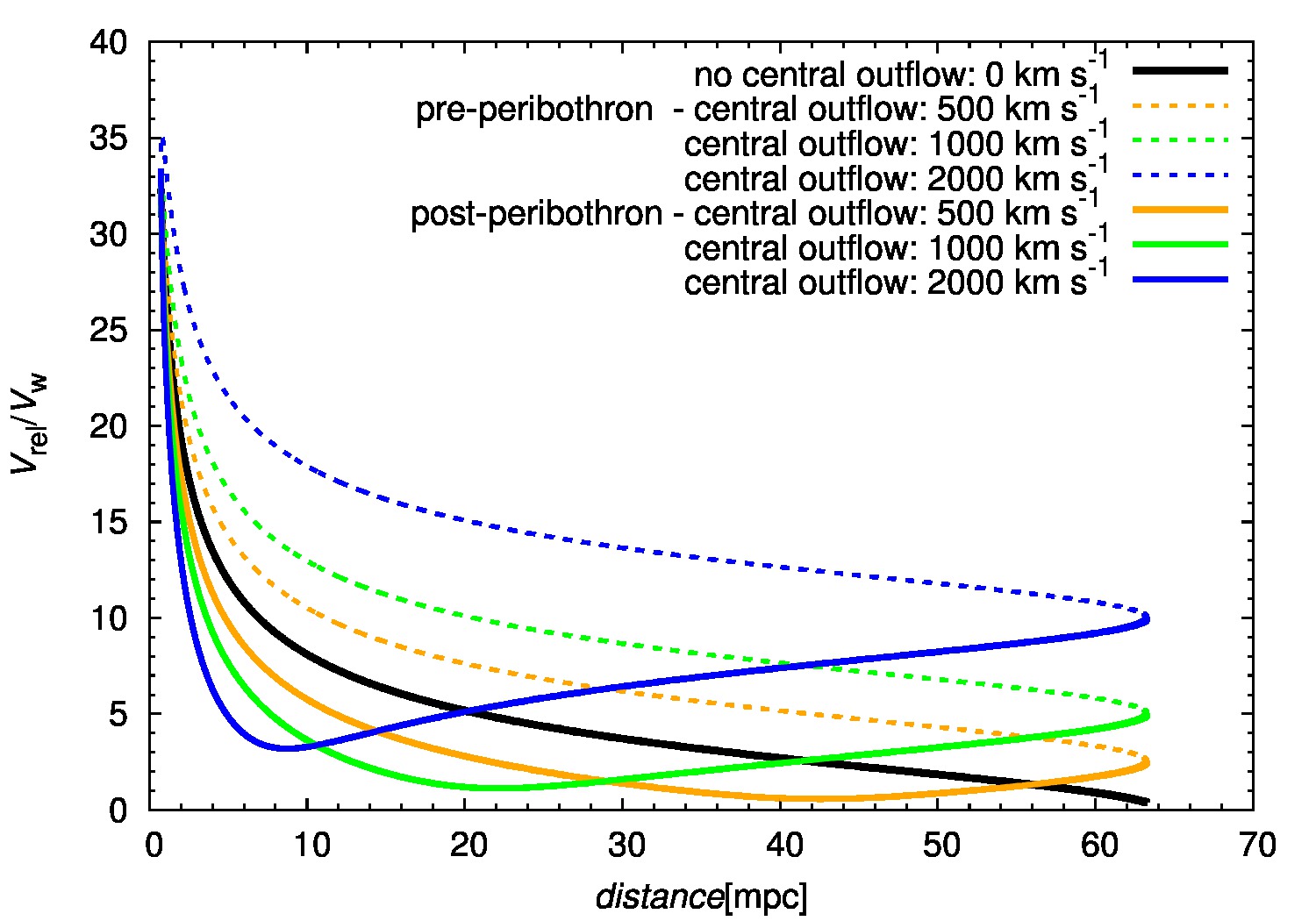}
  \end{tabular}
  \caption{\textit{Left}: Temporal evolution of the stagnation radius $R_0$, eq. \eqref{eq_stagnation_radius}, for different velocities of the outflow (see the legend). The solid lines represent the stagnation radius as measured in the orbital plane, whereas the dashed lines stand for the projected angular scale of the stagnation radius based on the inferred inclination of $113^{\circ}$ for the DSO source \citep{2015ApJ...800..125V}. \textit{Right}: The ratio of relative velocity and stellar wind velocity, $\alpha=v_{\rm{rel}}/v_{\rm{w}}$, as a function of the distance from the SMBH (in milliparsecs). The black line denotes the case for negligible outflow; for non-zero cases an asymmetry develops between the pre-peribothron (dashed lines) and post-peribothron evolution (solid lines). The quantities in both panels were computed for the inferred highly-elliptical orbit of the DSO source \citep{2015ApJ...800..125V}.}
  \label{fig_stagnation_radius_alpha}
\end{figure*}

Even for the symmetric outflow from the Galactic centre and an isotropic stellar wind, the star-SMBH system can behave asymmetrically along the trajectory of the source. This can be shown by the evaluation of the stagnation radius $R_0$, eq. \eqref{eq_stagnation_radius}, along the orbit, which is illustrated in the left panel of Fig. \ref{fig_stagnation_radius_alpha}. Unsurprisingly, for no outflow, the maximum of the stagnation radius, $R_0 \approx 100\,\rm{mas}$, occurs at the apobothron, where the ambient pressure is the smallest. The minimum of the stagnation radius, $R_0 \approx 0.1\,\rm{mas}$, is at the peribothron where the relative velocity as well as the ambient pressure are the highest. 

For increasing outflow velocities, the asymmetry between pre-peribothron and post-peribothron phases develops: the minimum of the stagnation radius stays at the peribothron; however, the maximum shifts from the apobothron towards the peribothron and occurs during the post-peribothron phase; see Fig. \ref{fig_stagnation_radius_alpha} for comparison. The solid lines stand for the values in the orbital plane, the dashed lines depict the projected values according to the current orbital solution of the DSO \citep{2015ApJ...800..125V}. 

The evolution of the stagnation radius of the bow shock is determined by the value of the $\alpha$-function, $\alpha=v_{\rm{rel}}/v_{\rm{w}}$. The plot of the function $\alpha$ for different outflow velocities as a function of distance from the SMBH is in Fig. \ref{fig_stagnation_radius_alpha} (right). It can be clearly seen that the case with zero outflow is symmetric (black solid line), whereas for the non-zero outflow the asymmetry is apparent between pre-peribothron phase (dashed lines) and post-peribothron phase (solid lines). At a fixed distance the difference becomes larger for stronger ambient wind; see Fig. \ref{fig_stagnation_radius_alpha} (right panel). 

\subsection{Shell velocity profiles}
\label{subsection_shell_velocity}

\begin{figure*}
  \centering
  \begin{tabular}{cc}
     \includegraphics[width=0.5\textwidth]{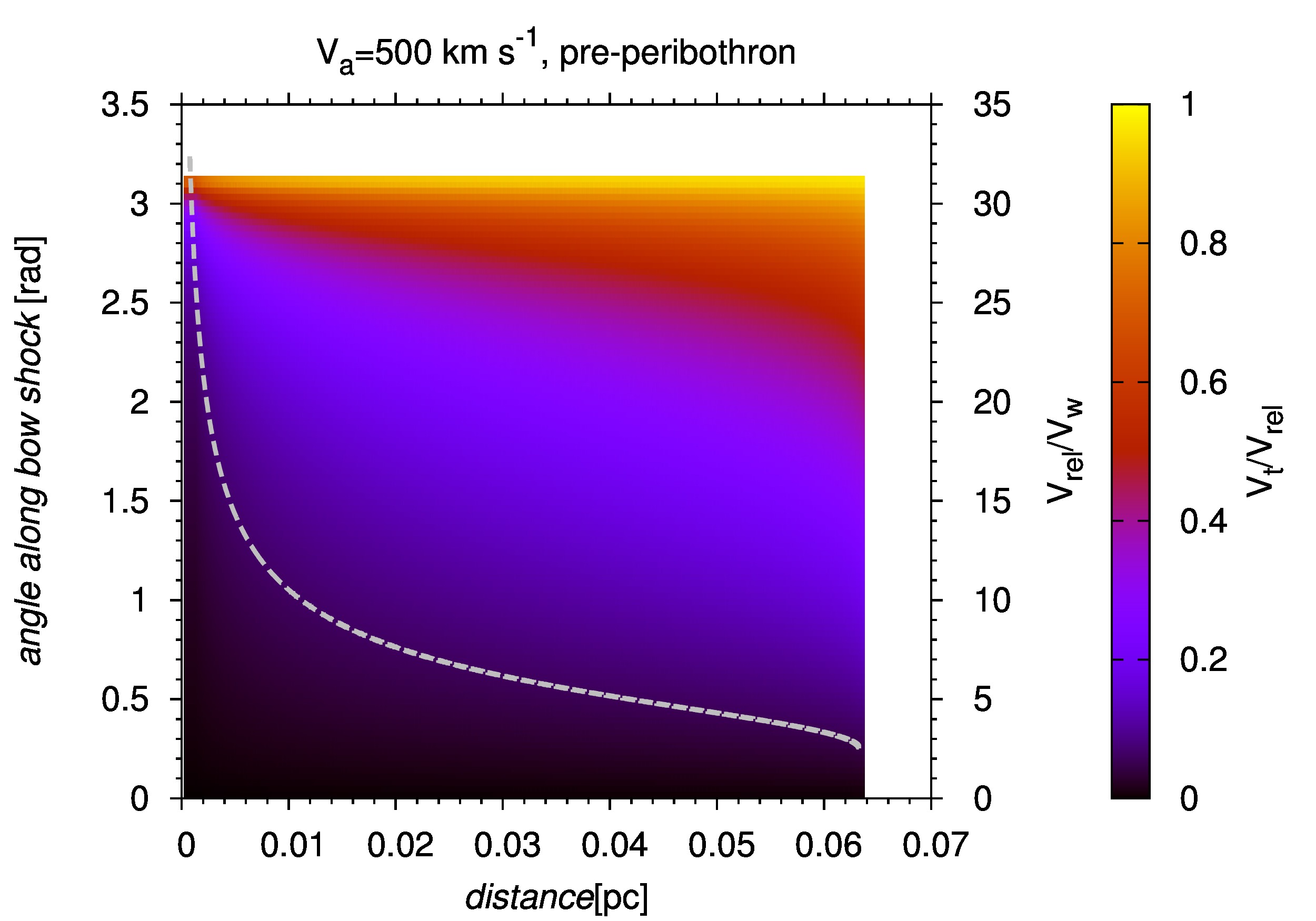} & \includegraphics[width=0.5\textwidth]{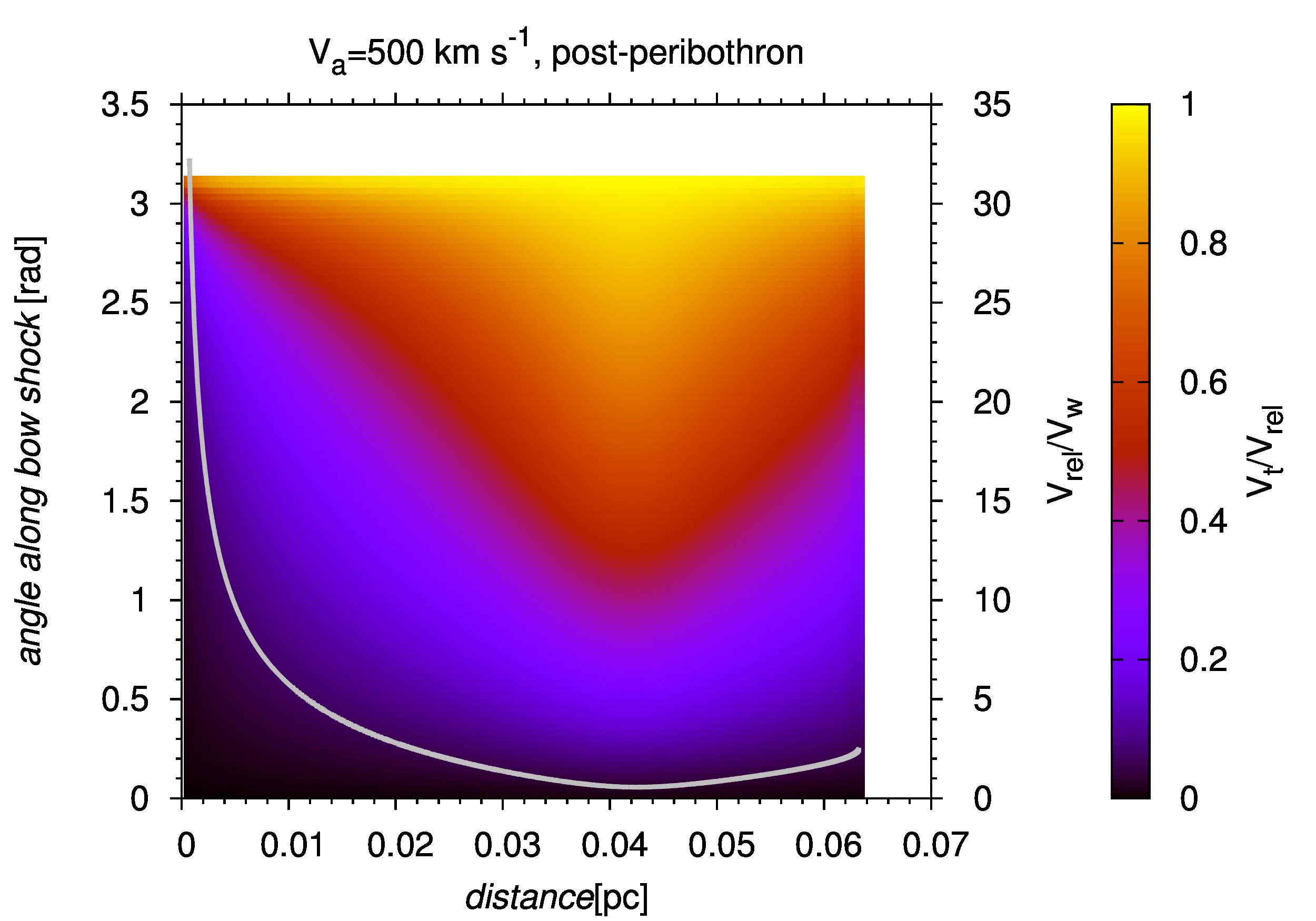}\\
     \includegraphics[width=0.5\textwidth]{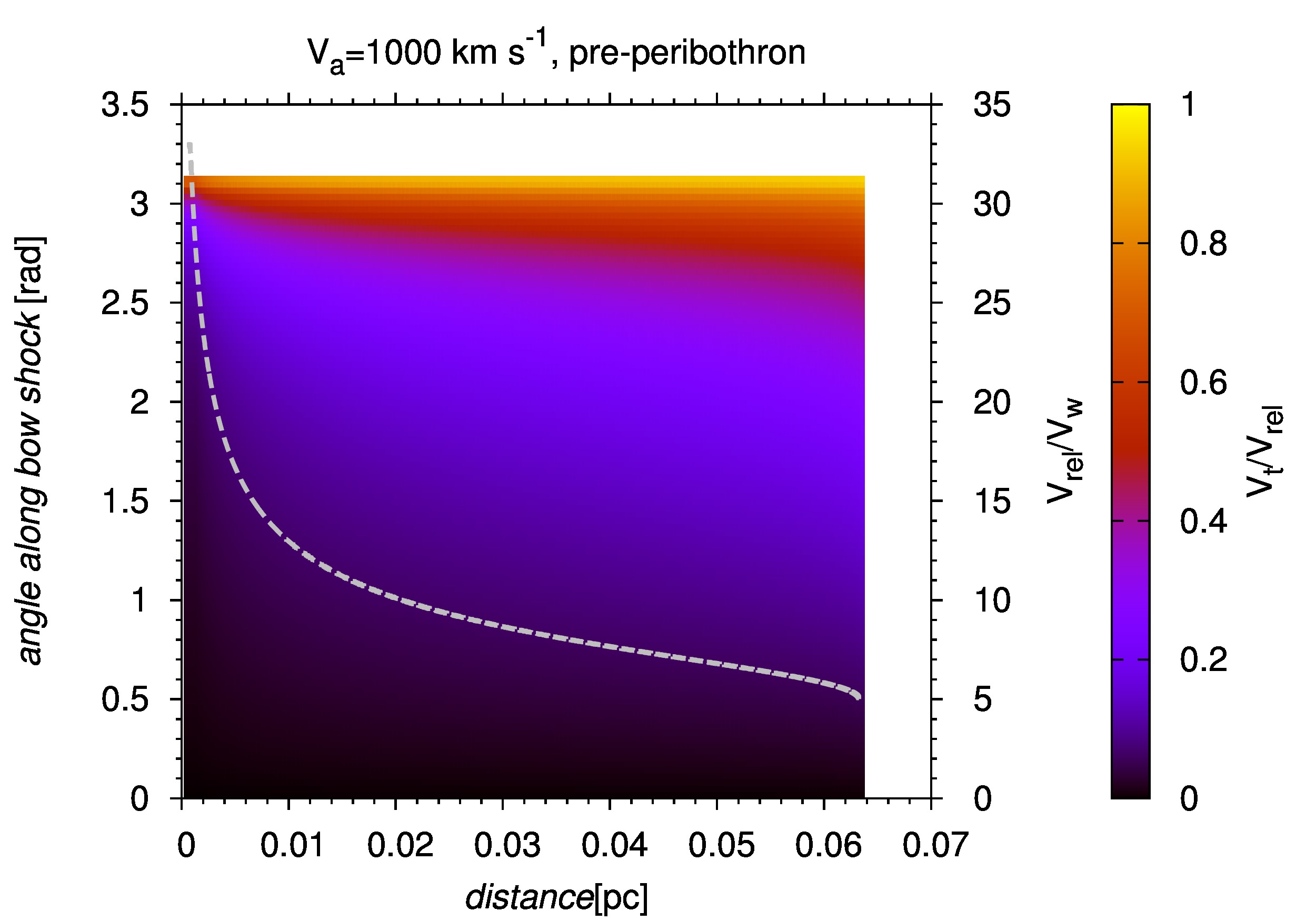} & \includegraphics[width=0.5\textwidth]{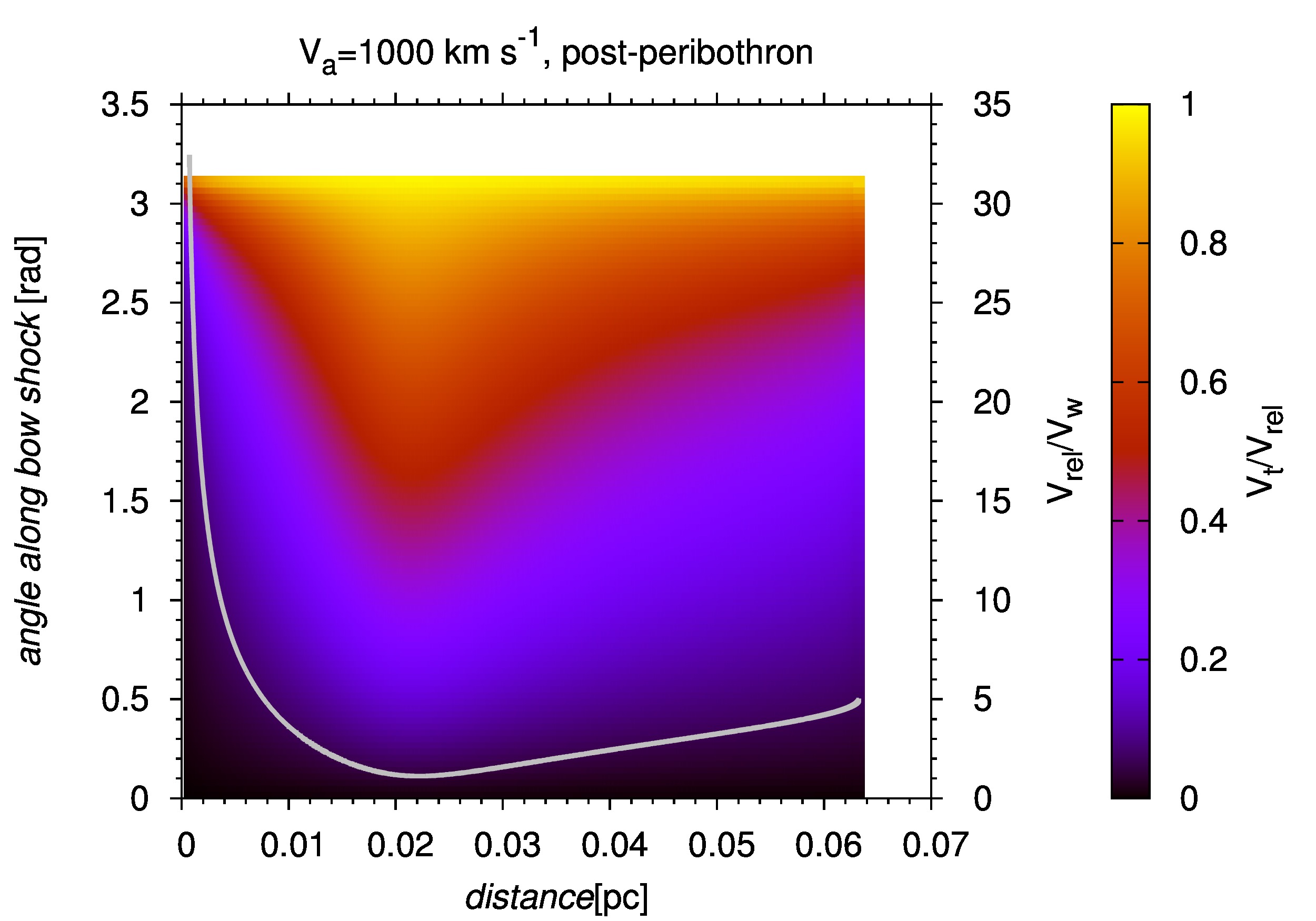}\\ 
     \includegraphics[width=0.5\textwidth]{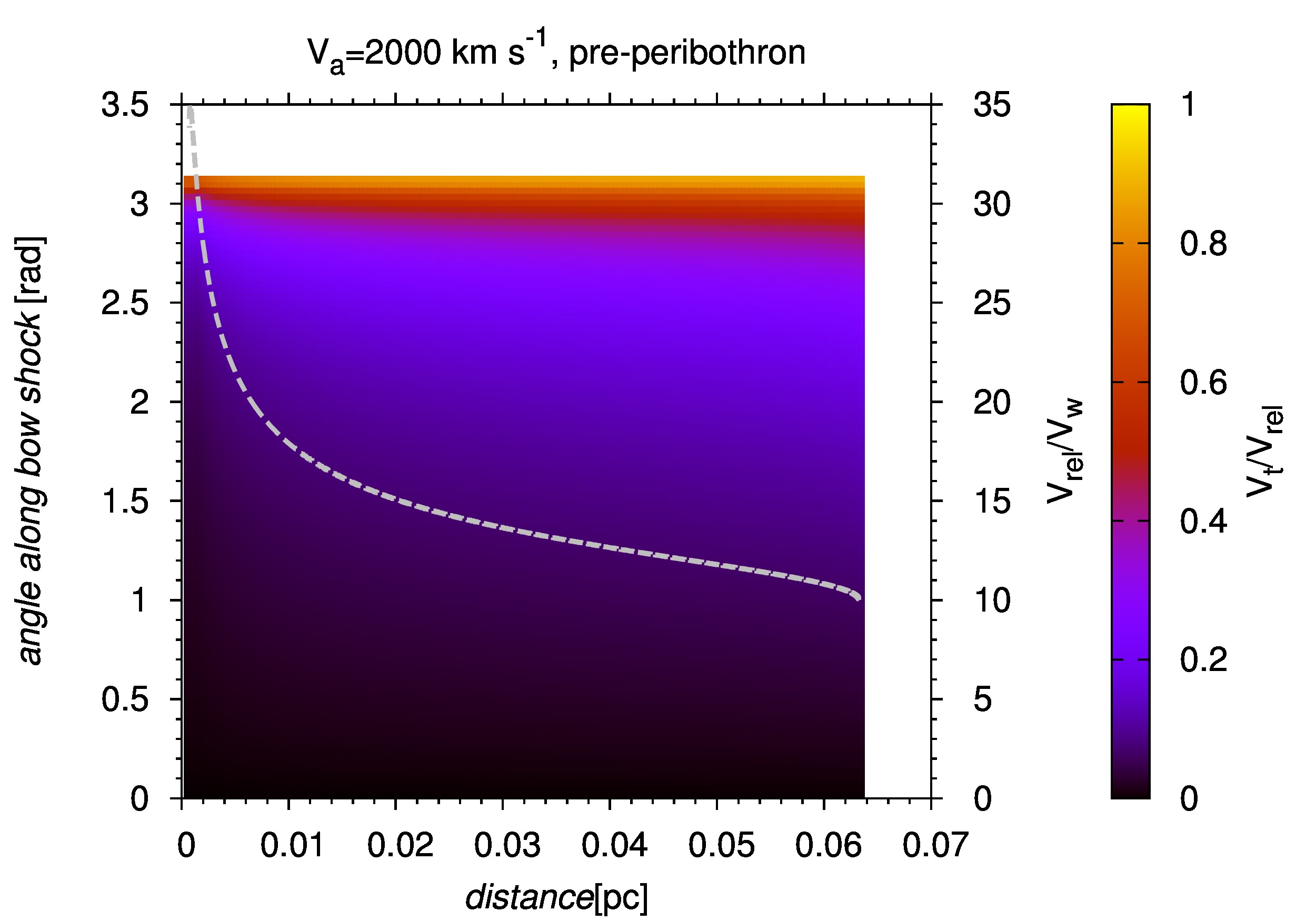} & \includegraphics[width=0.5\textwidth]{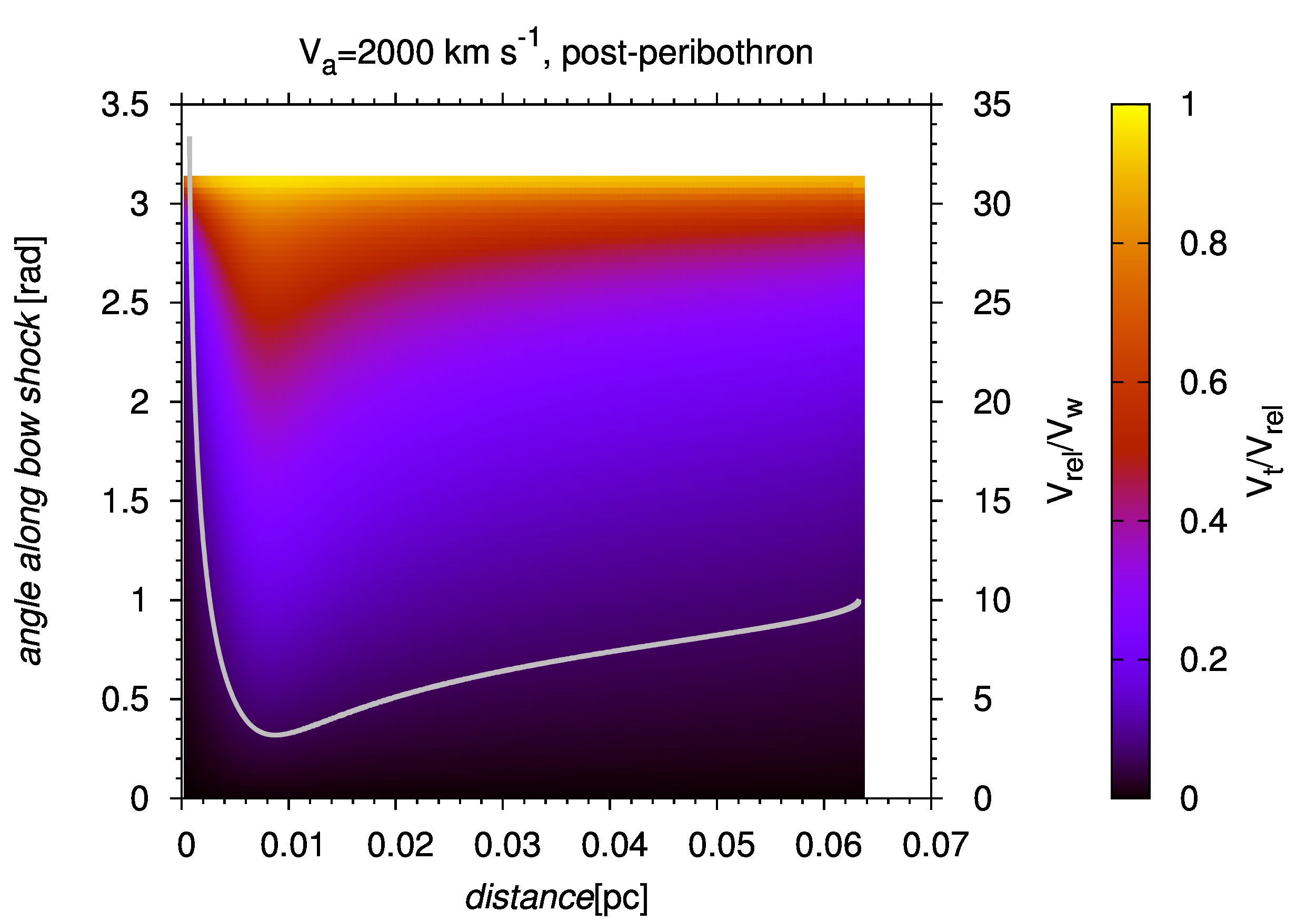}
  \end{tabular}
  \caption{The colour-coded profiles of the velocity along the bow shock shell (in units of the relative velocity of the star with respect to the ambient medium) as a function of the distance from the SMBH (horizontal axis, expressed in pc) and the spherical angle (left vertical axis, measured from the bow shock symmetry axis in radians). The left-hand column contains pre-peribothron velocity profiles, the right-hand column depicts post-peribothron velocity profiles.  The ratio of the relative velocity and the stellar wind velocity, $v_{\rm{rel}}/v_{\rm{w}}$, as a function of the distance is plotted as a grey line and its values are along the right vertical axis.}
  \label{fig_shell_velocity_profiles}
\end{figure*}

 The shocked gas layers -- ambient and stellar wind shocks -- are assumed to mix efficiently and quickly cool radiatively, which is typical of low-mass pre-main-sequence stars interacting with HII regions. Using eq. \eqref{eq_shell_velocity} we evaluate the velocity profile of the shocked flow both as a function of the spherical angle $\theta$ ($\theta=0\, {\rm rad}$ corresponds to the vertex of the bow shock and $\theta$ close to $\pi\,{\rm rad}$ covers the bow-shock downstream region, further beyond the star) and the distance from the SMBH. The colour-coded profiles are plotted in Fig. \ref{fig_shell_velocity_profiles}: the left column depicts the pre-peribothron profiles and the right column the post-peribothron ones. These maps show the values of the ratio between the shell velocity $v_{\rm{t}}$ and the relative velocity $v_{\rm{rel}}$, $v_{\rm{t}}/v_{\rm{rel}}$. The plots have two vertical axes: the left one shows the values of the spherical angle $\theta$ as measured along the bow shock and the right one depicts the values of the $\alpha$-function, $\alpha=v_{\rm{rel}}/v_{\rm{w}}$, which is plotted as a grey curve. The absolute value of the shell velocity for a given distance can be determined from the $\alpha$-function by knowing the stellar wind terminal velocity ($v_{\rm{w}}=200\,\rm{km\,s^{-1}}$) in our model calculations.

The basic feature is that the velocity is close to zero near the vertex of the bow shock across all distances from the SMBH as well as different ambient velocity outflows. Furthermore, there is an increase downstream towards higher angles and at $\theta \approx 3\,\rm{rad}$ the ratio $\alpha$ approaches unity, $v_{\rm{t}}/v_{\rm{rel}}\approx 1$.

 The velocity profile along the bow shock, however, varies with the distance of the source from the SMBH, which is caused by the change of the ratio $\alpha=v_{\rm{rel}}/v_{\rm{w}}$. Near the peribothron the ratio $v_{\rm{rel}}/v_{\rm{w}}$ is largest and the shell velocity remains below $0.4\,v_{\rm{rel}}$ up to $\theta\approx 3\,\rm{rad}$ and then rises steeply. The ambient outflow leads to the minimum of the ratio $v_{\rm{rel}}/v_{\rm{w}}$ in the post-peribothron phase. At this minimum the shell velocity reaches the values above $0.5\,v_{\rm{rel}}$ closer to the vertex of the bow shock and the overall profile is flattened downstream. The asymmetry between the pre- and post-peribothron phase is apparent: in the pre-pericentre part (left-hand side of Fig. \ref{fig_shell_velocity_profiles}), the region where $v_{\rm{t}}\approx v_{\rm{rel}}$ shrinks downstream towards the higher angle for stronger outflows. For comparison, in the post-peribothron part (right-hand panel in Fig. \ref{fig_shell_velocity_profiles}), the range of the distance from the SMBH, where the profile of the shell velocity $v_{{\rm t}}$ increases steeply close to the bow-shock vertex and then flattens out downstream, shifts from the apocentre closer to the SMBH. The stronger the outflow velocity is, the closer to the SMBH this flat profile (with the large shell velocity, $v_{\rm t}\approx v_{{\rm rel}}$) gets.

\subsection{Surface density profiles}
\label{subsection_surface_density}

\begin{figure*}
  \centering
  \begin{tabular}{cc}
     \includegraphics[width=0.5\textwidth]{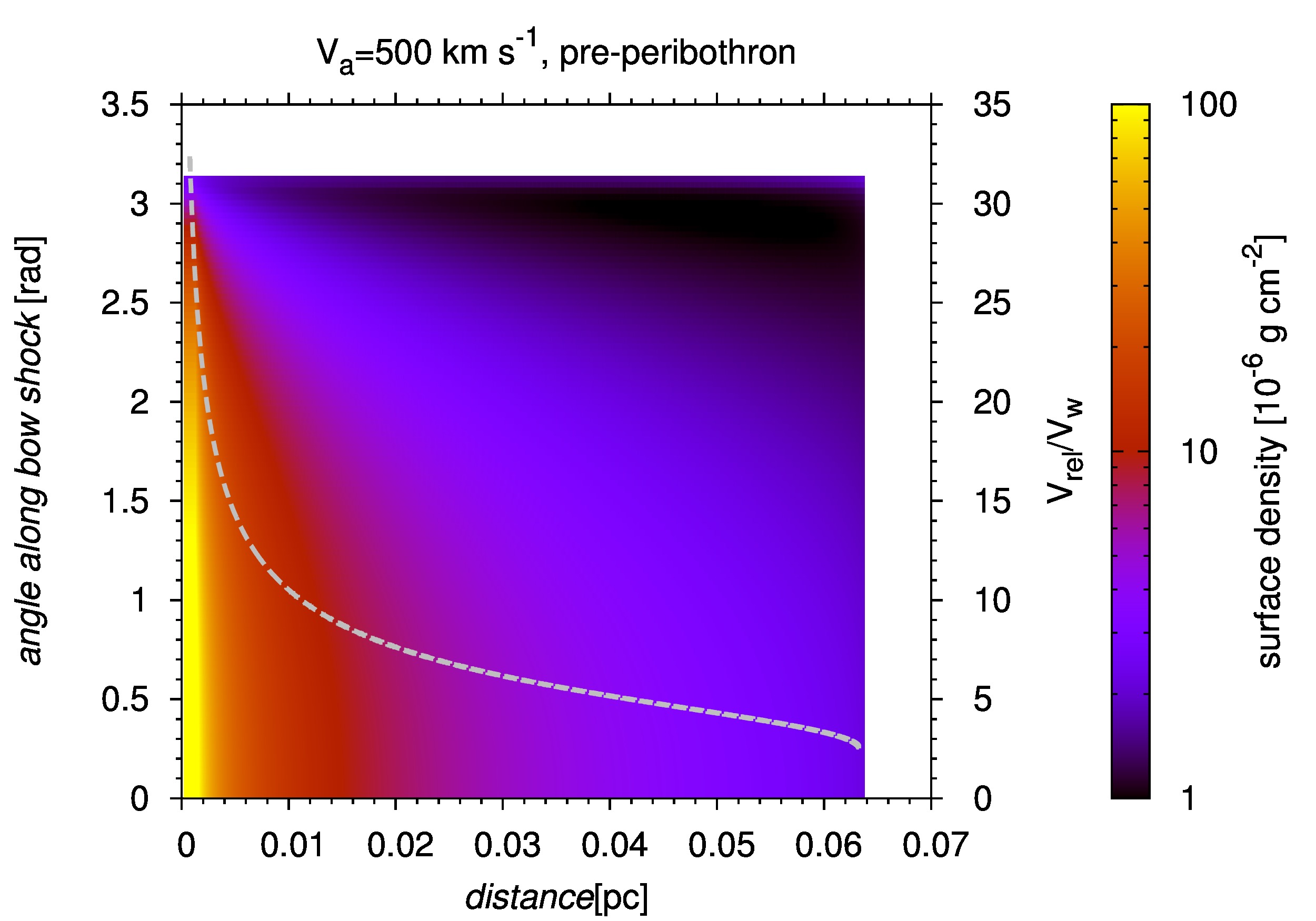} & \includegraphics[width=0.5\textwidth]{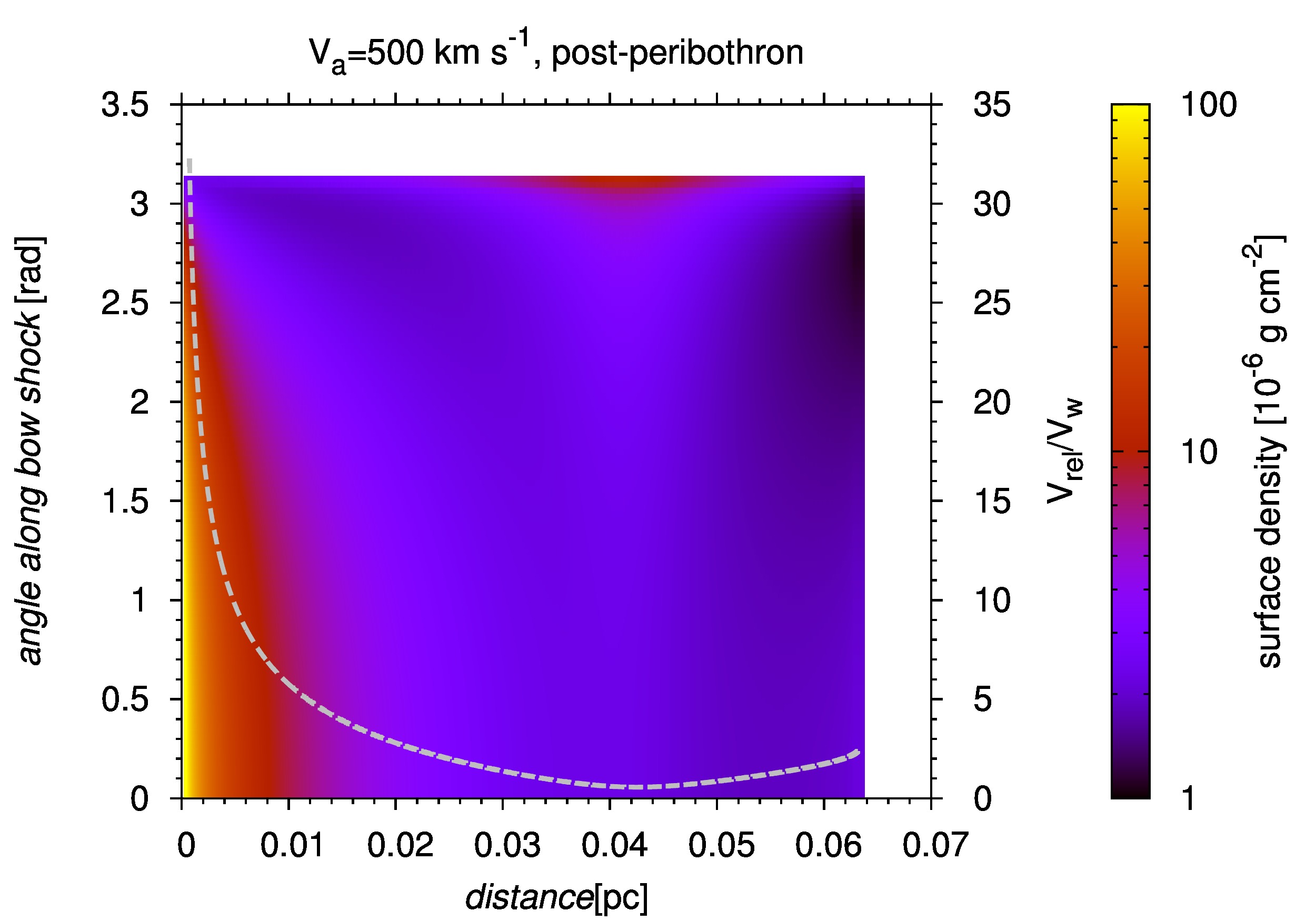}\\
     \includegraphics[width=0.5\textwidth]{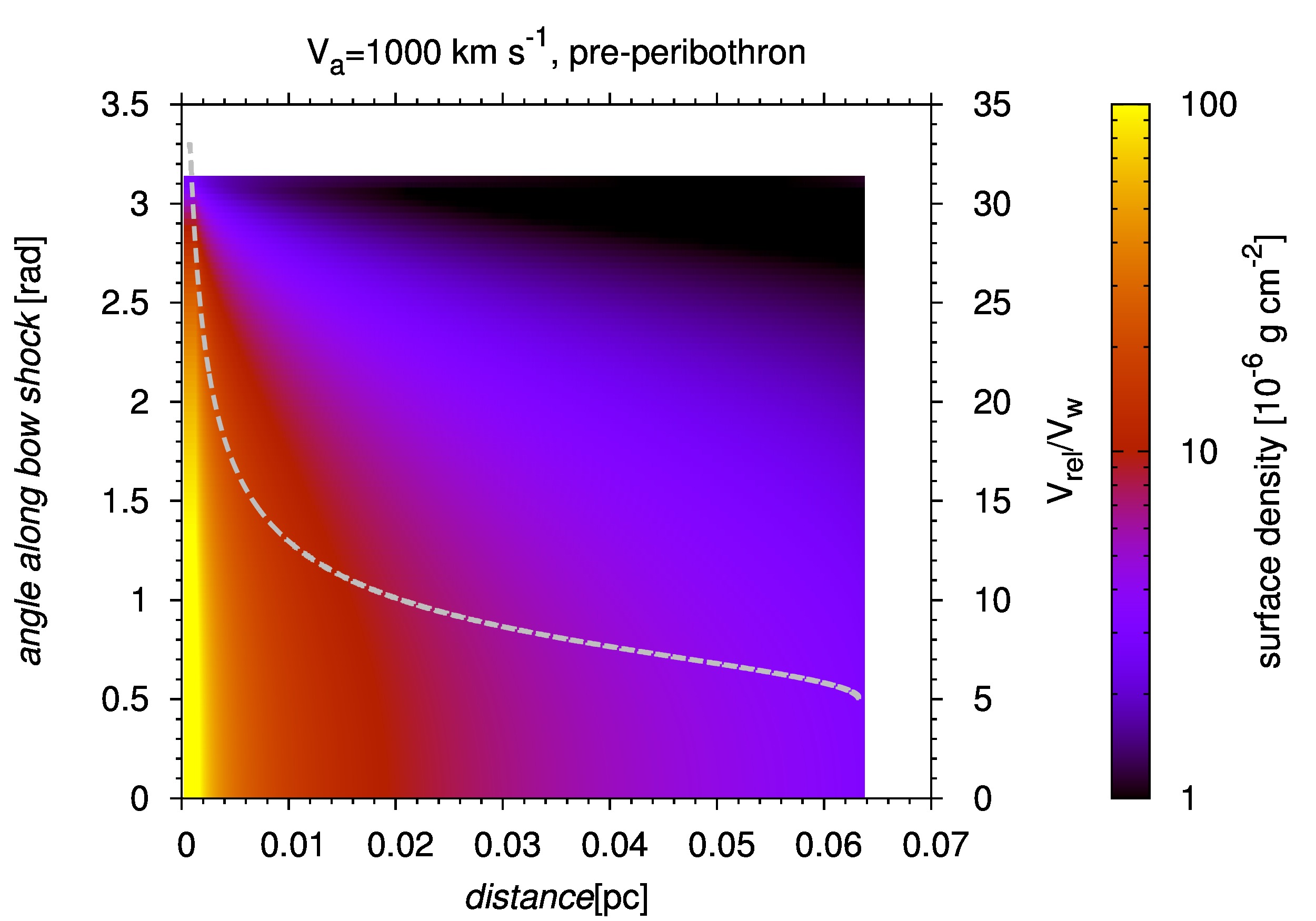} & \includegraphics[width=0.5\textwidth]{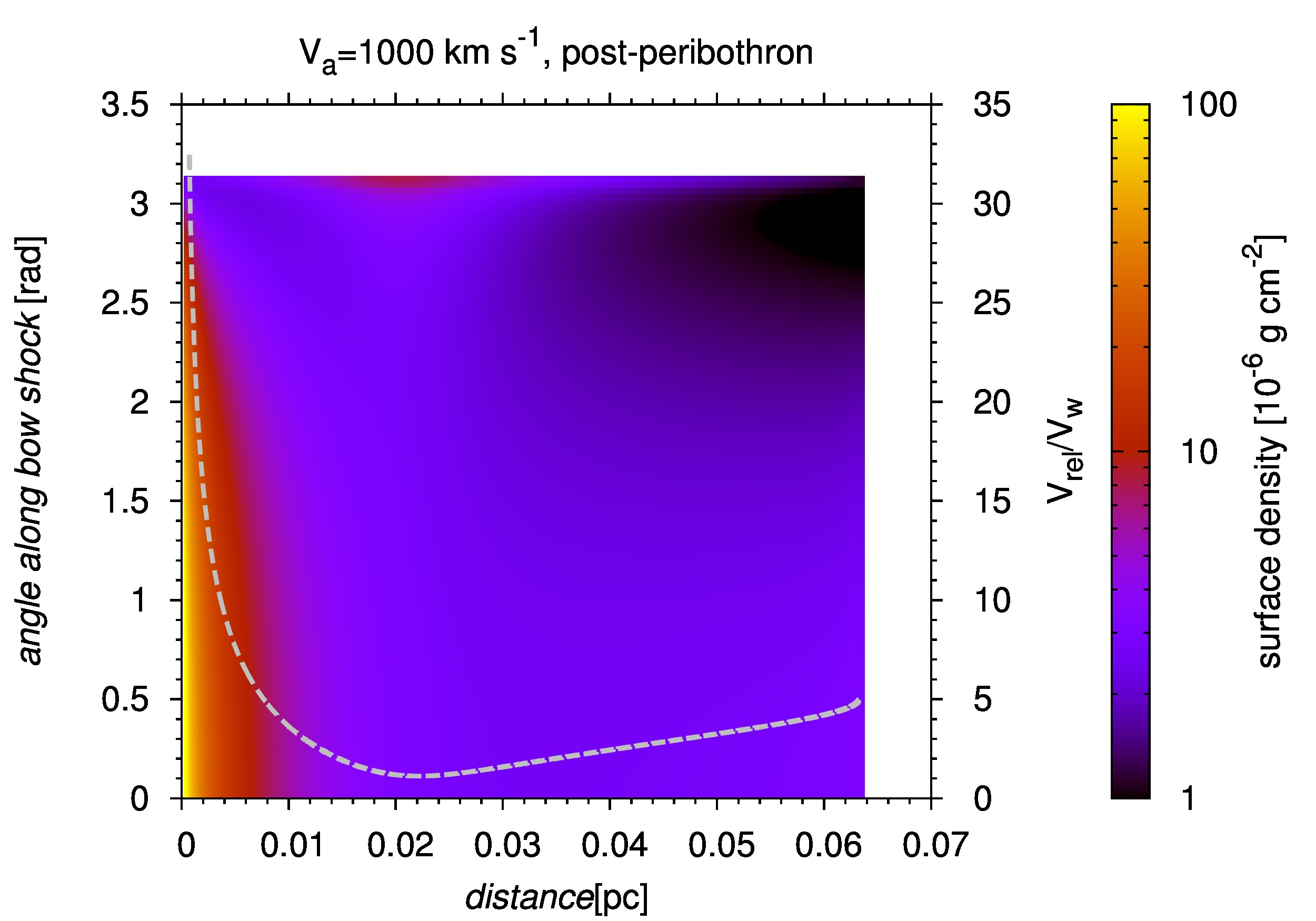}\\ 
     \includegraphics[width=0.5\textwidth]{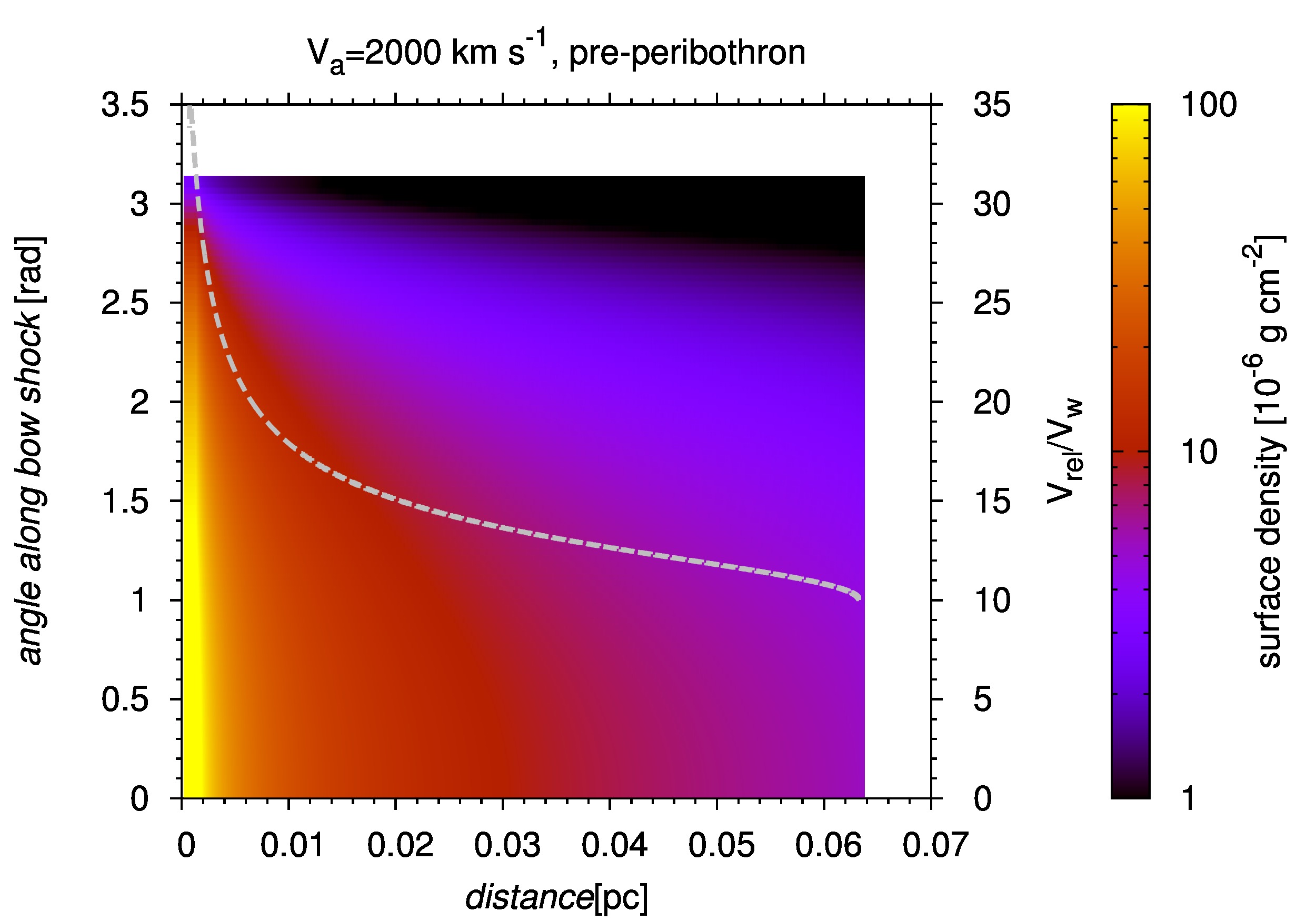} & \includegraphics[width=0.5\textwidth]{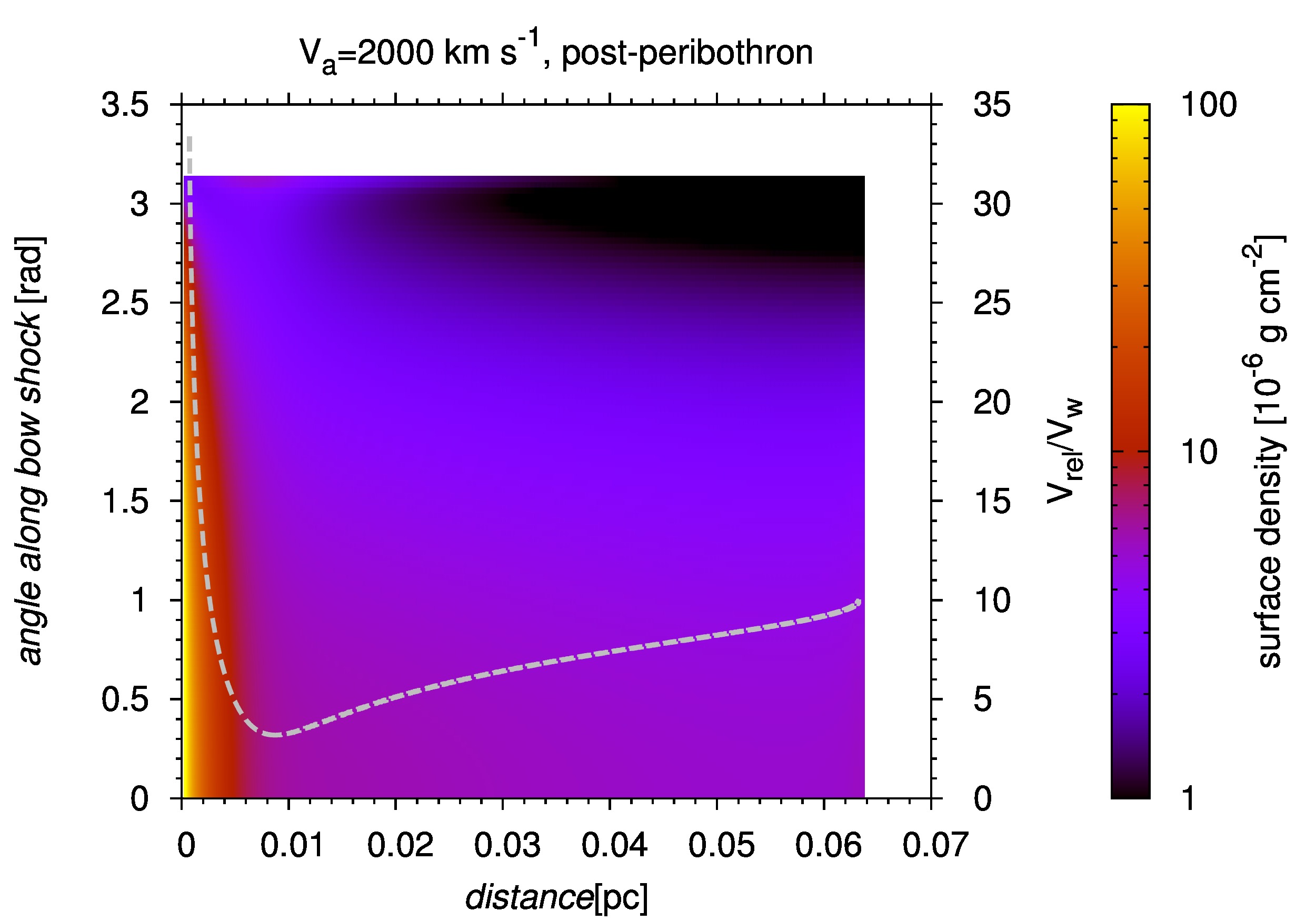} 
  \end{tabular}
  \caption{Colour-coded plots of the surface density of the stellar bow shock as a function of distance from the SMBH and the spherical angle measured from the symmetry axis of the bow shock. The surface density is expressed in $10^{-6}\,{\rm g\, cm}^{-2}$. The ratio of the relative velocity and the stellar wind velocity, $\alpha=v_{\rm{rel}}/v_{\rm{w}}$, is depicted as a grey line with the values along the right axis. The left-hand column corresponds to the pre-peribothron phase, while the right-hand column shows the post-peribothron part of the orbit.}
  \label{fig_surface_density}
\end{figure*}

The surface density along the stellar bow shock can be computed from eq. \eqref{eq_mass_flux1} combined with the shell velocity profile given by eq. \eqref{eq_shell_velocity} for the thin axisymmetric bow shock model. The colour-coded plots of the surface density are shown in Fig. \ref{fig_surface_density},  which are divided into the pre-peribothron phase in the left-hand column and the post-peribothron phase in the right-hand column as in the case of the shell velocity (see Fig. \ref{fig_shell_velocity_profiles}).   

The surface density is expressed in absolute units, specifically $10^{-6}\,{\rm g\, cm}^{-2}$, and is largest near the peribothron passage in all cases, where the bow shock shrinks in size significantly, see Fig. \ref{fig_stagnation_radius_alpha} (left column). The analysis of the angular dependence for a fixed distance during the pre-peribothron stage shows that the surface density drops for the down-stream part, as expected. However, the post-pericentre profile is different from the pre-pericentre evolution -- the decrease for larger angles is smaller and becomes apparent only for stronger outflows near the apocentre of the orbit. There is also a local maximum of the surface density at the distance where the minimum of the ratio $v_{\rm{rel}}/v_{\rm{w}}$ occurs. This shallower profile shifts closer to the Galactic centre for stronger outflows. 

\subsection{Relative change of emissivity along the shocked layer for different outflow velocities}
\label{subsection_emission_measure}

\begin{figure*}
  \centering
  \begin{tabular}{cc}
     \includegraphics[width=0.5\textwidth]{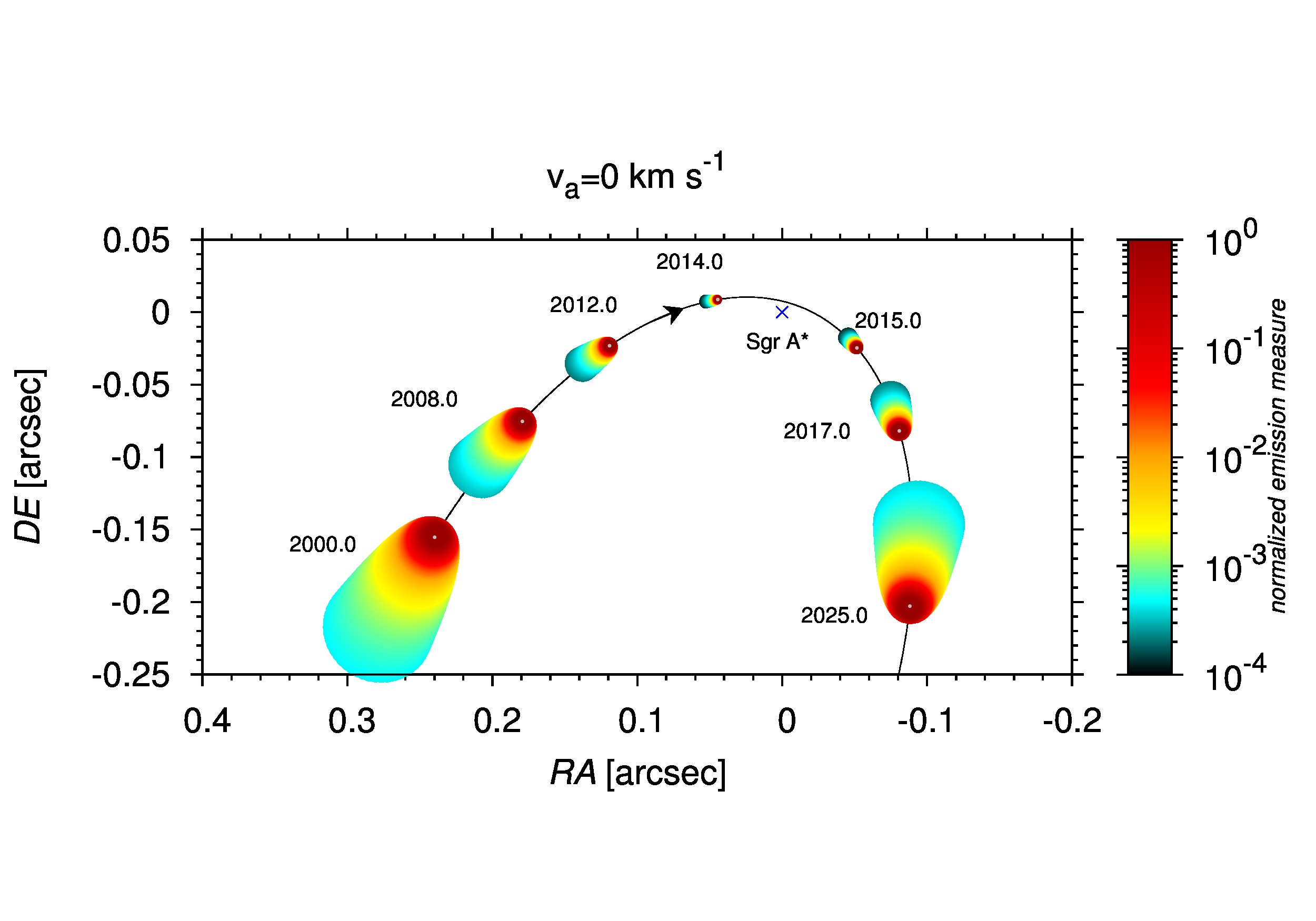} & \includegraphics[width=0.5\textwidth]{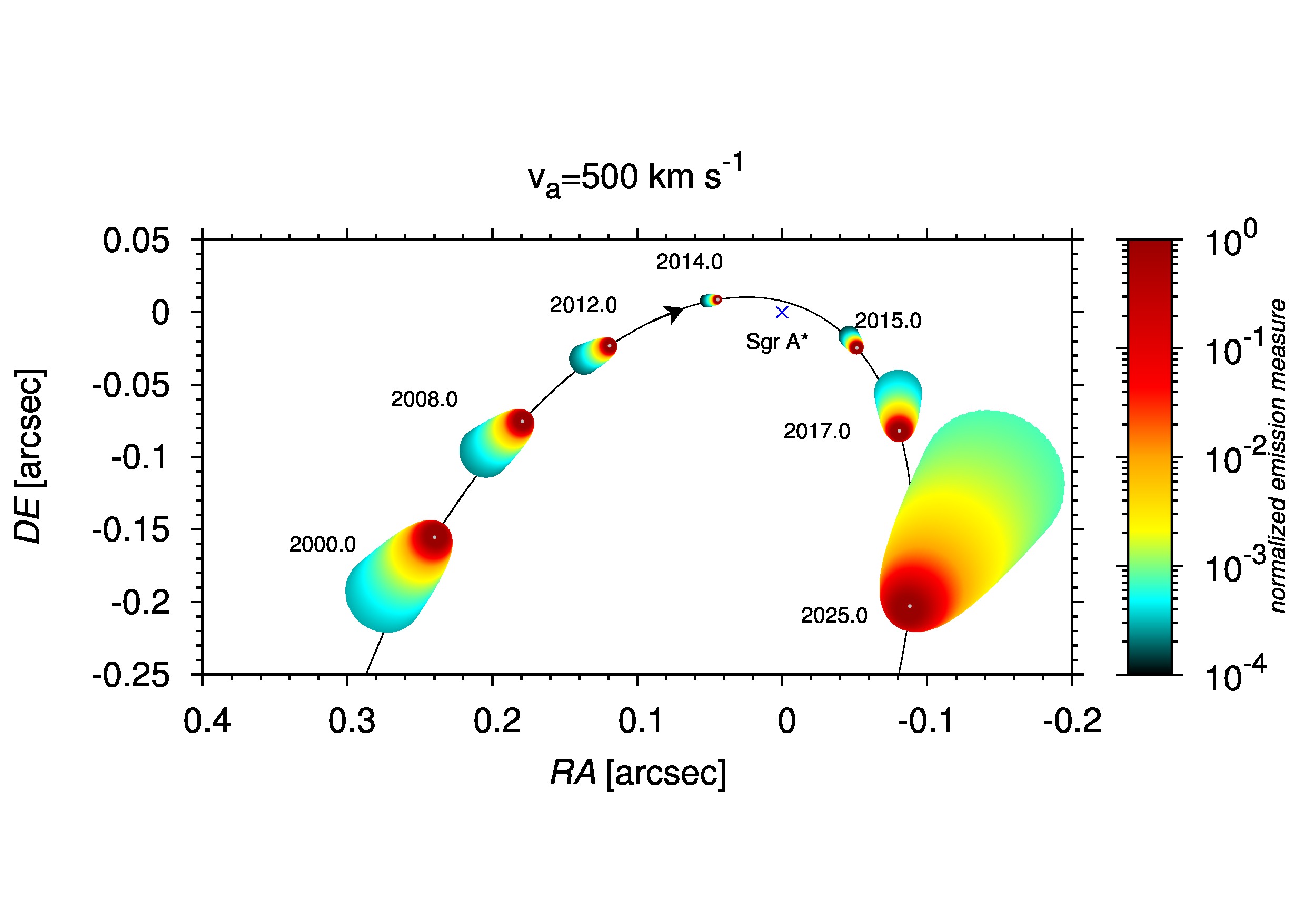}\\
     \includegraphics[width=0.5\textwidth]{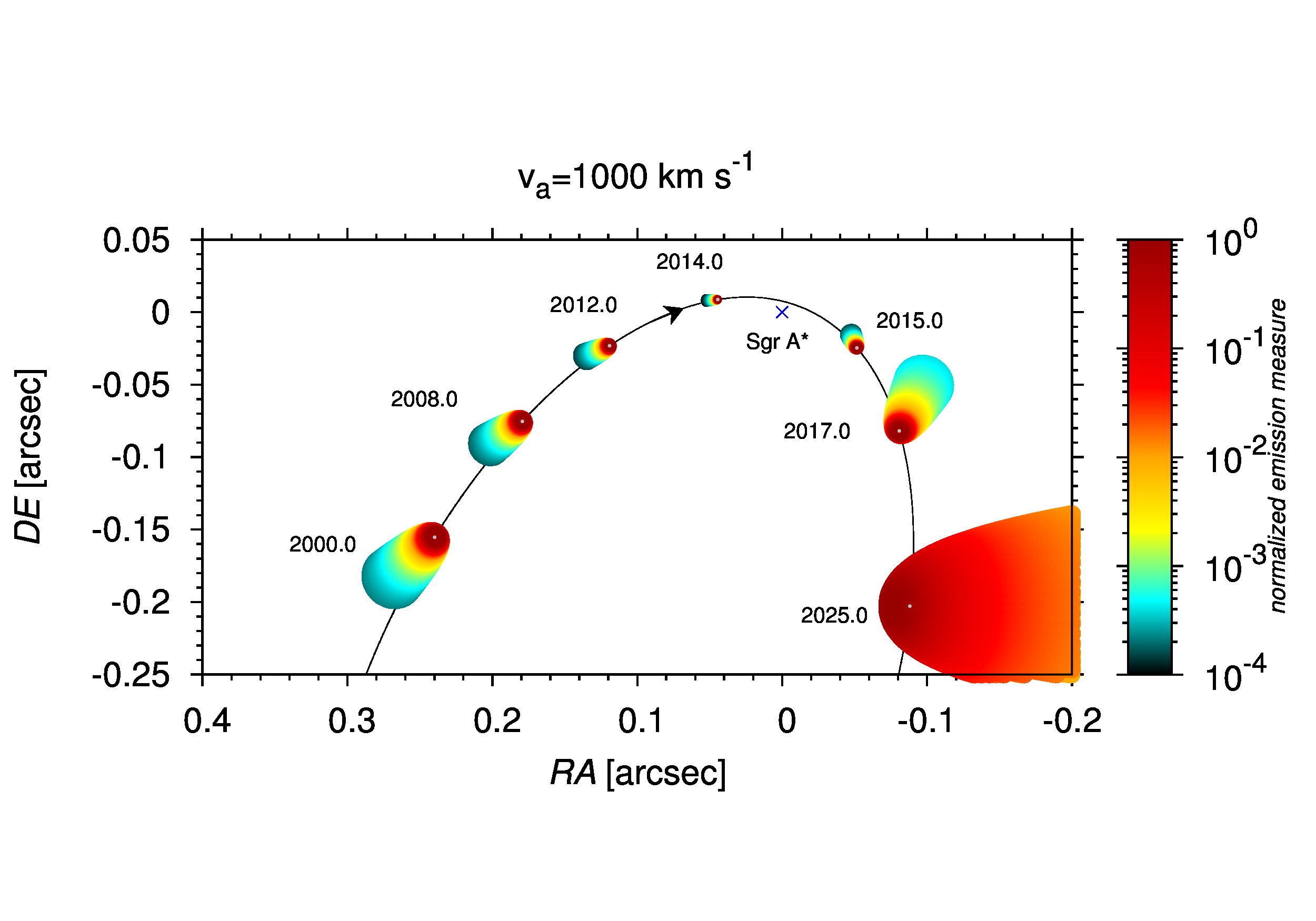} & \includegraphics[width=0.5\textwidth]{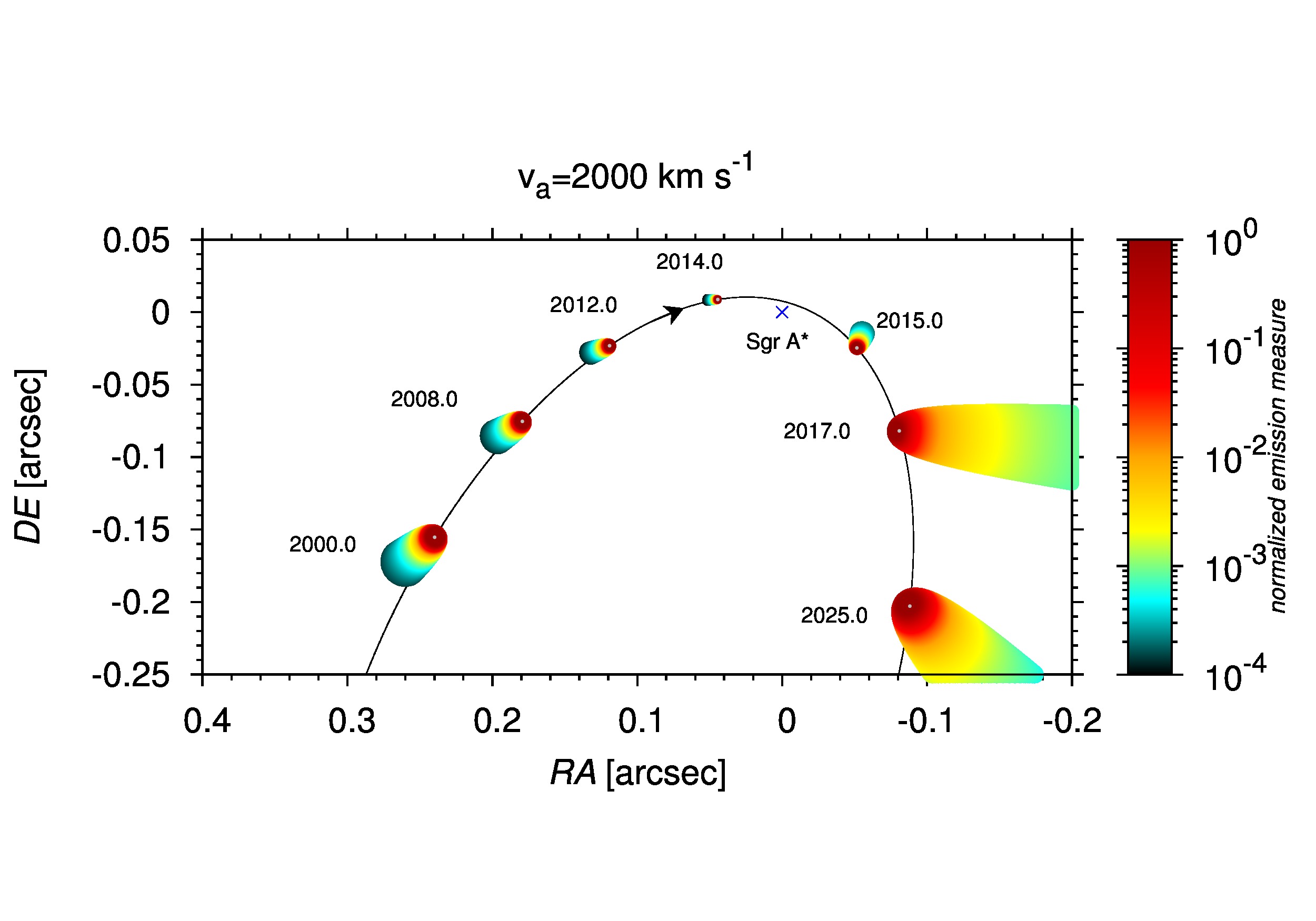}   
  \end{tabular}
  \caption{Emission measure maps of the DSO-like object normalised with respect to the maximum emission measure for each chosen epoch. The four cases correspond to four different outflow velocities: $0$, $500$, $1000$, and $2000\,{\rm km\, s}^{-1}$. The most apparent features are the change of the orientation and the size of the bow shock, and the differences in the emission measure distribution along the shocked shell in the post-pericentre part of the orbit.}
  \label{fig_emission_maps}
\end{figure*}

We compute emission maps for the wind-blowing source on an eccentric trajectory around the SMBH, which is a plausible scenario for the DSO. Using eq. \eqref{eq_emission_measure} we obtain the angular dependence for the emission measure. These maps are transformed to the observer's frame and normalized with respect to the maximum emission measure for each epoch (for the axisymmetric case the maximum is at the vertex of the bow shock). The results for different outflows are displayed in Fig. \ref{fig_emission_maps}. The most notable features are the change of the bow-shock size, orientation, and the emission measure distribution along the bow shock in the post-pericentre part of the orbit for different outflow velocities. The bow shock size scales with the stagnation radius $R_0$ whose evolution is in the left-hand part of Fig. \ref{fig_stagnation_radius_alpha}. For a stronger ambient wind, the maximum bow-shock size shifts from the apocentre closer to the SMBH. For negligible outflow the maximum angular size of the stagnation radius is $96\,{\rm mas}$, for $500\,{\rm km\,s}^{-1}$-outflow the radius decreases to $53\,{\rm mas}$, $1000\,{\rm km\,s^{-1}}$-outflow yields $19\,{\rm mas}$, and the stagnation radius shrinks to only $4\,{\rm mas}$ for an outflow of $2000\,{\rm km\,s^{-1}}$.  

\begin{figure*}
  \centering
  \begin{tabular}{cc}
  \includegraphics[width=0.5\textwidth]{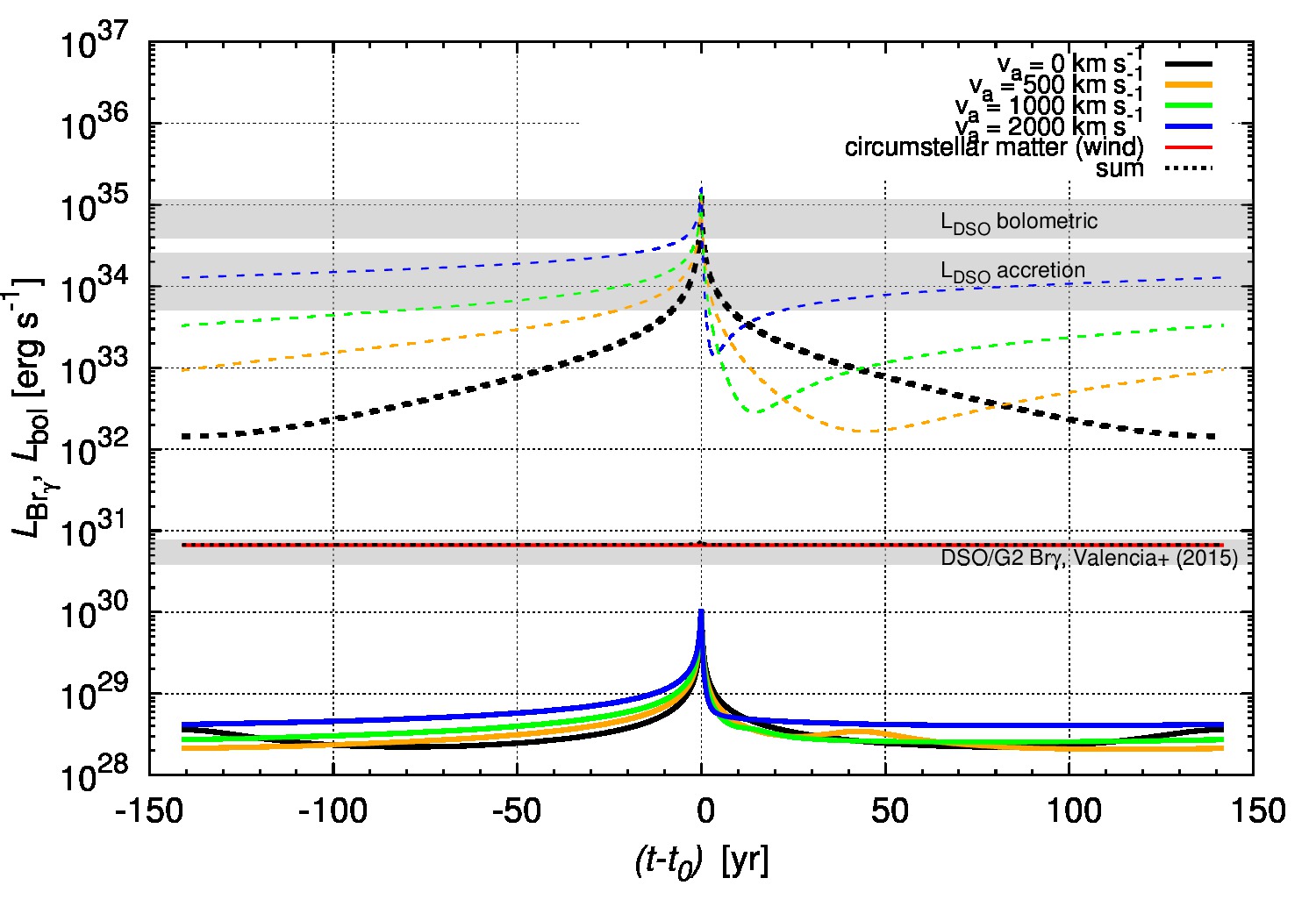} & \includegraphics[width=0.5\textwidth]{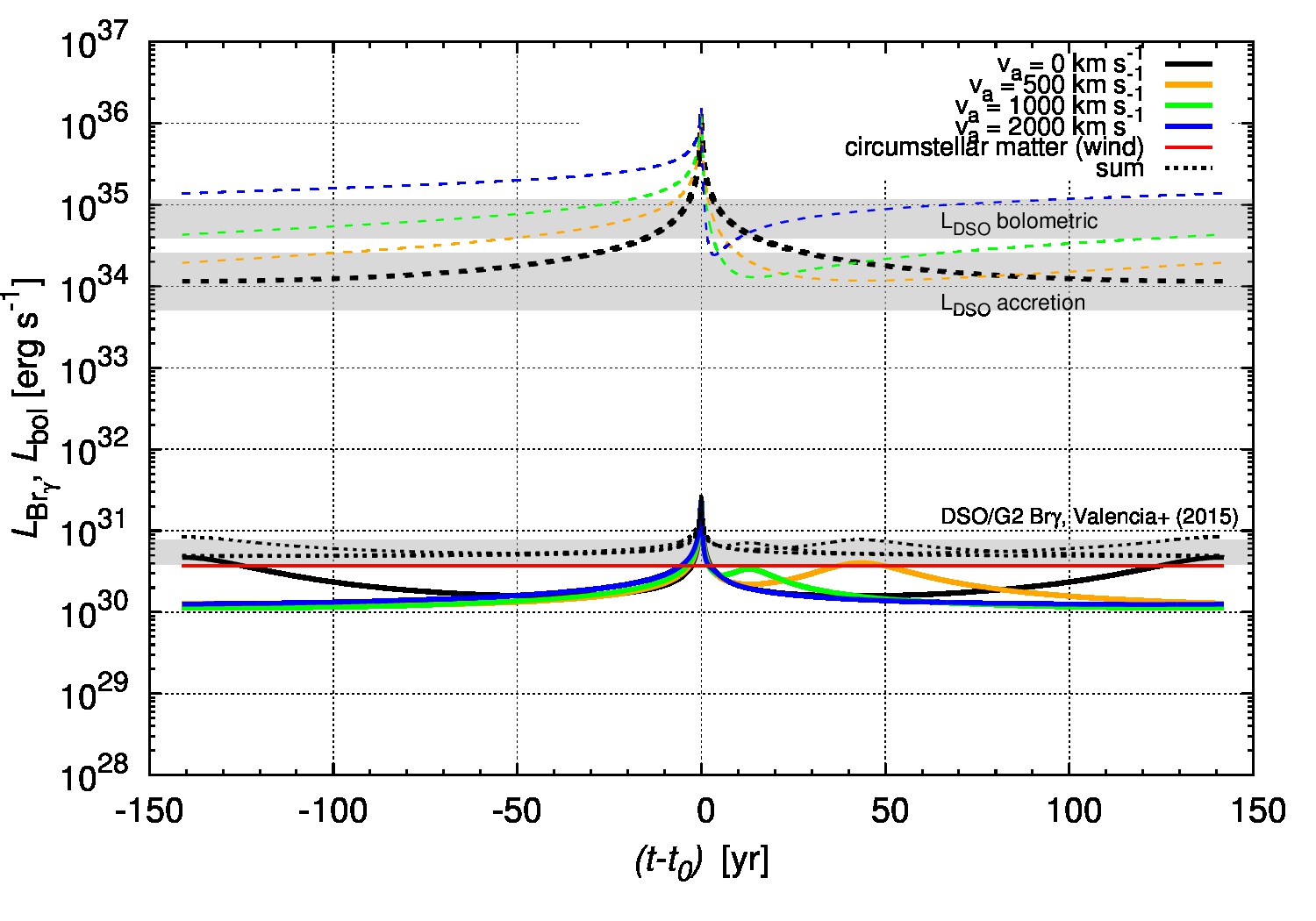}
  \end{tabular}
  \caption{\textit{Left}: Temporal evolution of $\rm{Br}\gamma$ luminosity for the following intrinsic stellar wind parameters: $\dot{m}_{\rm{w}}=10^{-8}\,{\rm M_{\odot}\,yr}^{-1}$, $v_{\rm{w}}=200\,\rm{km\,s^{-1}}$. The bow-shock luminosity is computed for different outflow velocities (see the legend). The contribution of the circumstellar matter (wind) and the total luminosity are also plotted for comparison. The grey areas mark the ranges of the observed Br$\gamma$ luminosity, bolometric luminosity, and the inferred accretion luminosity of the DSO source around the peribothron passage, see also \citet{2013A&A...551A..18E,2014ApJ...796L...8W,2015ApJ...800..125V}. Dashed lines represent an upper limit for total radiative luminosity for different outflow velocities. \textit{Right}: The same as the left panel for the following stellar wind parameters: $\dot{m}_{\rm{w}}=10^{-7}\,{\rm M_{\odot}\,yr}^{-1}$, $v_{\rm{w}}=600\,\rm{km\,s^{-1}}$.}
  \label{fig_luminosity_Brgamma}
\end{figure*}

When the layer of the shocked gas is externally ionised by the intense UV field and the shocked gas is approximately isothermal, the maps of the emission measure directly express the emissivity of hydrogen recombination lines and that of the free-free continuum radiation. This is often the case in expanding HII regions, such as the Orion nebula \citep{2000AJ....119.2919B,2001ApJ...546..299B}. The Galactic centre minicavity may be considered to be such a photoionised HII region, where massive OB/Wolf-Rayet stars provide Lyman continuum photons with the production rate of $N_{{\rm LyC}}\sim 10^{50}\,{\rm s}^{-1}$\citep{1985ApJ...293..445S}. 

\begin{figure*}
  \centering
  \begin{tabular}{cc}
    \includegraphics[width=0.5\textwidth]{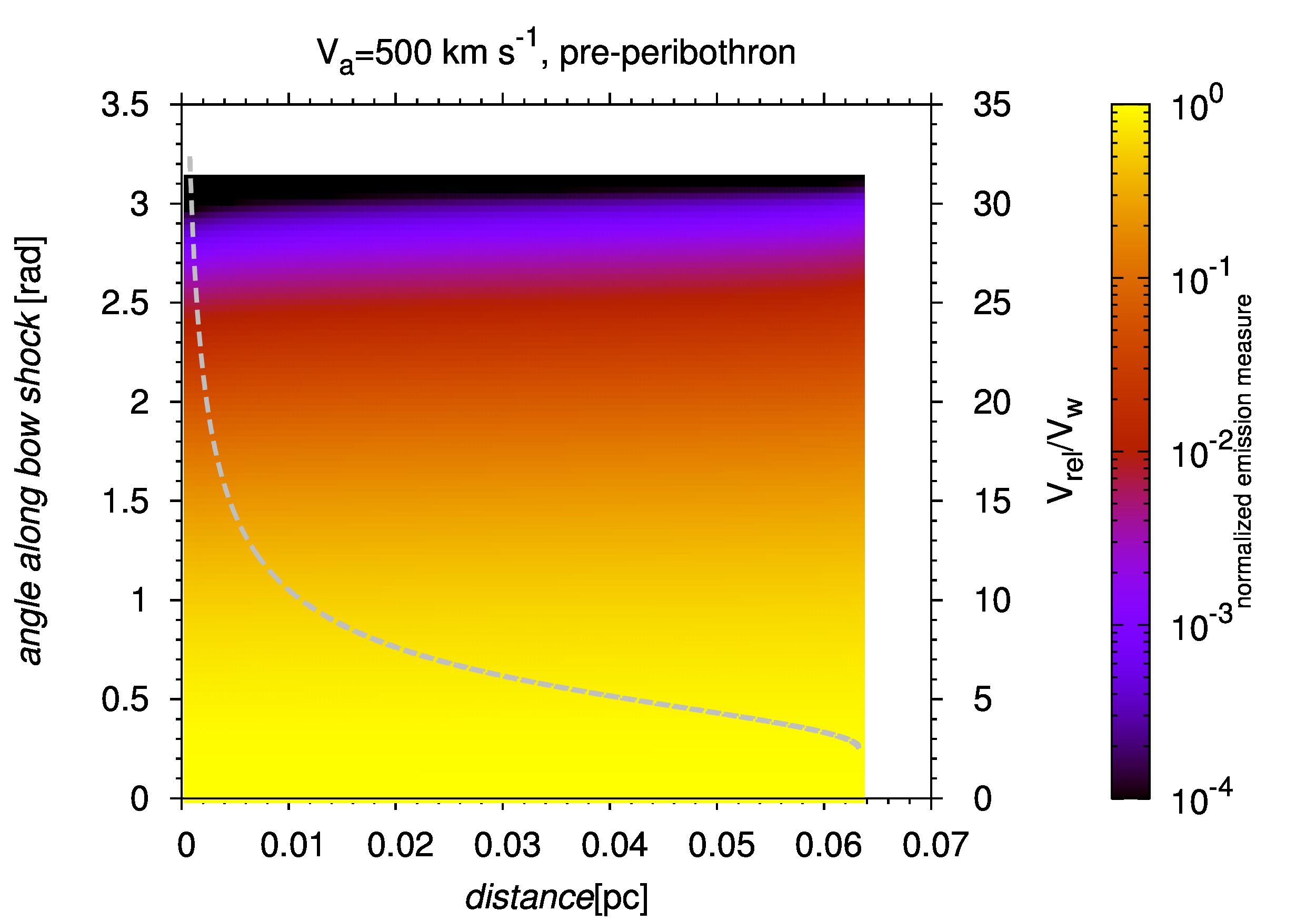} & \includegraphics[width=0.5\textwidth]{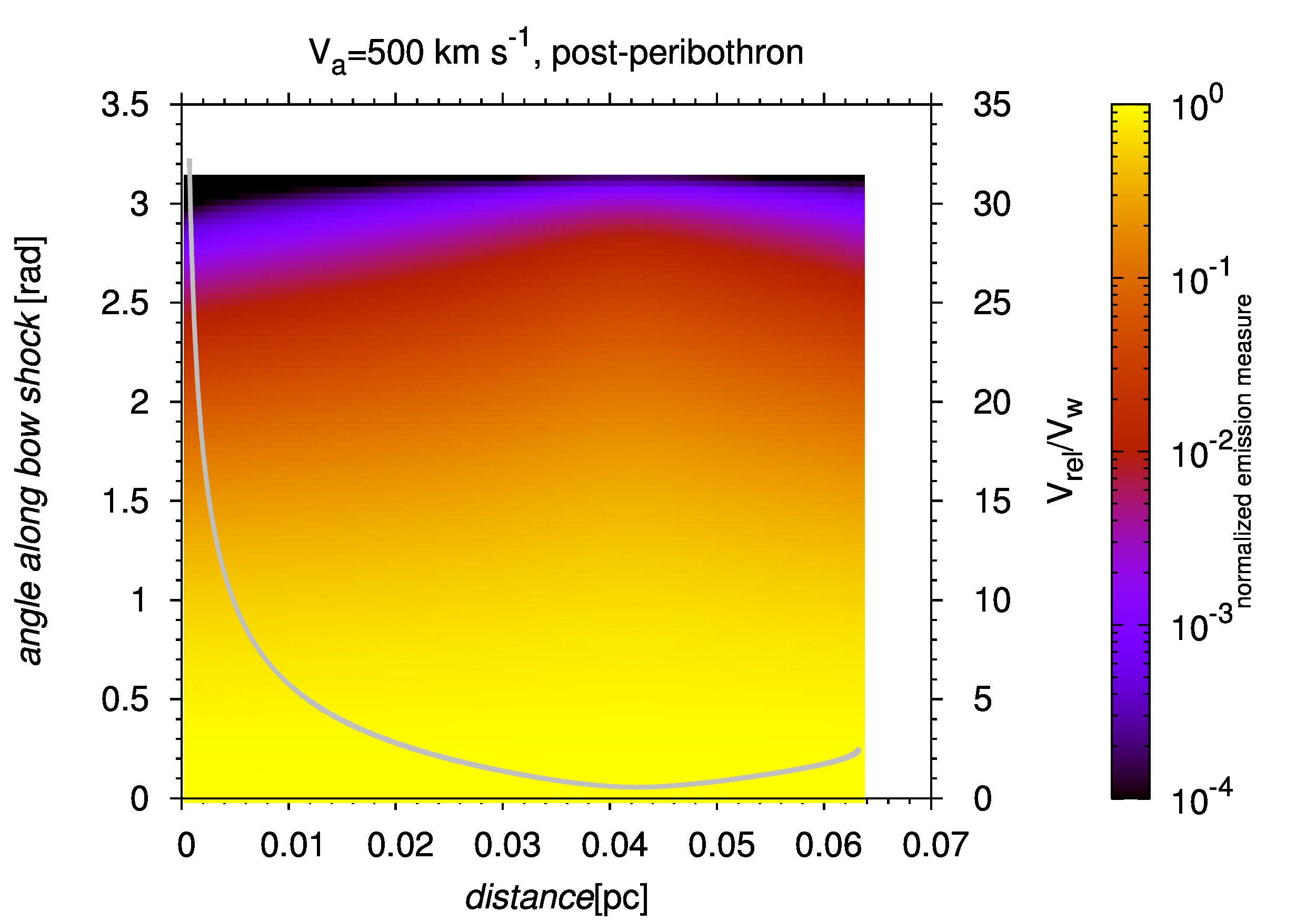}\\
    \includegraphics[width=0.5\textwidth]{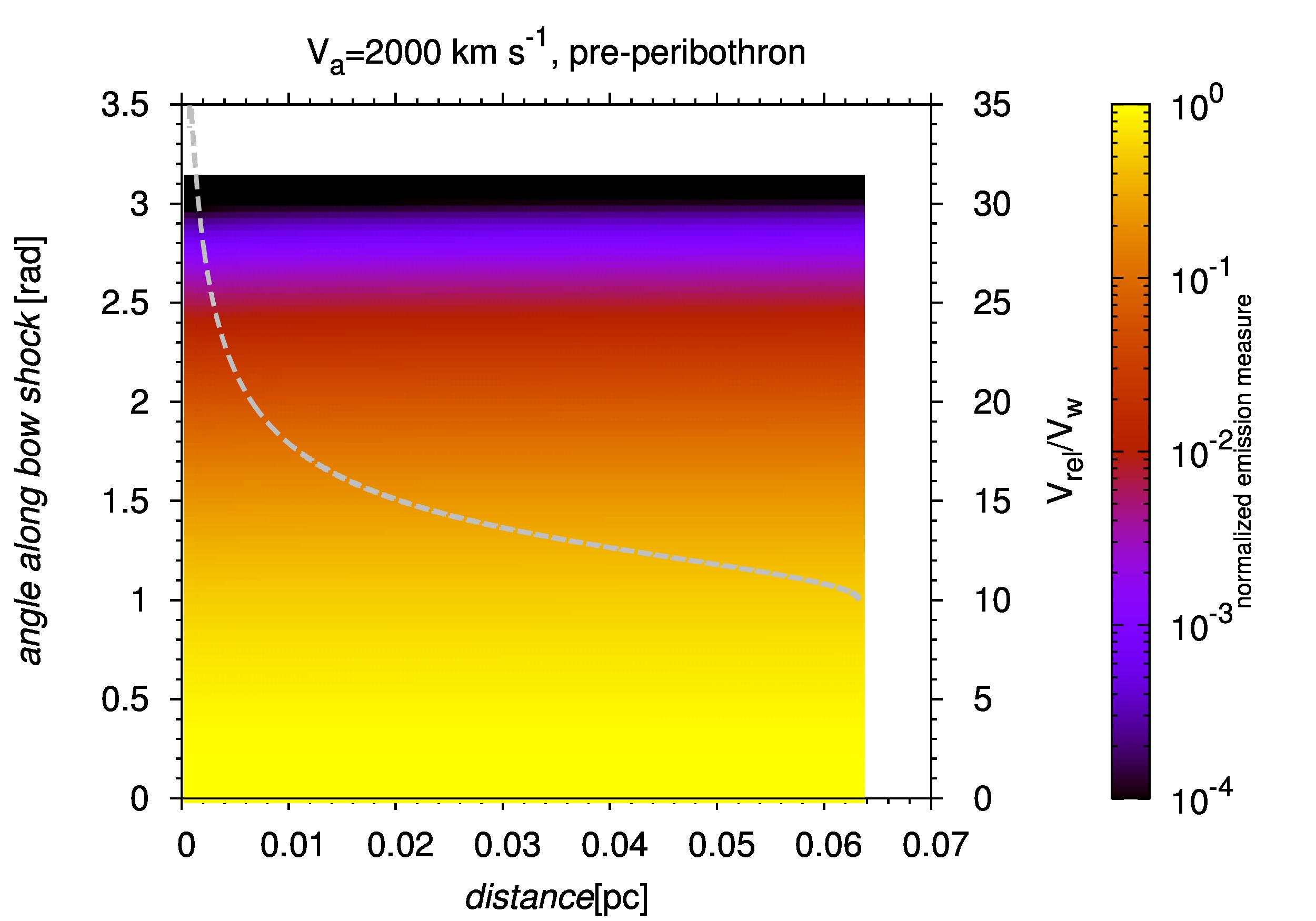} & \includegraphics[width=0.5\textwidth]{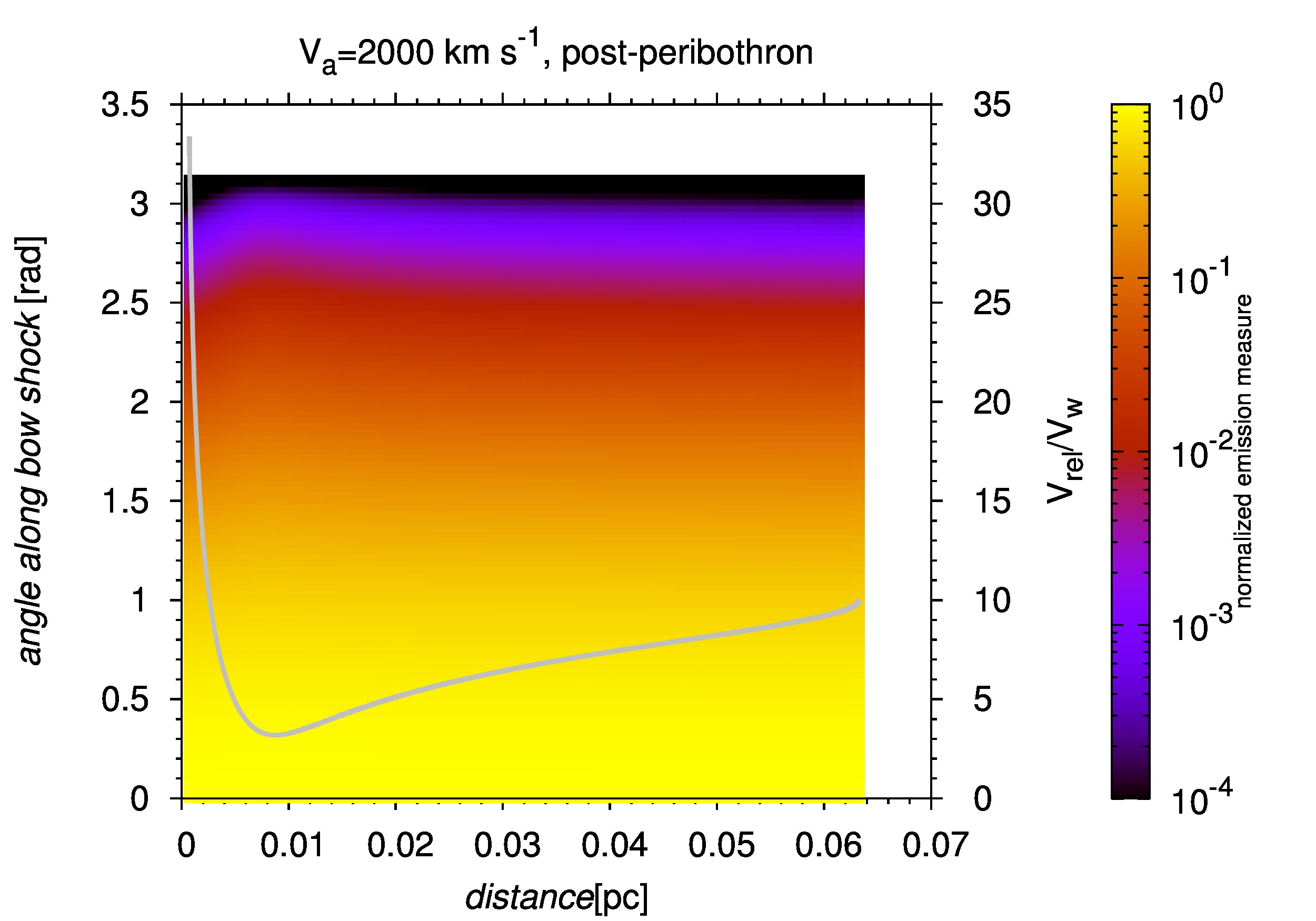}
  \end{tabular}
  \caption{The colour-coded angular profile of the normalized emission measure as a function of the distance from the SMBH. Top panels: $500\,{\rm km s}^{-1}$ outflow; the left-hand panel corresponds to the pre-peribothron phase, the right-hand panel represents the post-peribothron evolution. Bottom panels: $2000\,{\rm km s}^{-1}$ outflow.}
  \label{fig_emission_profs}
\end{figure*}

The emission profiles as functions of the spherical angle change with the distance from the SMBH as well as with the velocity of the ambient flow; see Fig.  \ref{fig_emission_profs} for $500\,{\rm{km\, s}^{-1}}$- and $2000\,{\rm km\, s}^{-1}$-outflow. For the pre-peribothron part of the orbital evolution the emission measure is largest near the axis of symmetry and then decreases to one thousandth of the peak value downstream for angles of $\lesssim 3\,\rm{rad}$. The comparison with the post-peribothron part (see the right-hand column in Fig. \ref{fig_emission_profs}) shows that there is a distance range  where the emission measure profile is flatter and decreases to one thousandth of the peak value for a larger angle, $\gtrsim 3\,\rm{rad}$. Again, this corresponds to the minimum of the ratio $v_{\rm{rel}}/v_{\rm{w}}$, which is plotted as a grey curve as in previous plots (Figs. \ref{fig_shell_velocity_profiles} and \ref{fig_surface_density}).

\subsection{Evolution of bow-shock luminosity and comparison with other sources of emission}
\label{subsection_comparison_emission}

 Furthermore, we compute the volumetric emission measure of the bow shock by multiplying the emission measure across the shocked layer, eq. \eqref{eq_emission_measure}, by the infinitesimal area of the bow shock at angle $\theta$, $EM_{\rm{V}}=EM \times 2\pi w \sqrt{\mathrm{d}z^2+\mathrm{d}w^2}$. By summing these contributions across the bow-shock layer we get the integrated emission measure that can be scaled to the luminosity for an optically thin case. In order to compare with the observations of Br$\gamma$ emission line of the DSO source, we transform the volumetric emission measure to Br$\gamma$ luminosity using the Case B recombination factor \citep[][see also \citet{2014ApJ...789L..33D} for comparison]{2013ApJ...776...13B}:

\begin{equation}
   L_{\rm{Br}\gamma}\approx 3.44 \times 10^{-27} (T/10^{4}\,\rm{K})^{-1.09} \textit{EM}_{\rm{V}}\,\rm{erg\,s^{-1}}\,,
   \label{eq_br_lum}
\end{equation}   
where the temperature of the shocked layer is taken to be approximately constant, $T=10^{4}\,\rm{K}$, due to the heating rate of UV field of massive OB stars.
We plot the temporal evolution of this luminosity for different outflow velocities from the Galactic centre in the left panel of Fig. \ref{fig_luminosity_Brgamma}. For all cases of the ambient outflow there is an increase of the bow-shock luminosity from the apobothron to the peribothron. For comparison, we compute the contribution from the isotropic stellar wind that is being launched at $0.01 \,\rm{AU}$ (in accordance with the analysis of stellar winds of young stars, \citeauthor{2006A&A...453..785F}, \citeyear{2006A&A...453..785F}). This comparison shows that the circumstellar matter can dominate for the typical parameters of a young star adopted here and the contribution of the bow shock to the overall emission measure is of the order of $1\%$ at the pericentre and $\gtrsim 0.01\%$ at the apocentre (see the contribution of the circumstellar matter and the overall luminosity in Fig. \ref{fig_luminosity_Brgamma}). The total luminosity of the wind and the bow shock is stable and our adopted parameters for the mass-loss rate,  $\dot{m}_{\rm{w}}=10^{-8}\,{\rm M_{\odot}\,yr}^{-1}$, and the terminal wind velocity, $v_{\rm{w}}=200\,\rm{km\,s^{-1}}$, approximately match the observed luminosity of the DSO source, as seen in the range of observed values in Fig. \ref{fig_luminosity_Brgamma} according to \citet{2015ApJ...800..125V} (grey band).  

In general, the ratio of the bow-shock luminosity to the luminosity associated with the outflow (and also inflow) depends on the mass-loss rate and velocity field, as well as temperature and density of the circumstellar flow (see eq. \eqref{eq_br_lum}). Hence, the observability of bow shocks varies for different stages of stellar evolution \citep[see][especially their Table 3]{2014MNRAS.444.2754M}. As an exemplary case, we compute the temporal evolution of Br$\gamma$ luminosity for an order of magnitude larger mass-loss rate, $\dot{m}_{\rm{w}}=10^{-7}\,{\rm M_{\odot}\,yr}^{-1}$, and larger terminal wind velocity, $v_{\rm{w}}=600\,\rm{km\,s^{-1}}$ (right panel of Fig. \ref{fig_luminosity_Brgamma}). In this case the bow-shock luminosity is comparable to the stellar wind luminosity and is even greater at the pericentre of the orbit. The total $Br\gamma$ luminosity can then increase by a factor of a few during the pericentre passage. This is not detected by \citep{2015ApJ...800..125V} but \citet{2015ApJ...798..111P} report an indication of a small increase.      

An upper limit for the radiative bow-shock luminosity may be simply derived from the sum of kinetic terms of colliding ambient and stellar winds, since a fraction of the kinetic energy is thermalised \citep{1997IAUS..182..343W,2012A&A...541A...1M}:

\begin{equation}
  \dot{E}_{\rm{tot}}\approx \dot{E}_{\rm{amb}}+\dot{E}_{\rm{wind}}=\frac{1}{2}\dot{m}_{\rm{w}}(v_{\rm{rel}}^2+v_{\rm{w}}^2)\,.
  \label{eq_radiative_lum}
\end{equation}

The temporal evolution of $\dot{E}_{\rm{tot}}$ is shown in Fig. \ref{fig_luminosity_Brgamma} (dashed lines). The maximum value for the case with the mass-loss rate of $\dot{m}_{\rm{w}}=10^{-8}\,{\rm M_{\odot}\,yr^{-1}}$ (left panel of Fig. \ref{fig_luminosity_Brgamma}) is $\sim 10^{35}\,{\rm erg\,s^{-1}}$ and the maximum for the case with $\dot{m}_{\rm{w}}=10^{-7}\,{\rm M_{\odot}\,yr^{-1}}$ (right panel of  Fig. \ref{fig_luminosity_Brgamma}) is one order of magnitude larger. One should take into account that this upper limit is associated with the bolometric luminosity of the bow shock (not only Br$\gamma$ luminosity) and only about $10\%$ of the kinetic energy is thermalised, as inferred from the comparison with hydrodynamic simulations \citep{2012A&A...541A...1M}. For completeness, the upper limit for the bolometric luminosity of the DSO is estimated to be $\sim 10\,L_{\odot}$ -- $\sim 30\,L_{\odot}$ \citep{2013A&A...551A..18E,2014ApJ...796L...8W,2015ApJ...800..125V}, which means that the potential overall contribution of the bow-shock luminosity to the luminosity of the DSO is of the order of $\lesssim 10\,\%$, as inferred from the left panel of Fig. \ref{fig_luminosity_Brgamma}, taking into account the conversion factor of $10\%$ between the kinetic energy and the radiative luminosity of the bow shock.

 Another plausible contribution for the case of a young star is the accretion flow onto the stellar surface whose origin is an accretion disc surrounding the star, see Fig. \ref{fig_star_dso_illustration} (left panel) for illustration. The density and the emission measure of material flowing along accretion funnels was already computed and discussed in \citet{2015ApJ...800..125V} for an axisymmetric magnetospheric model of the accretion flow around the DSO. In the framework of the model of a pre-main-sequence star there is a known correlation between the line luminosity $L({\rm Br}\gamma)$ and the accretion luminosity $L_{\rm acc}$ \citep{2014A&A...561A...2A},
 
 \begin{equation}
 \log{(L_{\rm{acc}}/L_{\odot})}=1.16 (0.07)\log{[L(\rm{Br} \gamma) / L_{\odot}]} + 3.60 (0.38)\,.
 \label{eq_correlation_luminosities}
 \end{equation}
For the DSO source, its measured Br$\gamma$ emission-line luminosity is of the order of $L(Br\gamma)= f_{\rm{acc}} \times 10^{-3}\, L_{\odot}$ and stays approximately constant. The factor $f_{\rm{acc}}$ is of the order of unity. From eq. \eqref{eq_correlation_luminosities} we get the following values for the accretion luminosity, $(1.3,2.9,4.7,6.6)\,L_{\odot}$ for $f_{\rm{acc}}=\{1,2,3,4\}$, see Fig. \ref{fig_luminosity_Brgamma} for the comparison with the bolometric luminosity of the DSO source and the range of Br$\gamma$ luminosity. 
 

 The comparison of various contributions (bow shock, stellar wind, accretion flow) indicates that the circumstellar matter can dominate over the bow-shock emission. Hence, the overall luminosity from the stellar bow-shock source can remain constant within measurement uncertainties along its trajectory around the SMBH unless the intrinsic properties of the star, i.e. inflow and outflow rates, change considerably. Significant changes in both the inflow and the outflow take place due to stellar evolution, which is expected on the timescale of $10^{3}$--$10^{4}$ orbital periods ($P_{\rm{orb}}\approx 100\,\rm{yr}$ for the source closely bound to the SMBH in the S-cluster).

\subsection{Contribution to the Doppler broadening of emission lines}
\label{subsection_Doppler}

The bow shocks of low-mass pre-main-sequence stars can be analysed using hydrogen recombination lines. It was shown \citep{2013ApJ...768..108S} that the dense shocked layer of colder gas in the bow shock could significantly contribute to the hydrogen line emission of the DSO if it is a compact stellar source.
 The DSO has been monitored using the ${\rm Br}\gamma$ emission line and its line width is reported to increase upon approaching Sgr~A* \citep{2015ApJ...798..111P,2015ApJ...800..125V}. 

\begin{figure*}
  \centering
\begin{tabular}{cc}  
  \includegraphics[width=0.5\textwidth]{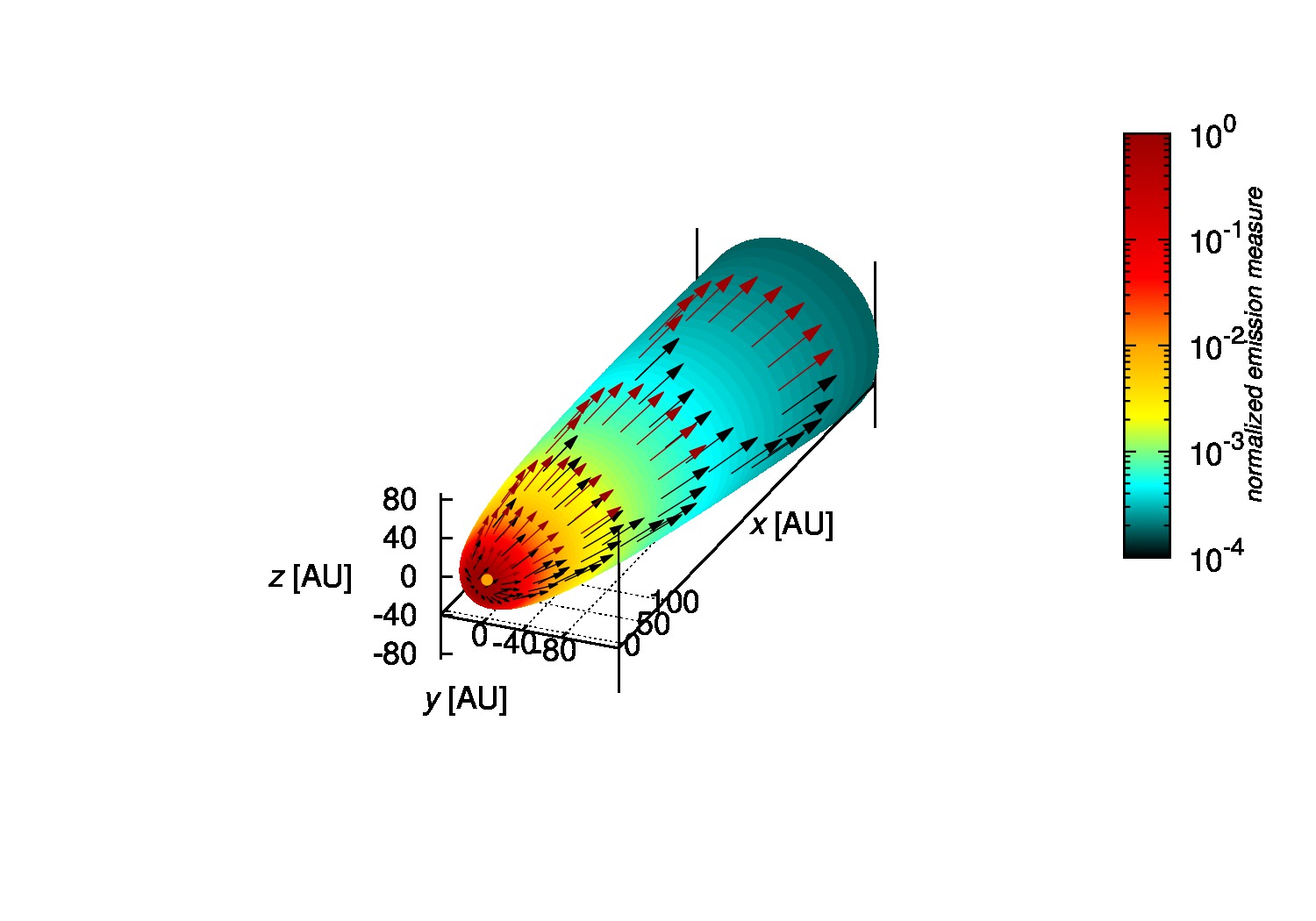}  &   \includegraphics[width=0.5\textwidth]{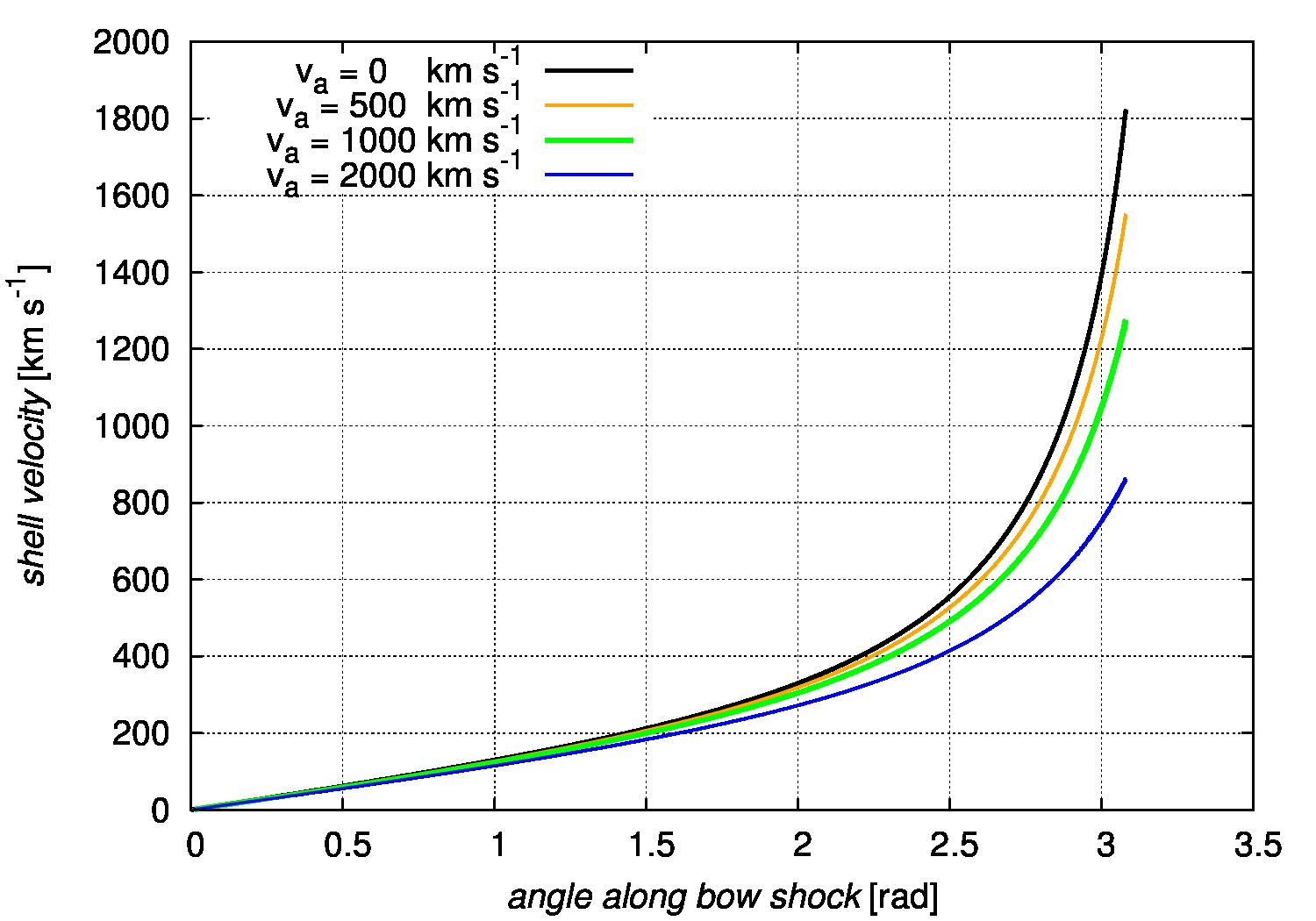}
\end{tabular}  
  \caption{\textit{Left}: 3D velocity field of the shocked gas in the stellar bow shock. The units along axes are in astronomical units with the origin at the position of the star (orange point). The colour wedge labels the normalized emission measure with respect to its maximum. \textit{Right}: Velocity profile along the shocked shell for different velocities of the outflow: 0, 500, 1000, and 2000 ${\rm km\,s^{-1}}$, respectively (see the legend). The parameters used for computing both the emission map and the velocity field were adopted from the DSO model ($\dot{m}_{\rm w}=10^{-8}\,M_{\odot}\,{\rm yr}^{-1}$, $v_{\rm w}=200\,{\rm km \, s}^{-1}$) for the epoch of $2016.0$.}
  \label{fig_velocity_field}
\end{figure*}

The shell velocity increases from the stagnation point towards the rear part of the bow shock, as seen in the shell velocity maps in Fig. \ref{fig_shell_velocity_profiles}. The 3D velocity field along the shocked layer is displayed in the left panel of Fig. \ref{fig_velocity_field} in the coordinate system centred on the stellar source. The shell velocity increases from $0\,{\rm km\,s}^{-1}$ at the vertex of the bow shock up to $\approx 1\,000\,{\rm km\,s}^{-1}$ downstream (the right panel of Fig. \ref{fig_velocity_field}). The velocity along the bow shock decreases at stronger outflow velocities. 

\begin{figure*}
  \centering
  \begin{tabular}{cc}
     \includegraphics[width=0.5\textwidth]{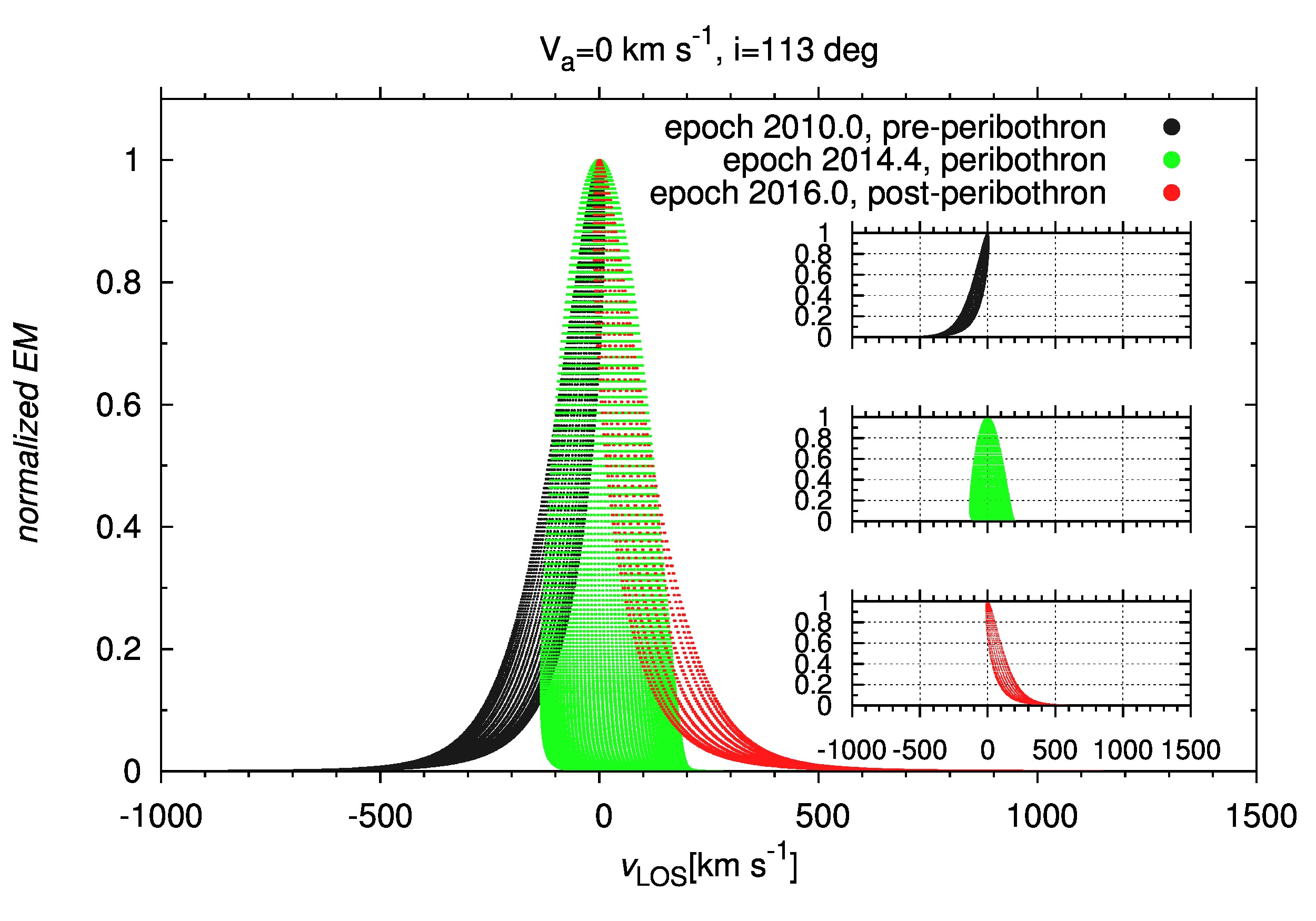} & \includegraphics[width=0.5\textwidth]{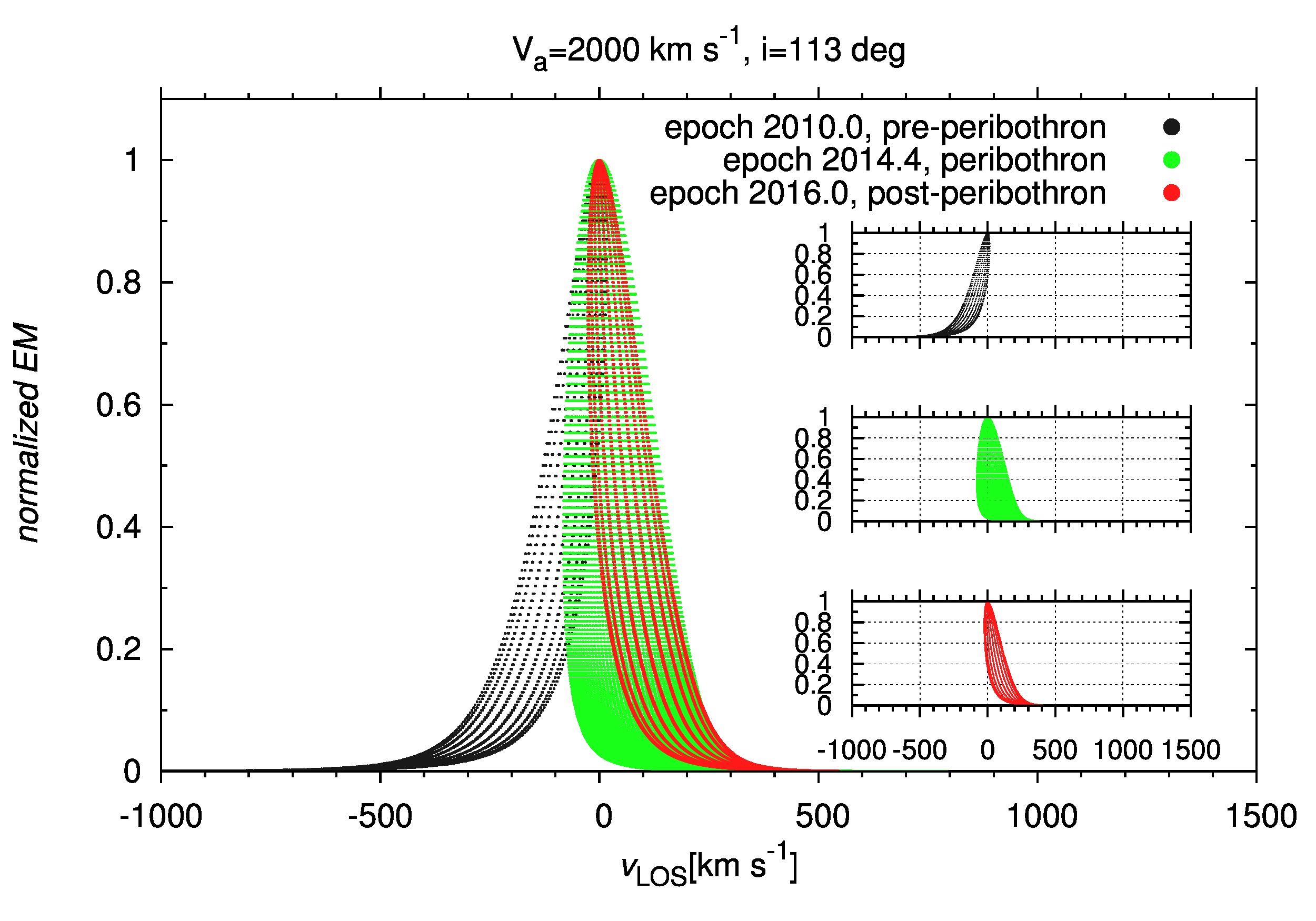}  
  \end{tabular}
  \caption{The plots of normalised emission measure versus the line-of-sight component of the velocity field of the shocked shell material. The depicted coloured regions show the Doppler contribution to the emission line width from the bow shock of a stellar source moving around the SMBH. The left-hand panel corresponds to the set-up with no outflow from the centre. The right-hand panel shows the model with a strong outflow of $\rm 2000\,{km s}^{-1}$ from the Galactic centre. The three bands of different colour correspond to the Doppler contributions for three different epochs: the black band indicates the pre-peribothron phase ($\sim 2010.0$ for the DSO source), green points mark the peribothron passage ($\sim 2014.4$ for the DSO), and the red points correspond to the immediate post-peribothron phase ($\sim 2016.0$ for the DSO). These contributions are separately plotted in the figure insets. The orbital elements used for the calculation correspond to the inferred orbit of the DSO source with the orbital inclination of $i=113^{\circ}$ \citep{2015ApJ...800..125V}.}
  \label{fig_doppler_contribution}
\end{figure*}

To illustrate the effect of the bow shock flow on the width and the shape of emission lines, we calculate the Doppler contribution of the shell velocity field along the line of sight for pre-peribothron, peribothron, and immediate post-peribothron phases. As a specific example, we adopt the parameters of the DSO assuming it is a young star \citep{2015ApJ...800..125V}. Plots of the emission measure as a function of the line-of-sight shell velocity are depicted in Fig. \ref{fig_doppler_contribution} for two outflow models: no outflow (left-hand side) and a strong outflow of $2000\,{\rm km s}^{-1}$ (right-hand side). We assume an optically thin bow-shock flow, i.e. the whole shocked layer contributes to the observed line emission and there is no shielding effect included in the computation. In Fig. \ref{fig_doppler_contribution} the line-of-sight velocity of the shocked flow is corrected for the motion of the star around the SMBH.

We determine the size of the velocity span for an arbitrary value of the normalised emission measure, which we set to $0.5$. The comparison of the velocity span for the same value of the emission measure at all epochs can be considered as measure of the change of the line width of hydrogen recombination lines produced in the bow shock layer. The values of the velocity span for a given outflow model and the epoch are summarised in Table \ref{tab_velocity_range}.

\begin{table*}
  \centering
  \resizebox{0.7\textwidth}{!}{
  \begin{tabular}{|c|ccccccc}
      \hline
      \hline
      Outflow velocity & \multicolumn{6}{ |c }{Velocity span at $EM=0.5$}\\ \hline     
                       & 2008.0 & 2010.0 & 2012.0 & \textbf{2014.4} & 2016.0 & 2018.0 & 2020.0 \\
       $[{\rm km\, s}^{-1}]$ & $[{\rm km\, s}^{-1}]$ & $[{\rm km\, s}^{-1}]$ & $[{\rm km\, s}^{-1}]$ & $[{\rm km\, s}^{-1}]$ &  $[{\rm km\,s}^{-1}]$  & $[{\rm km\,s^{-1}}]$ & $[{\rm km\,s}^{-1}]$ \\
      \hline
      $0$  &  $91$  &   $97$   & $109$    &   $\mathbf{233}$         &     $106$     & $91$ & $86$ \\
      $2\,000$ & $116$  &  $123$ & $138$   &  $\mathbf{218}$        &     $137$     & $182$ & $162$  \\
      \hline   
  \end{tabular}
  }
  \caption{The size of the line-of-sight velocity span for two ambient outflow scenarios (no outflow and $2000\,{\rm km s}^{-1}$) and several epochs in the pre-peribothron (2008.0, 2010.0, 2012.0), peribothron (2014.4), and post-peribothron phase (2016.0, 2018.0, 2020.0). The numbers in bold stand for the velocity span at the peribothron of the DSO source. The corresponding profiles of the normalised emission measure with respect to the line-of-sight velocity are plotted in Fig. \ref{fig_doppler_contribution} for three chosen epochs. The parameters for the calculation were adopted from the stellar scenario of the DSO. }
  \label{tab_velocity_range}
\end{table*}

The basic feature of the evolution of the velocity span is that in both cases of the outflow model the width increases by approximately a factor of $2$ towards the pericentre and then decreases by the same factor. Qualitatively, a similar tendency is observed for the FWHM of ${\rm Br}\gamma$ line for the DSO source: before pericentre the FWHM was $\sim 200\,{\rm km s}^{-1}$, at the pericentre it increased up to $\sim 500\,{\rm km s}^{-1}$, and the first observations after the pericentre indicate its decrease \citep{2015ApJ...800..125V}. Although the observed values of the line width seem to be larger in comparison with our model, the Doppler contribution of the bow shock is modulated by the terminal speed of the stellar wind, which is highly uncertain. For the young stellar sources the terminal wind speed could be also larger by a factor of a few than the value used in our calculation (we adopted $v_{{\rm w}}=200\,{\rm km s}^{-1}$). 

The outflow from the centre causes the line width to increase again in the post-peribothron phase at a certain epoch (see epoch $2018.0$ in Table \ref{tab_velocity_range} for the case of 2000 ${\rm km\, s}^{-1}$ outflow). This is caused by the change of the orientation of the bow shock in the post-peribothron case, see the epoch $2017.0$ in Fig. \ref{fig_emission_maps} for 2000 ${\rm km\, s}^{-1}$-outflow. The velocity span increases because of the contribution of both approaching and receding flows in the shell.

\begin{figure}
  \includegraphics[width=0.5\textwidth]{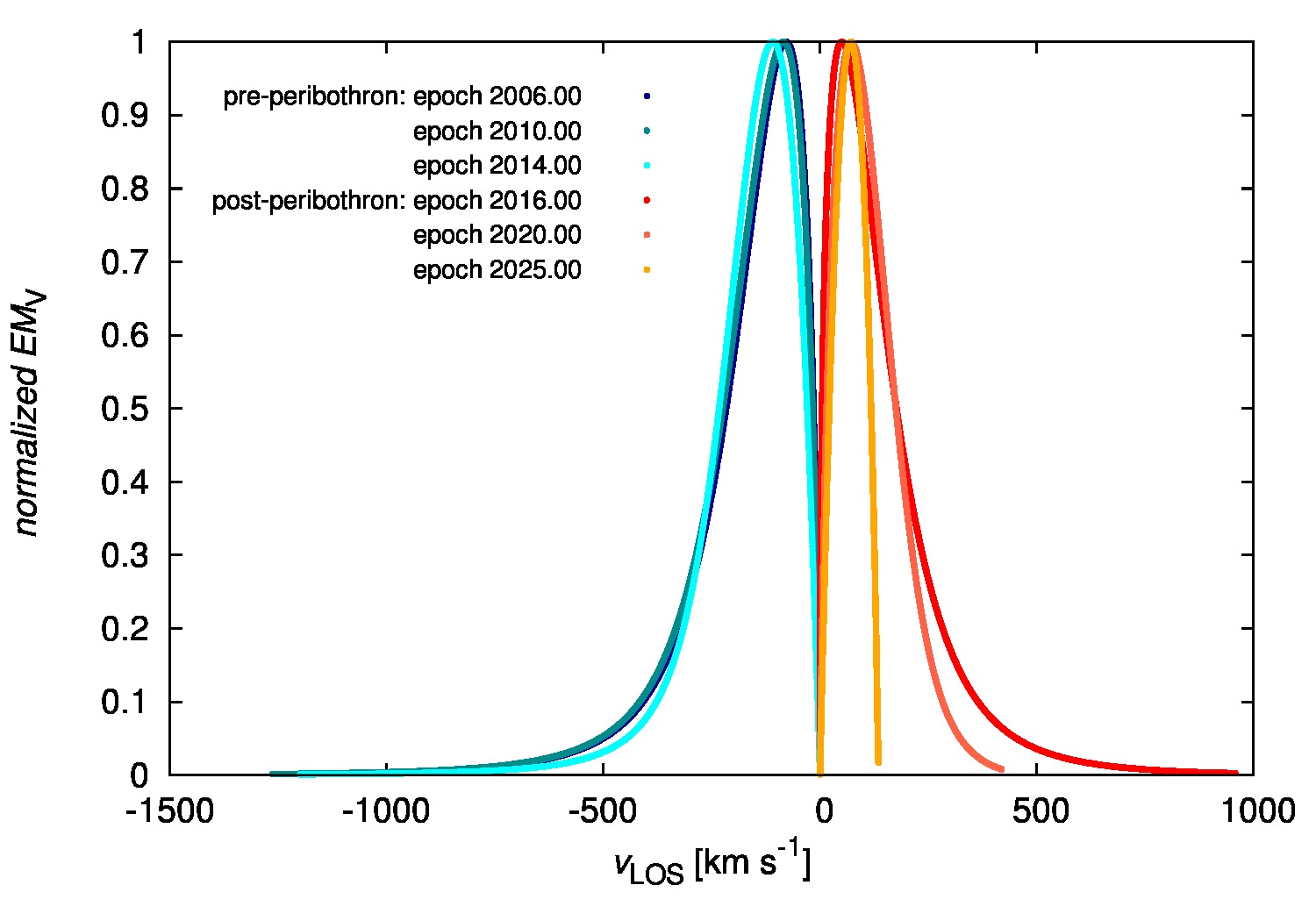}
  \caption{Computed emission line profiles for the outflow of $1000\,{\rm km\,s^{-1}}$ and different epochs in the pre-peribothron and post-peribothron part (see the legend).}
  \label{fig_line_profile}
\end{figure}

The plots in Fig. \ref{fig_doppler_contribution} also indicate that the pre-peribothron emission measure profile is blueshifted and the post-pericentre profile redshifted, which can skew the resulting line profile or at least make it asymmetric. We demonstrate this by constructing synthetic line profiles of bow-shock emission. First, we weigh the emission measure calculated using eq. \eqref{eq_emission_measure} by the area of infinitesimal bow-shock slices and these contributions are further normalised with respect to the maximum. Consequently, we plot the weighted normalised emission measure as a function of the line-of-sight velocity of the shocked gas corresponding to the given slice (see Fig. \ref{fig_line_profile} for the outflow velocity of $1000\,{\rm km\,s^{-1}}$). In the pre-peribothron part the flow is intrinsically blue-shifted, whereas in the post-pericentre part the emission is red-shifted. The profile is highly asymmetric and variable in all cases. This is consistent with the observed spectra of bow-shock knots present in Herbig-Haro objects \citep[see e.g.][]{2010ApJ...719.1565G}.   

 \subsection{Non-thermal emission of bow shocks near the Galactic centre}
 \label{subsection_nonthermal}
 
 The non-thermal emission of Sgr A* in the radio/sub-mm regime arises due to synchrotron radiation from 
relativistic electrons, and is highly variable on timescales ranging from minutes to days and months. 
The pericentre passage of the DSO was predicted to lead to an increase in radio/sub-mm emission in two 
ways: by an increase in accretion of matter into the central black hole, and by interactions with the hot 
plasma near the black hole leading to the formation of a bow-shock. The bow-shock interaction with the 
accretion flow would lead to the acceleration of electrons to relativistic energies, producing synchrotron 
emission peaking at $\sim$1\,GHz \citep{narayan2012,sadowski2013,crumley2013}. 
Depending on the models used, this excess flux density, which scales linearly with the size of the cloud, 
ranges from 0.02-22\,Jy. For a pure cloud scenario, \citet{sadowski2013} predict fluxes of 1.4--22\,Jy at 
1.4\,GHz for a cross-section of 10$^{30}\,{\rm cm}^{2}$ while \citet{shcherbakov2014} predict lower fluxes of $\sim$0.3\,Jy 
for a magnetically arrested cloud model of smaller cross-section ($\sim10^{29}\,{\rm cm}^{2}$). 
For the case of a bow-shock arising from the wind driven by a star, \citet{crumley2013} predict a 
radio flux density of 0.02\,Jy. 

\begin{figure*}
  \centering
\begin{tabular}{cc}  
  \includegraphics[width=0.5\textwidth]{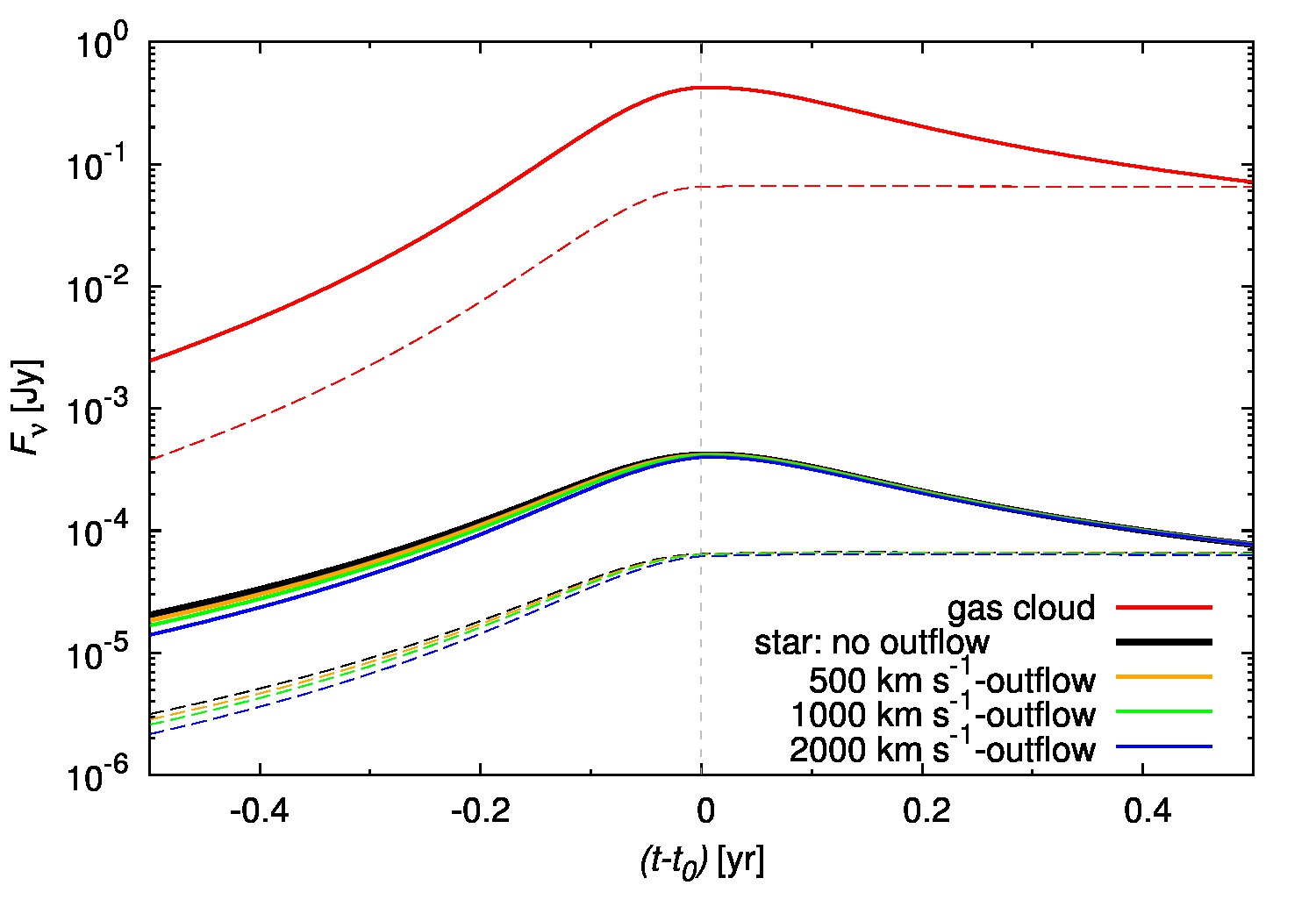}  &   \includegraphics[width=0.5\textwidth]{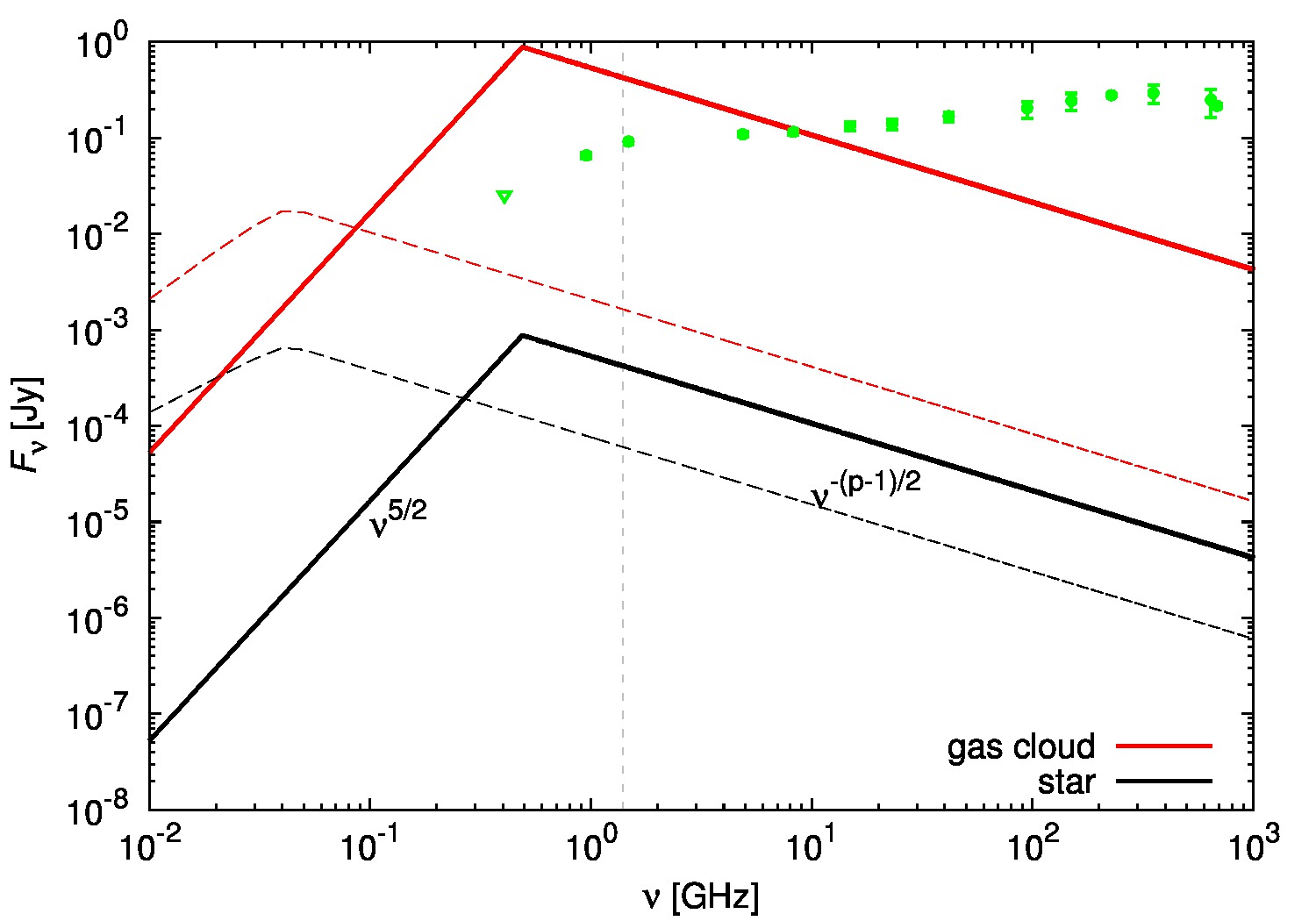}
\end{tabular}  
  \caption{\textit{Left}: Light curve profile at $1.4\,\rm{GHz}$ for the non-thermal emission of a bow shock formed by a star or a core-less gas cloud moving supersonically at low Mach numbers close to the Galactic centre. In the calculation we considered both the plowing limit of synchrotron emission (solid lines) as well as the local model (dashed lines). Different outflow velocities were also taken into account for the passages of a star (see the legend). The vertical line denotes the peribothron passage.  \textit{Right}: Spectrum profiles of synchrotron emission for a star and a cloud. The vertical line represents the frequency of $1.4\,{\rm GHz}$ considered in the calculation of the light curve. The high-energy power-law spectrum of accelerated electrons has a slope of $p=2.4$ in our calculations, in accordance with the simulations of \citet{sadowski2013}. The green points represent the measurements of the intrinsic flux of Sgr~A* (see the text for references). The bow-shock spectral profile and the spectrum of Sgr~A* have a different spectral index for $\nu\gtrsim 1\,{\rm GHz}$. The distance to the Galactic centre was taken to be $8\,{\rm kpc}$.}
  \label{fig_light_curve}
\end{figure*}

Here we revisit calculations of synchrotron emission to compare the scenarios of a star and a cloud for updated orbital elements \citep{2015ApJ...800..125V} and our model set-up (ambient medium and parameters of a young star). Moreover, we assess scenarios with different outflows from the centre. Following the theory outlined in \citet{crumley2013} and \citet{sadowski2013} \citep[see also][]{ryb} we compute the synchrotron flux around the peribothron passage at a fixed frequency of $1.4\,\rm{GHz}$ (light curve; see Fig. \ref{fig_light_curve}, left panel) and subsequently we calculate the spectrum of the passage for the peribothron epoch taking into account synchrotron self-absorption (Fig. \ref{fig_light_curve}, right panel). For the light curve we also compare two possible limits of synchrotron emission: \textit{plowing model} (where all accelerated electrons are kept in the shocked region and radiate in the shocked magnetic field; solid lines) and \textit{local model} (where accelerated electrons leave the shocked region and radiate in the unshocked magnetic field; dashed lines). For both scenarios, a star and a gas cloud, the flux peaks close to the peribothron, with the peak flux of $\sim 0.4\,{\rm Jy}$ for the gas cloud with the fixed cross-section of $A_{{\rm cloud}}=\pi \times 10^{30}\,{\rm cm}^{2}$ as in \citet{crumley2013} and the maximum flux of $\sim 0.45\,{\rm mJy}$ for the stellar model in this work. The difference is as much as three orders of magnitude, which is also apparent in the spectrum in Fig. \ref{fig_light_curve} (right panel), where the non-thermal flux associated with the star is below the intrinsic emission of Sgr~A* across all frequencies for which reliable measurements were performed \citep[observed flux denoted by green points for particular frequencies was collected from][]{1976MNRAS.177..319D,2000A&A...362..113F,2003ApJ...586L..29Z,2008ApJ...682..373M}. This is also consistent with the findings of \citet{crumley2013}. We also find a small difference for different outflow velocities in the pre-peribothron phase, with the smaller flux for the stronger outflow. This difference is, however, negligible and in the post-peribothron phase the light curves have the same profile for all outflow velocities considered here.       

Long-term observations of Sgr A* in the radio/sub-mm regime in 2013 and 2014 have not detected any 
significant increase in flux density beyond the intrinsic variability of Sgr A* \citep{2013ATel.5025....1B,bower2015,park2015}. There was also no significant change in the spectral index reported, as would be expected by differences in spectra between intrinsic flux of Sgr~A* and that of a bow shock, see the right panel of Fig. \ref{fig_light_curve} for comparison. \citet{bower2015} estimate an upper size limit of $\sim2\times10^{29}{\rm cm}^{2}$ to the DSO 
based on the absence of bow-shock emission at 1.4\,GHz. 
\citet{peri2015} discuss the possibility of non-thermal radio emission by electrons accelerated in a bow-shock 
of runaway/wind-blowing stars. Both the synchrotron and inverse-Compton mechanisms are viable candidates 
depending on the (largely uncertain) magnetic field intensity in the bow-shock region. The latter process is more 
likely to be relevant for the production of high-energy photons. However in this case 
the expected signal is generally weak compared to the infrared band and many bow-shock stellar sources 
exhibit no detectable radio emission. In particular, observed peribothron passages of short-period stars S2 and S102 did not produce any flare (within uncertainties) above the quiescent emission of Sgr~A* that would be typical of synchrotron emission associated with a stellar bow shock (with the power-law spectral distribution $S\propto \nu^\alpha$, where $\alpha<0$ \citep[, see light curves at 15 and 23 GHz in][]{2004AJ....127.3399H}. However, \citet{2015arXiv150906251G} analyse the possibility of the detection of non-thermal bow-shock radiation and the synchrotron emission associated with the bow shock of S2 star should be detectable for larger stellar mass-loss rates of the order of $10^{-5}\,M_{\odot}\rm{yr^{-1}}$ and faster stellar winds of the order of $10^3\,\rm{km\,s^{-1}}$.

In conclusion, the lack of enhancement in the radio domain of the flaring activity during the 2014 DSO peribothron passage turns out to be consistent with the compact stellar scenario, where the bow-shock non-thermal emission associated with the moving body is expected to be weak, rather than a core-less gas cloud, where a more significant increase has been expected (see Fig. \ref{fig_light_curve} for a comparison of both scenarios).

\section{Discussion}
\label{sec_discussion}

The basic feature studied in this paper is the formation of temporal asymmetry of the variation of bow-shock properties along the orbit of a star moving through a spherically symmetric outflow. Even if the ambient medium obeys perfect spherical symmetry of temperature and density profiles and the interaction of the star with its environment is approximately axially symmetric \citep{1996ApJ...459L..31W}, the coupled system of the star-Galactic centre behaves highly asymmetrically when the post-pericentre and the pre-pericentre phases of the orbit are compared for different outflow velocities.

 Although the presented toy model uses several assumptions (see section \ref{sec_model_setup}) that may be violated to a certain extent in the complex Galactic centre region, it enables one to carry out a fast calculation of the basic trends of bow-shock features, namely the size and the orientation of the bow shock, change of its velocity profile, the Doppler contribution to its line emission, and the shocked layer density profile as functions of distance and time. The most important part of the analysis can be mainly performed by comparing the observed hydrogen line emission or free-free continuum maps with emission maps calculated from this model. 

Since the first observations of the stellar proper motions in the S-cluster \citep{1996Natur.383..415E,1997MNRAS.284..576E}, orbits of several stars have been well-constrained \citep{2009ApJ...692.1075G}. Given the condition that the bow-shock emission is resolved, the observed signal could be studied and investigated for the basic signs of asymmetry. The upcoming interferometric observations with the 6-baseline interferometer GRAVITY in NIR K-band \citep{2011Msngr.143...16E} may provide the evidence for these trends of stellar bow shock evolution, especially for the case of the post-peribothron evolution of the DSO and the pericentre passages of S2 and S102 stars, which have orbital periods of the order of 10 years.       

The current monitoring of the DSO infrared source is ideally suited for testing the properties of the ambient medium. The source size based on the analysis of the $\rm{Br}\gamma$ emission line maps \citep{2015ApJ...800..125V} is limited to the region of $\lesssim 20\,{\rm mas}$, which is consistent with the outflow (see the temporal evolution of the stagnation radius in Fig. \ref{fig_stagnation_radius_alpha}). However, the source has been monitored only close to its pericentre passage so far. If the bow shock contributes to the observed emission of the DSO and the ambient outflow is non-negligible, we expect the following trends during its post-peribothron evolution:

\begin{itemize}
  \item the bow shock size increases and the source reaches the point along its orbit where the minimum of the ratio $v_{\rm rel}/v_{\rm w}$ occurs,
  \item at this point the bow shock reaches its maximum size (see the stagnation radius evolution in Fig. \ref{fig_stagnation_radius_alpha}); for the DSO source and an outflow velocity of $2000\,{\rm km s}^{-1}$ the maximum occurs at year $\sim 2019$, for  $v_{{\rm a}}=1000\,{\rm km s}^{-1}$ at the epoch of $\sim 2030$, and for $v_{{\rm a}}=500\,{\rm km s}^{-1}$ at $\sim 2060$, 
  \item the stronger the spherical outflow, the sooner the source reaches this domain after the peribothron, 
  \item this phase is also characterised by a different velocity, density, and emission profiles (see Figs. \ref{fig_shell_velocity_profiles}, \ref{fig_surface_density}, and \ref{fig_emission_profs}, respectively). It is especially notable that at the maximum of the bow shock size the emissivity drops downstream more slowly than before or after this phase,
  \item the broadening of the bow shock during the post-pericentre phase causes a drop in surface density, which can help to detect the central star, unless it is further shielded by the circumstellar dusty envelope and a disc with accretion streams. 
\end{itemize} 

It is not yet clear if a stable bow shock structure has developed in case of the DSO source. There can be different sources for its observed hydrogen emission lines, such as the accretion funnels from a circumstellar disc, as is discussed in \citet{2015ApJ...800..125V} in detail. However, even if the outlined trends corresponding to the stellar source--outflow interaction are not observed, it could have one or more of the following implications for the ambient medium and the source itself:

\begin{itemize}
  \item there is no significant gas outflow from Sgr~A*, which would be in contradiction with the analysis of X3/X7 comet-shaped sources and the minicavity formation theory,
  \item or outflow from Sgr~A* is not isotropic,
  \item or the outflow is launched at larger distances due to the stellar winds of massive OB/WR stars,
  \item DSO is a very compact source and/or it has no significant stellar wind (which has implications on the type of the central star), which would cause the density along the shocked layer to be low and the resulting emissivity of the bow shock below the detection limit. Consequently, the contribution of the shocked gas layer  to the line width would be negligible, as well,
  \item the ambient medium around the Galactic centre is more diluted than assumed, see eq. \eqref{eq_density}, which causes the ambient pressure to decrease. Hence, the bow-shock emission also drops, see eq. \eqref{eq_emission_measure},
  \item  if the ambient medium contains inhomogeneities of the length-scale $H$, which is of the order of $H/R_0 \approx 1$, it will lead to the degeneracy in the observed asymmetry, i.e. it will be difficult to disentangle the outflow and the inhomogeneity contribution. Moreover, these inhomogeneities would increase the density gradient locally, leading to the intrinsic asymmetry of the bow shock, for which the generalised solution of \citet{2000ApJ...532..400W} applies.       
\end{itemize}

 A stellar bow shock is also susceptible to several hydrodynamic instabilities, namely
\begin{itemize}
  \item Rayleigh-Taylor (RT) instability due to centrifugal acceleration of the flow,
  \item Kelvin-Helmholtz (KH) instability due to relative velocity shear between the forward and the reverse shock layers,
  \item Radiatively cooled thin shells (with cooling parameter $\chi=t_{\rm{cool}}/t_{\rm{dyn}} \ll 1$) are also susceptible to non-linear thin-shell instability \citep{1994ApJ...428..186V} and transverse acceleration instability \citep{1993A&A...267..155D}.
\end{itemize}

Moreover, \citet{1996ApJ...461..927D} investigate the stability of stellar bow shocks using linear stability analysis. They find that the bow-shock stability in the thin-shell limit depends on the single parameter $\alpha$, which is the ratio of the relative velocity of a star with respect to the ambient medium and the stellar wind velocity, $v_{{\rm rel}}/v_{{\rm w}}$. Bow shocks with $v_{{\rm rel}}/v_{{\rm w}} \ll 1$ are more stable than those with $v_{{\rm rel}}/v_{{\rm w}} \gg 1$. Based on this result bow shocks associated with stellar sources around the SMBH are expected to be more stable near the apobothron, where the ratio $\alpha$ is smaller in comparison with the peribothron value, where the ratio reaches $v_{{\rm rel}}/v_{{\rm w}}\sim 35$ (right panel of Fig. \ref{fig_stagnation_radius_alpha}).

On the other hand, there are several mechanisms that inhibit the growth of hydrodynamic instabilities, namely,
 \begin{itemize}
  \item[(a)] the presence of a warm interstellar medium ($\sim 8\,000$--$10\,000\,\rm{K}$) \citep{2012A&A...548A.113D}\,,
  \item[(b)] ionising radiation from an external source \citep{2014MNRAS.437..843G}\,,
  \item[(c)] interstellar magnetic field \citep{2014A&A...561A.152V}.
\end{itemize}
All of these factors play a role in the Galactic centre region and may contribute to the apparent stability of several observed bow-shock sources.

  Taking into consideration the implications of the detection of the bow-shock asymmetrical evolution or its absence, it is worthwhile observing the DSO and the S-stars as well as prominent bow-shock sources along at least part of their orbits to search for the changes typical of a wind-wind bow-shock interaction and its evolution along the orbit.

The change of the line width of the DSO source, specifically related to Br$\gamma$ emission lines, on its way towards the pericentre was attributed to the tidal stretching of the gas cloud. It has been claimed that this is proof of the unbound nature of the source \citep{2015ApJ...798..111P}. However, the calculations presented in section \ref{subsection_Doppler} show that qualitatively the same increase is caused by the line-of-sight component of the shell velocity field due to the projection of the elliptical orbit. Combined with the possible contribution of the winds and accretion funnels \citep{2015ApJ...800..125V}, the compact stellar model of the DSO can alternatively explain the observed characteristics. 

 It is also useful to compare the mass of the shell and that of the putative star associated with the DSO. Spectral decomposition puts an upper on the bolometric luminosity of the DSO, $L_{\rm{DSO}}\lesssim 30\,L_{\odot}$ \citep{2013A&A...551A..18E,2015ApJ...800..125V}, which automatically constrains the mass as well as the radius of the star, $M_{\rm{DSO}}\lesssim 3\,M_{\odot}$ and $R_{\rm{DSO}}\lesssim 10 R_{\odot}$, respectively (see \citet{2015arXiv150700237Z} for the analysis and discussion). The mass of the shell $M_{\rm{shell}}$ may be simply estimated from the integral over 4$\pi$ sterad: $M_{\rm{shell}}=\int_0^{R_{0}} \int_0^{4\pi} \rho_{\rm{w}}R^{2}\mathrm{d}\Omega=\dot{M}_{\rm{w}}R_0/v_{\rm{w}}$. For the stellar parameters, $\dot{m}_{\rm{w}}=10^{-8}\,M_{\odot}\,{\rm yr}^{-1}$ and $v_{\rm{w}}=200\,{\rm km\,s^{-1}}$, the shell mass lies in the range $\sim 10^{-10}\,M_{\odot}$ up to $10^{-7}\,M_{\odot}$, being smaller at the peribothron. The mass estimate of the shell should be considered as a lower limit since it takes into account only the stellar wind contribution; see \citet{2012A&A...541A...1M} and \citet{2012A&A...548A.113D} for a more detailed discussion for $\alpha$ Ori -- Betelgeuse. Finally, the total mass deposited along the orbit during one orbital period is of the order of $\dot{m}_{\rm{w}}P_{\rm{orb}}\approx 10^{-8}\,M_{\odot}\,{\rm yr}^{-1}\times 100\,\rm{yr}=10^{-6}\,M_{\odot}$. 

Observations of several infrared-excess, dusty objects provided the evidence for ongoing star-formation in the very central region of our Galaxy, as seen in the analysis of mid-infrared sources in \citet{2004ApJ...602..760E} and \citet{2004A&A...425..529M,2005A&A...443..163M}, as well as the recent detection of proplyd-like bow-shock sources using radio continuum observations \citep{2015ApJ...801L..26Y}.
Although the \textit{in situ} star formation and the presence of young stars of T Tauri and LL Ori type was originally considered rather difficult due to the strong shearing tidal field in the central region \citep{1993ApJ...408..496M}, it has been shown that the infall of low-angular-momentum clumps from the circumnuclear disk makes it possible due to the tidal compression at the peribothron \citep{2014MNRAS.444.1205J} or strong shocks provided by the collision of streams in tidally stretched filaments \citep{1998MNRAS.294...35S}. These processes cause the gas to overcome the critical Jeans density and leads to the formation of a cluster of protostellar cores. These pre-main-sequence stellar sources are characterized by continuing accretion of matter from a circumstellar envelope or disc that is accompanied by outflows, which is also indicated by recent radio and infrared observations of the inner $\sim 2$--$5\,{\rm pc}$ from Sgr~A* \citep{2015arXiv150505177Y}. These can drive shocks into the ambient medium and combined with the potential supersonic motion of a star, a bow shock is formed. Therefore a model of the DSO as a young star that is in the phase of both accretion and outflows is fully consistent with these findings.

\section{Conclusions}
\label{sec_conclusions}

We studied the effect of a quasi-spherical outflow from the Galactic centre on moving stellar sources that develop bow-shock structures due to their supersonic motion with respect to the surrounding ambient medium. We showed that the density and the velocity of this outflow can be constrained through the change of bow-shock characteristics along the orbit. These include mainly the stagnation radius, shell velocity profile, density profile, emission measure maps, and the Doppler broadening of emission lines. We described the changes both qualitatively and quantitatively with respect to the distance of the source from the SMBH as well as along the bow-shock shell.

We demonstrated that the temporal evolution of the bow shock of a stellar source is fully symmetric with respect to the peribothron passage when the effect of the ambient outflow is negligible in comparison with the effect of the motion of the source. For stronger outflows the asymmetry develops between the pre-peribothron and post-peribothron orbital phases. We applied the model on the orbital configuration and parameters of the DSO source assuming to be a wind-blowing star rather than a core-less ionised cloud. The line-of-sight component of the shell velocity field can contribute to the observed increase of the emission line width. The calculations of the bow shock evolution show that the observations of the DSO source during its post-pericentre phase are equally valuable as the pre-pericentre observations due to the potential asymmetry of both parts. 

The results concerning the evolution of bow-shock characteristics along the orbit are relevant for all fast-moving stars closely bound to the Galactic centre. They can also serve as a motivation for further, more detailed numerical studies.

\section*{Acknowledgments}

We thank two anonymous referees for critical, constructive comments that helped to improve the manuscript. We are grateful to Rhys Taylor for helpful comments and discussion.
The research leading to these results received funding from
the European Union Seventh Framework Program (FP7/2007-2013) under grant agreement no 312789 -- Strong gravity: Probing Strong Gravity by Black Holes Across the Range of Masses.  This work was supported in part by the Deutsche
Forschungsgemeinschaft (DFG) via the Cologne Bonn Graduate School (BCGS), the Max Planck Society through the International Max Planck Research School (IMPRS) for Astronomy and Astrophysics, as well as special funds through the University
of Cologne and SFB 956--Conditions and Impact of Star Formation. M. Zaja\v{c}ek and B. Shahzamanian are members of the IMPRS. Part of this work was supported by fruitful discussions with members of
the Czech-DAAD collaboration program ``Effects of Albert Einstein's Theory of General Relativity Revealed with the Instruments of European Southern Observatory" and the Czech Science Foundation -- DFG project No. 13-00070J.

\bibliographystyle{mn2e} 
\bibliography{zajacek} 

\appendix

\section[]{Equilibrium of a stellar bow shock along the orbit}
\label{appendix_eq}

\begin{figure}
  \centering
  \includegraphics[width=0.5\textwidth]{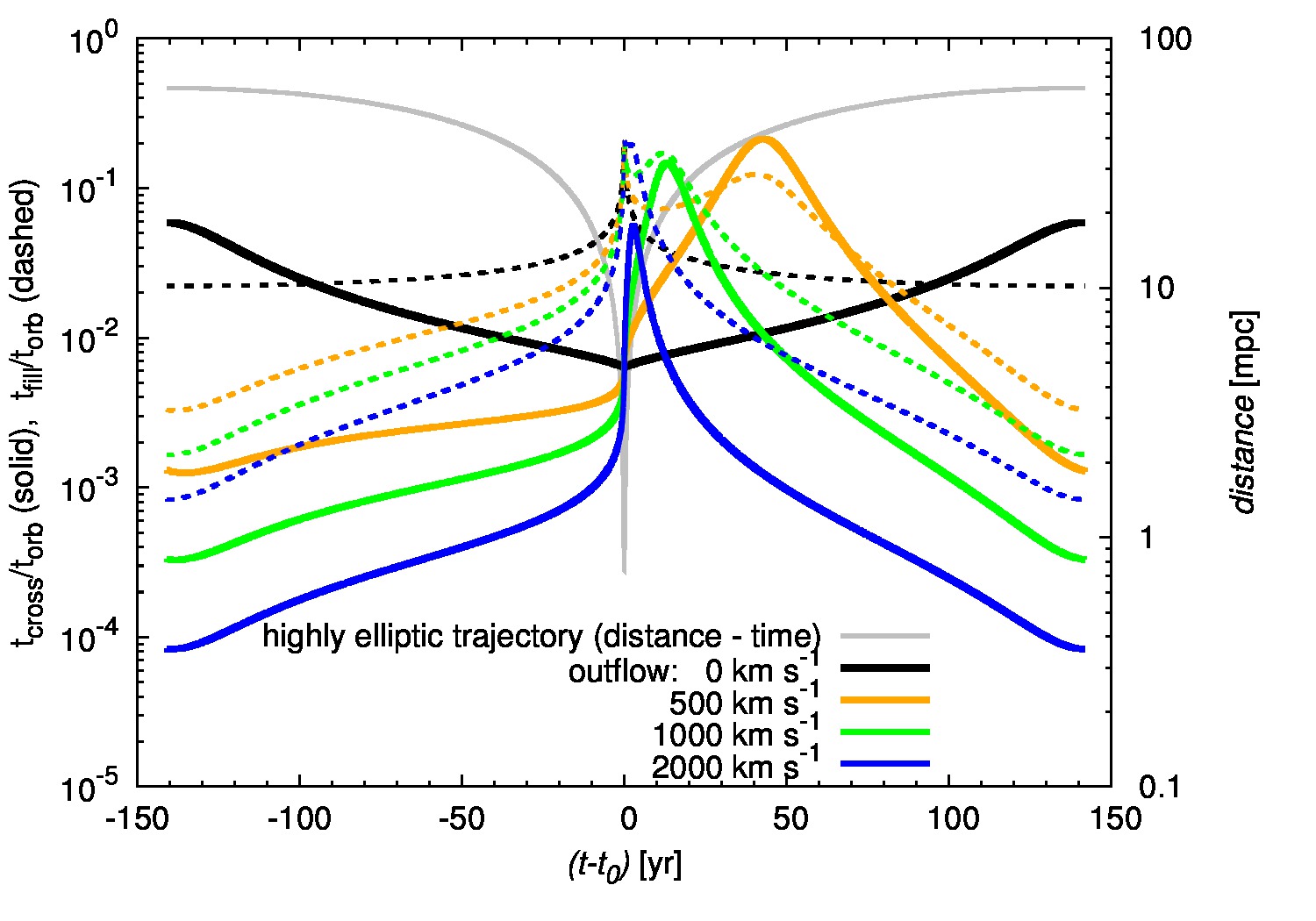}  
  \caption{The evolution of ratios $t_{\rm{cross}}/t_{\rm{orb}}$ (solid lines; eqs. \eqref{eq_crossing_timescale} and \eqref{eq_orbtimescale}) and $t_{\rm{fill}}/t_{\rm{orb}}$ (dashed lines; eqs. \eqref{eq_wind_filling} and \eqref{eq_orbtimescale}) for different outflow velocities (see the legend) along the whole orbit (left vertical axis). The grey line represents the temporal evolution of the distance of the star from the SMBH for the assumed highly elliptical orbit (right vertical axis).} 
  \label{fig_timescales}
\end{figure}

 An equilibrium, momentum-supported bow-shock model applied in this work clearly has its limitations. An equilibrium can be reached only when the crossing timescale, 
 \begin{equation}
 t_{\rm{cross}}=R_0/v_{\rm{rel}}\,,
 \label{eq_crossing_timescale}
 \end{equation}
  which determines the ability of the ambient medium to shape the bow shock, and the wind-filling timescale,
  \begin{equation}
    t_{\rm{fill}}=R_0/v_{\rm{w}}\,,
   \label{eq_wind_filling}
   \end{equation} 
     which is related to the ability of the stellar wind to fill the bow-shock cavity, are smaller than the orbital timescale,
    \begin{equation} 
      t_{\rm{orb}}=D/v_{\rm{orb}}\,,
     \label{eq_orbtimescale}
     \end{equation} 
       where $D$ is the distance of the star from the SMBH and $v_{\rm{orb}}$ is the stellar orbital velocity. We plot the ratio of these timescales, $t_{\rm{cross}}/t_{\rm{orb}}$ (solid lines) and $t_{\rm{fill}}/t_{\rm{orb}}$ (dashed lines), for different velocities of the central outflow, see Fig. \ref{fig_timescales}. For the terminal stellar wind velocity of $v_{\rm{w}}=200\,{\rm km\,s}^{-1}$ and the mass-loss rate of $\dot{m}_{\rm{w}}=10^{-8}\,{\rm M_{\odot}\,yr^{-1}}$, the ratios are smaller than unity along the whole orbit, so the equilibrium bow-shock structure should be approximately reached and it may serve as a basis for our qualitative studies of temporal asymmetry.

\section[]{Velocity divergence and density gradient at a fixed position of the bow-shock source from the Galactic centre}
\label{appendix_divgrad}

\begin{figure}
  \centering 
  \includegraphics[width=0.5\textwidth]{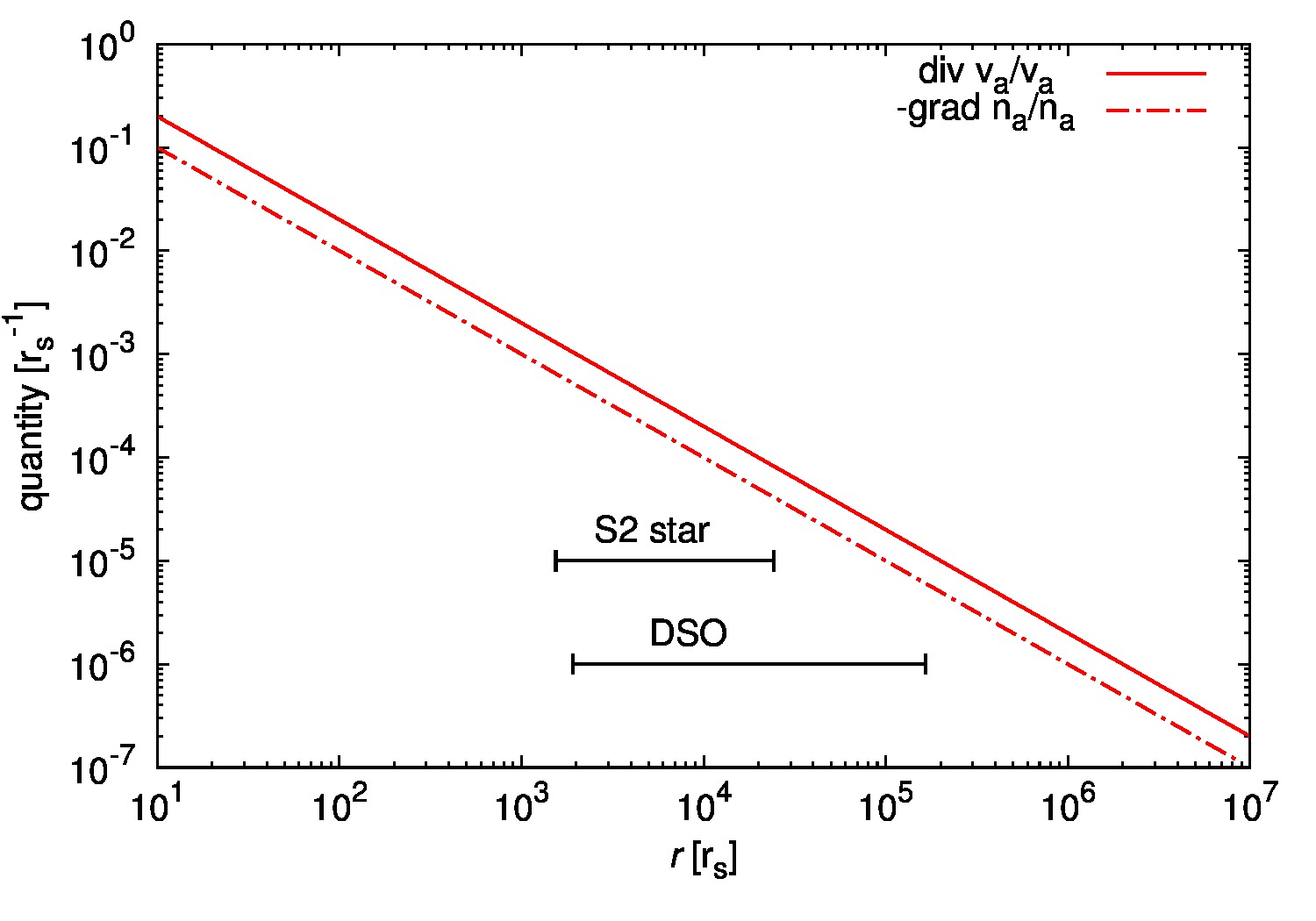}
  \caption{Profile of the divergence of the spherical central outflow divided by the velocity magnitude with respect to the distance from the Galactic centre (solid line). The distance is expressed in Schwarzschild radii and the ratio of the divergence and the velocity magnitude is in $r_{\rm{s}}^{-1}$. In the same plot we show the distance profile of the density gradient divided by the density at the given distance (dash-dotted line; we take its negative value for the plot). The horizontal lines in both panels mark the distance range for S2 star and the DSO source and are positioned at arbitrary values of the vertical axis.}
  \label{fig_appendix_divgrad}
\end{figure}

For the spherical central ambient outflow we have $\mathbf{v}_{{\rm a}}=(v_{r}, v_{\theta}, v_{\phi})=(v_{r},0,0)$. The divergence of this flow in spherical coordinates yields,

\begin{align*}
  \nabla \cdot \mathbf{v_{{\rm a}}} &= \frac{1}{r^2}\frac{\mathrm{d}}{\mathrm{d}r}(r^2 v_{r})\\
                                   &= \frac{2v_{r}}{r_{{\rm s}}}\left(\frac{r}{r_{{\rm s}}}\right)^{-1}\,, 
\end{align*}
where $v_{r}$ is constant. The velocity divergence as a function of the distance from the Galactic centre (expressed in Schwarzschild radii) is plotted in Fig. \ref{fig_appendix_divgrad} (solid line). The values along the vertical axis represent the velocity divergence divided by the velocity magnitude with respect to the distance of Schwarzschild radius. For the distance range of our interest (S-cluster) the values of the divergence are of the order of $10^{-3}$  up to $10^{-5}\,r_{\rm{s}}^{-1}$.

Using the spherical density profile, eq. \eqref{eq_density}, we compute the density gradient as a function of the distance, 

\begin{align*}
  \nabla  n_{\rm a} &= -\frac{n_{{\rm a}}^0}{r_{\rm s}}\left(\frac{r}{r_{{\rm s}}}\right)^{-2}\,. 
\end{align*}
The density gradient profile (divided by the density at the given distance) is plotted in Fig. \ref{fig_appendix_divgrad} (dash-dotted line) and the corresponding values for the S-cluster distance range are of the order of $10^{-4}$  up to $10^{-5}\,r_{\rm{s}}^{-1}$. 

\end{document}